\documentclass[journal=mamobx,manuscript=article,layout=twocolumn]{achemso}

\usepackage{chemformula} 
\usepackage[T1]{fontenc} 
\usepackage{natbib}
\usepackage{widetext}

\author{Hsiao-Ping Hsu}
\affiliation{Max-Planck-Institut f\"ur Polymerforschung, Ackermannweg 10, 55128, Mainz, Germany}
\email{hsu@mpip-mainz.mpg.de}
\author{Kurt Kremer}
\affiliation{Max-Planck-Institut f\"ur Polymerforschung, Ackermannweg 10, 55128, Mainz, Germany}
\email{kremer@mpip-mainz.mpg.de}

\title[]{Clustering of entanglement points in highly strained polymer melts}

\abbreviations{}
\date{}
\newcommand*{\sometext}{
   Polymer melts undergoing large deformation by uniaxial elongation
are studied by molecular dynamics simulations of bead-spring chains in melts.
Applying a primitive path analysis to strongly deformed polymer melts, the
role of topological constrains in highly entangled polymer melts
is investigated and quantified. We show that the
over-all, large scale conformations of the primitive paths (PPs) of stretched chains
follow affine deformation while the number and the distribution of entanglement points along the PPs
do not. Right after deformation, PPs of chains retract in
both directions parallel and perpendicular to the elongation.
Upon further relaxation we observe a long-lived clustering of entanglement points.
Together with the delayed relaxation time this leads to a metastable inhomogeneous
distribution of topological constraints in the melts.
}

\let\oldmaketitle\maketitle
\let\maketitle\relax

\begin{document}

\twocolumn[
\begin{@twocolumnfalse}
\oldmaketitle
\begin{abstract}
\sometext
\end{abstract}
\end{@twocolumnfalse}
]

\section{Introduction}

Materials based on polymeric compounds are essential for many areas of modern technology. Their outstanding properties originate from chemical details determining local interactions and the fact that polymers typically are long chain molecules. The latter, generic aspect is of interest for the present work, where the fact that chains cannot cross through each other plays the central role.
Such complex topological constraints resulting in entanglements
play an essential role for dynamical and rheological properties
of polymer melts~\cite{deGennes1979,Doi1986,Mcleish2003,Everaers2004,Sukumaran2005,Padding2011,Qin2012,Terentjev2015}. 
In the linear viscoelastic regime these are well described by reptation theory (the tube model)~\cite{deGennes1979,Doi1980,Doi1986},
and have been confirmed by many simulations~\cite{Kremer1988,Kremer1990,Paul1991,Wittmer1992,Kremer1992,Kopf1997,Puetz2000,Harmandaris2003,Tsolou2005,Hsu2016,Hsu2017} and experiments~\cite{Fetters1994,Wischnewski2002,Wischnewski2006,Graessley2008,Herrmann2012}. 
However, despite this remarkable achievement, a precise definition of an entanglement within the reptation concept still is lacking and attempts to include multi-chain effects analytically were of very limited success only~\cite{Iwata1989}.
Everaers et al~\cite{Everaers2004} introduced the primitive path analysis (PPA)
based on the concept of Edwards' tube model~\cite{Edwards1967} to identify the  
backbone of the tube, i.e. the primitive path (PP) of each polymer chain 
in a melt, and applied it to bead-spring chains~\cite{Kremer1990}.
A detailed discussion regarding self-entanglements, local self-knot effect, and finite-size effect is
given in Refs.~\citenum{Sukumaran2005,Moreira2015}. 
Kr\"oger et al.~\cite{Kroeger2005,Shanbhag2007} subsequently developed the Z-code and its updated version, Z1-code, 
where PPs are treated as infinitely thin and tensionless lines 
using geometrical operations rather than
multi-bead chains. 
Once all PPs, represented by the shortest (optimal) paths (SPs), are obtained the entanglement molecular weight $N_e$ is simply obtained from the length of the path. Alternatively one can also count the kinks of adjacent segments along the chain to get a first rough estimate.
Tzoumanekas et al.~\cite{Tzoumanekas2006} implemented another algorithm, CReTA, which is
capable of reducing the atomistic configuration of a computational polymer sample
to a network of corresponding PPs where the topological constraints are conserved.
Other similar methods dealing with entanglements in polymer melts are given in 
Refs.~\citenum{Shanbhag2005,Zhou2005,Hoy2007}.
Thus, there is a number of methods available, which analyze the role of topological
constraints, i.e. the role of entanglements in a polymer melt. Some of them,
such as PPA are reproducing the entanglement length quantitatively correctly~\cite{Everaers2004,Everaers2012,Hsu2016}.

Beyond the analysis of constraints themselves there are many discussions in the literature~\cite{Terentjev2015}
dealing with the motion of polymers due to entanglement constraints
and the time such constraints last. For entangled
chains, processes of contour length fluctuation
(CLF)~\cite{Doi1983,Likhtman2002,Likhtman2014} of the PPs,
and constraint release (CR)~\cite{Likhtman2002, Klein1978, Daoud1979, Rubinstein1988}
also contribute in addition to pure reptation, i.e. the motion of the polymer along
the hypothetical reptation tube, and have to be considered.
However, it is not yet clear in which way all the above mentioned concepts can be employed for
describing the behavior of polymer melts in the non-linear viscoelastic regime. 
To shed light on this problem, we recently have started to approach this problem from the computational side~\cite{Hsu2018a,Hsu2018b}. We have compared the predictions of chain conformations 
given by the Doi-Edwards tube model~\cite{Doi1986} 
and its extensions based on the Graham-Likhtman-McLeish-Milner (GLaMM) tube model~\cite{Graham2003} to extensive simulations of polymer melts 
in the non-linear viscoelastic regime. For this we decided to 
concentrate on isochoric elongation of polymer melts.
The chain retraction mechanism, as predicted by the GLaMM concept, which sets in right after deformation, has
been investigated with contradicting results, based on both
experimental and simulation data~\cite{Blanchard2005,Graham2006,Wang2017,Xu2018,Zhou2018,Hsu2018b} 
while our recent results for longer chains support this general scheme~\cite{Hsu2018b}.
Although primitive path network models can account for viscosity
changes in entangled polymers upon elongational and shear 
flow~\cite{Nielsen2009,Yaoita2012,Yaoita2011,Bhattacharjee2017} 
in the linear viscoelastic regime, they seem to fail for strain rates $\dot{\varepsilon}$ 
faster than the inverse Rouse time $\tau_{R,N}$ of the whole chains. 
For example, the monotonic shear thinning behavior of entangled polymer melts
in elongational flow observed from the experiment~\cite{Bach2003} at $\dot{\varepsilon}>1/\tau_{R,N}$ cannot be 
described by simulations using the primitive path network models~\cite{Yaoita2011}.
The present work intends to make a contribution to filling a gap in this research field.

Here we start from well equilibrated and highly entangled polymer melts
composed of weakly semiflexible bead-spring chains at a monomer
density $\rho=0.85\sigma^{-3}$, prepared by a new, 
efficient hierarchical methodology~\cite{Zhang2014,Moreira2015,Hsu2016}.
These melts are subject to strong deformation by isochoric elongation in 
the non-linear rheological regime. Following this deformation we investigate in detail
the subsequent relaxation.
Applying the primitive path analysis (PPA)~\cite{Everaers2004,Sukumaran2005} 
to strongly deformed polymer melts in the non-linear viscoelastic regime,
we recently have shown that the force pattern along
the primitive paths (PPs) qualitatively match that of the corresponding
original paths (OPs)~\cite{Hsu2018a}. 
This indicates that the conformations of OPs of chains
within fuzzy tube-like regimes are well represented by their corresponding PPs.
We directly can relate sign switches of the
tension force to kinks,``effective entanglement points'', 
of high curvature along the PPs.
Based on these findings we use 
the relaxation of PPs of deformed chains in a
melt in order to shed some light on the role of topological
constraints for the relaxation of highly entangled deformed polymer melts.

The outline of this paper is as follows: in the next section, we summarize
the main features about our model and the simulation techniques. In the
third section we describe the conformational changes, the characteristics 
of topological constraints, and the stress relaxation of the deformed polymer melts 
followed by our conclusions in section four.

\section{Model and simulation methods}

\subsection{Melts of bead-spring chains with a weak bending stiffness}
  
  For our simulations, a polymer melt consisting of
$n_c$ polymer chains of chain size $N$, i.e. the number of monomers,
is described by a 
standard bead-spring model~\cite{Kremer1990} at a monomer density
$\rho=0.85 \sigma^{-3}=n_cN/V$ where $\sigma=1$ is the unit of length and the size
of a monomer. $V=L_x L_y L_z$ is the volume of the simulation box
with three orthogonal linear dimensions, $L_x$, $L_y$, and $L_z$.
Any pair of bonded and non-bonded monomers located at a distance $r$ apart interact
via a shifted, purely repulsive Lennard-Jones (LJ) 
potential~\cite{Bird1977,Ceperly1978,Bishop1982,Kremer1986} $U_{\rm LJ}(r)$,
\begin{eqnarray}
&U_{\rm LJ}(r) & \nonumber \\
&=& \left\{\begin{array}{ll}
 4\epsilon\left[\left(\frac{\sigma}{r}\right)^{12}-\left(\frac{\sigma}{r}\right)^{6}
+\frac{1}{4} \right] & , \, r \le r_{\rm cut} \\
0 &, \, r>r_{\rm cut} 
\end{array} \right.  \nonumber \\
\end{eqnarray}
\noindent
where $\epsilon$ is the energy unit of the pairwise interaction,
and $r_{\rm cut}=2^{1/6}\sigma$ is
the cutoff in the minimum of the potential such that force and potential are zero 
at $r_{\rm cut}$.
Any pair of bonded monomers interacts via the finitely extensible nonlinear elastic (FENE)
binding potential~\cite{Bird1977} $U_{\rm FENE}(r)$,
\begin{eqnarray}
&U_{\rm FENE}(r) & \nonumber \\ 
&= &\left\{ 
\begin{array}{ll}
 -\frac{k}{2}R_0^2 \ln \left[1-\left(\frac{r}{R_0}\right)^2 \right] & , r \le R_0 \\
 \infty &,  r>R_0 
\end{array} \right . \nonumber \\ &&
\end{eqnarray}
where $k=30 \epsilon/\sigma^2$ is the force constant and $R_0=1.5 \sigma$ is
the maximum value of bond length. These Lennard-Jones units $\sigma$ and $\epsilon$
also provide a natural time definition via $\tau=\sigma \sqrt{m/\epsilon}$ where
$m=1$ is the mass of the particles.
In addition, a weak bond-bending potential~\cite{Everaers2004} $U_{\rm BEND}(\theta)$ with a chain
stiffness parameter $k_\theta$ is introduced,
\begin{equation}
  U_{\rm BEND}(\theta)=k_\theta(1-\cos \theta) \
\end{equation}
where $\theta$ is the angle between two subsequent bonds, i.e.,
$\theta = \cos^{-1}\left(\frac{{\bf b}_j \cdot {\bf b}_{j+1}}{\mid {\bf b}_j\mid \mid {\bf b}_{j+1} \mid}
\right )$ and ${\bf b}_j={\bf r}_{j}-{\bf r}_{j-1}$ being the bond vector between
monomers $j$ and $j-1$ along the chain.
Choosing $k_\theta=1.5 \epsilon$, where chains become weekly semiflexible,
the mean square radius of gyration for unperturbed (i.e. fully equilibrated) chains in a melt is
$\langle R_g^2 \rangle_0 \approx \langle R_e^2 \rangle_0/6 \approx 0.484N\ell_b^2$
where $\langle R_e^2 \rangle_0$ is the mean square end-to-end distance 
and, $\ell_b=\langle {\bf b}^2 \rangle_0^{1/2} \approx 0.964\sigma$ is the 
root-mean-square (rms) bond length.
This gives $\langle R_{e}^2 \rangle_0$, $\langle R_g^2 \rangle_0 \propto N^{2\nu}$ with the Flory exponent 
$\nu=1/2$. Here the average $\langle \ldots \rangle_0$ denotes the average over all chains and over all
independent configurations of unperturbed melts. 
Polymer chains in a melt behave nearly as Gaussian chains~\cite{Hsu2016}.
For the above parameters the corresponding entanglement length 
$N_e=N_{e,PPA}^{(0)} \approx 28$, estimated both from 
the plateau modulus $G_N^0 = (4/5)\rho k_BT/N_e$ as well 
as from the primitive path analysis (PPA)~\cite{Everaers2004,Sukumaran2005,Moreira2015,Hsu2016}.
The Rouse relaxation time of a subchain of entanglement length $N_e$, and of the overall chain of size $N$ 
is $\tau_e=\tau_0N_e^2$ and $\tau_{R,N}=\tau_0N^2$,
respectively. Here the characteristic time prefactor $\tau_0 \approx 2.89\tau$ is determined from
the estimate of the mean square displacement of inner monomers~\cite{Hsu2016}
for $n_c=1000$ chains containing $18 \le Z \equiv N/N_e \le 72$ entanglements, $g_1(t)$.
We here focus on the above mentioned cases
and use the molecular dynamics simulations package~\cite{Espressopp} ESPResSo++ for all runs with simulation time
step set to $\Delta t =0.01\tau$. The temperature is kept constant 
($T=1\epsilon/k_B$, $k_B$ being the Boltzmann factor) through a Langevin thermostat with a
weak friction constant $\Gamma=0.5\tau^{-1}$.

\subsection{Deformation mechanism: isochoric elongation}

We start from fully equilibrated, highly entangled polymer melts 
(unperturbed polymer melts) originally in a cubic simulation box, i.e. $L_x=L_y=L_z=L_0$,
with periodic boundary conditions along the three orthogonal directions. 
We apply a simple ``isochoric elongation'' deformation mechanism.
At each elongation step the whole simulation box is instantaneously
stretched by a factor of $1.02$ along the $x$-direction,
contracted in the $y$-, $z$-directions by a factor of $1/ \sqrt{1.02}$
such that the box size $V=L_xL_yL_z=L_0^3$ is kept as a constant.
This deformation step is so small that it does not induce any
instabilities in the simulation. Then the system is given a short time to relax. By that an
average fixed strain rate $\dot{\varepsilon}$ in the range 
$\tau_{R,N}^{-1}<\dot{\varepsilon}<\tau_{e}^{-1}$ is introduced, namely
\begin{equation}
\dot{\varepsilon}=\frac{d\varepsilon}{dt}=\frac{1}{L}\frac{dL}{dt}=\frac{d\ln(L/L_0)}{dt} \,,
\end{equation}
i.e., {$L=L_0\exp(\dot{\varepsilon} t)$}. We set the strain rate $\dot{\varepsilon}$ to 
$\dot{\varepsilon}\tau_{R,N}=77$
that is $\dot{\varepsilon}\tau_e=77(N_e/N)^2$ (e.g. $\dot{\varepsilon}\tau_e\approx 0.015$ for $N=2000$)
at each elongation step. To obtain this deformation rate the system can relax for $(0.02\tau_{R,N}/77)\tau$ 
between two such steps.
Thus one could expect that locally, i.e. below a few $N_e$ $(\approx \sqrt{1/0.015}N_e \approx 8.2N_e$), 
the chain can fully relax during the
elongation while globally, the chain follows the affine deformation implying that
the macroscopic chain deformation follows the macroscopic strain.
 Altogether $81$ elongation steps are performed, leading to a total strain of~\cite{Hsu2018a,Hsu2018b}
$\lambda \approx 5.0$. As a result the deformed polymer melts are deep in the non-linear viscoelastic regime.

\subsection{Primitive path analysis (PPA) and definition of significant kinks}

According to the original PPA procedure~\cite{Everaers2004,Sukumaran2005},
the two ends of all the chains in the polymer melt are fixed at their actual position in space. Then
the intrachain excluded volume as well as the bond bending interaction are
switched off, i.e., $U_{\rm LJ}^{\rm (intra)}(r)=U_{\rm BEND}(\theta)=0$, while
interchain excluded volume interactions remain in order to prevent 
bond crossing and to preserve interchain topological constraints. Intrachain topological constraints were shown to be insignificant~\cite{Sukumaran2005} within the error bars achievable here.
Then the temperature is set to zero, so that the chains contract to
the shortest paths between the two ends, observing all interchain topological constraints.
Practically, the temperature is set to $T=0.001\epsilon/k_B$ (close to zero), 
and the basic time step $\Delta t$ is reduced to $0.006\tau$.
The friction constant is set to $\Gamma=20 \tau^{-1}$ during the first $10^3$ MD
steps, and $\Gamma=0.5\tau^{-1}$ afterwards~\cite{Sukumaran2005,Moreira2015,Hsu2016}.
Thus, chains straighten out when the bond springs try to reduce the
average bond length in order to minimize the energy from $\ell_b \approx 0.964 \sigma$ of the OPs to
$\langle b_{PP} \rangle_0=0.31\sigma$ of the PPs for unperturbed polymer melts and
$\langle b_{PP} \rangle_\lambda \approx 0.57 \sigma$ for strongly deformed polymer melts
at $\lambda \approx 5.0$, respectively.

\begin{figure*}[t]
\begin{center}
\includegraphics[width=1.00\textwidth,angle=0]{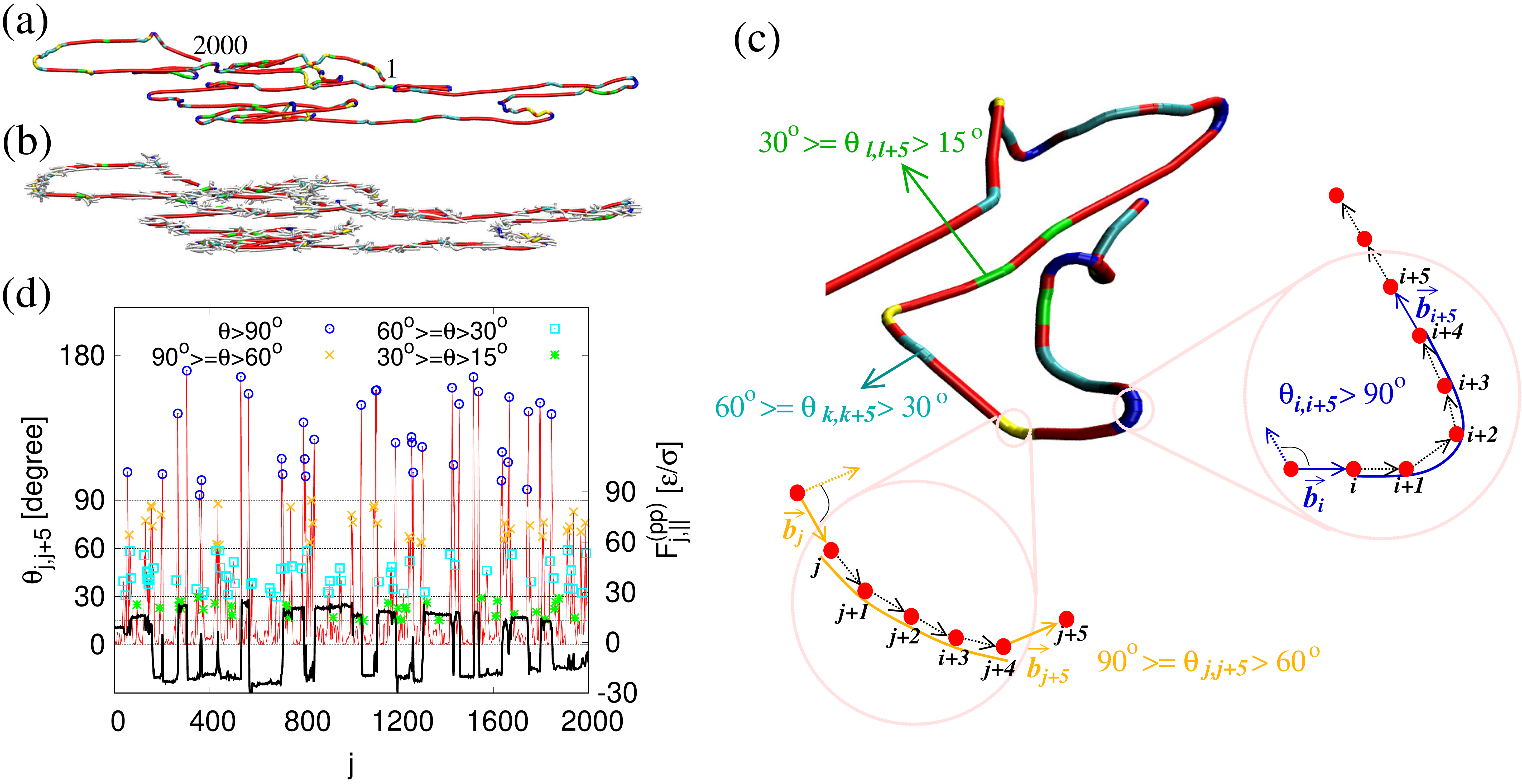}
\caption{(a) Snapshot of the PP of chain $i=251$
of size $N=2000$ in a deformed melt after uniaxial elongation
to $\lambda \approx 5.0$ at a rate of $\dot{\varepsilon}\tau_{R,N}=77$. 
The two ends of the chain are labeled 
by $1$ and $2000$.
(b) Same as in (a) but
short segments of other confining chains close to the selected PP are also included.
(c) Definition of bond angle $\theta_{j,j+5}$ between the $j$th bond
${\bf b}_j$ and the $(j+5)$th bond ${\bf b}_{j+5}$ describing the curvature
along the PP.
(d) Estimates of the bond angle $\theta_{j,j+5}$ plotted against $j$
with $j=2,3,\ldots,N-5$ along the PP as shown in (a).
In (d), local maxima of $\theta_{j,j+5}$ larger than $15^o$ are
marked by different symbols, as indicated. The tension force
$F_{j,||}$ with $j=2,3,\ldots,N$ derived from the FENE potential
($U_{\rm FENE}(r)$) along the PP in the direction parallel
to the stretching direction is shown by the black curve.
Similar results for another chosen chain and the definition of curvature are 
already given in Ref.~\citenum{Hsu2018a} and its supporting information.}
\label{fig-curvature}
\end{center}
\end{figure*}

 A typical snapshot of the PP of one selected chain of size $N=2000$ in a deformed
melt is shown in Figure~\ref{fig-curvature}a. 
The PP consists of straight pieces with relatively sharp kinks
created by the excluded volume interactions with other chains.
The distribution of kinks along the selected chain strongly depends on the surrounding
chains (see Figure~\ref{fig-curvature}b where only short segments
of surrounding chains near the test chain are shown).
Therefore, a direct way of recognizing ``entanglement points'' 
is to analyze the curvature along the PP to identify these kinks.
Since the sharp kinks are still rounded off
due to the resulting short bonds and the remaining interchain
excluded volume, we have chosen the bond angle $\theta_{j,j+5}$ between
bonds {${\bf b}_j$} and {${\bf b}_{j+5}$} for $j=2$, $3$, $\ldots$, $N-5$.
along the chain as shown in Figure~\ref{fig-curvature}c. 
For identifying significant kinks (entanglement points), all local maxima marked by symbols 
along the path denoted by $j$ are sorted into four categories, $\theta_{j,j+5}>90^o$, 
$90^o \ge \theta_{j,j+5} > 60^o$, $60^o \ge \theta_{j,j+5} > 30^o$, and
$30^o \ge \theta_{j,j+5} > 15^o$ (Figure~\ref{fig-curvature}d). 
Comparing to the force pattern along the same PP, 
we define an entanglement point to be located at {${\bf r}_{j+2}$} if the curvature of 
a kink corresponding to~\cite{Hsu2018a} $\theta_{j,j+5} \ge 60^o$. 
There is, however, an ambiguity related to this analysis. While the number of 
topological constraints up to some end effects is strictly conserved, 
their physically relevant number will vary with time. 
When two kinks come very close along the backbone of the constraining chain 
they probably act like one. Because of that we always present results, where we 
count them as one, 
if the distance along the same chain is less than $1 \sigma$.

\section{Simulation results}

\begin{figure*}[thb!]
\begin{center}
(a)\includegraphics[width=0.32\textwidth,angle=270]{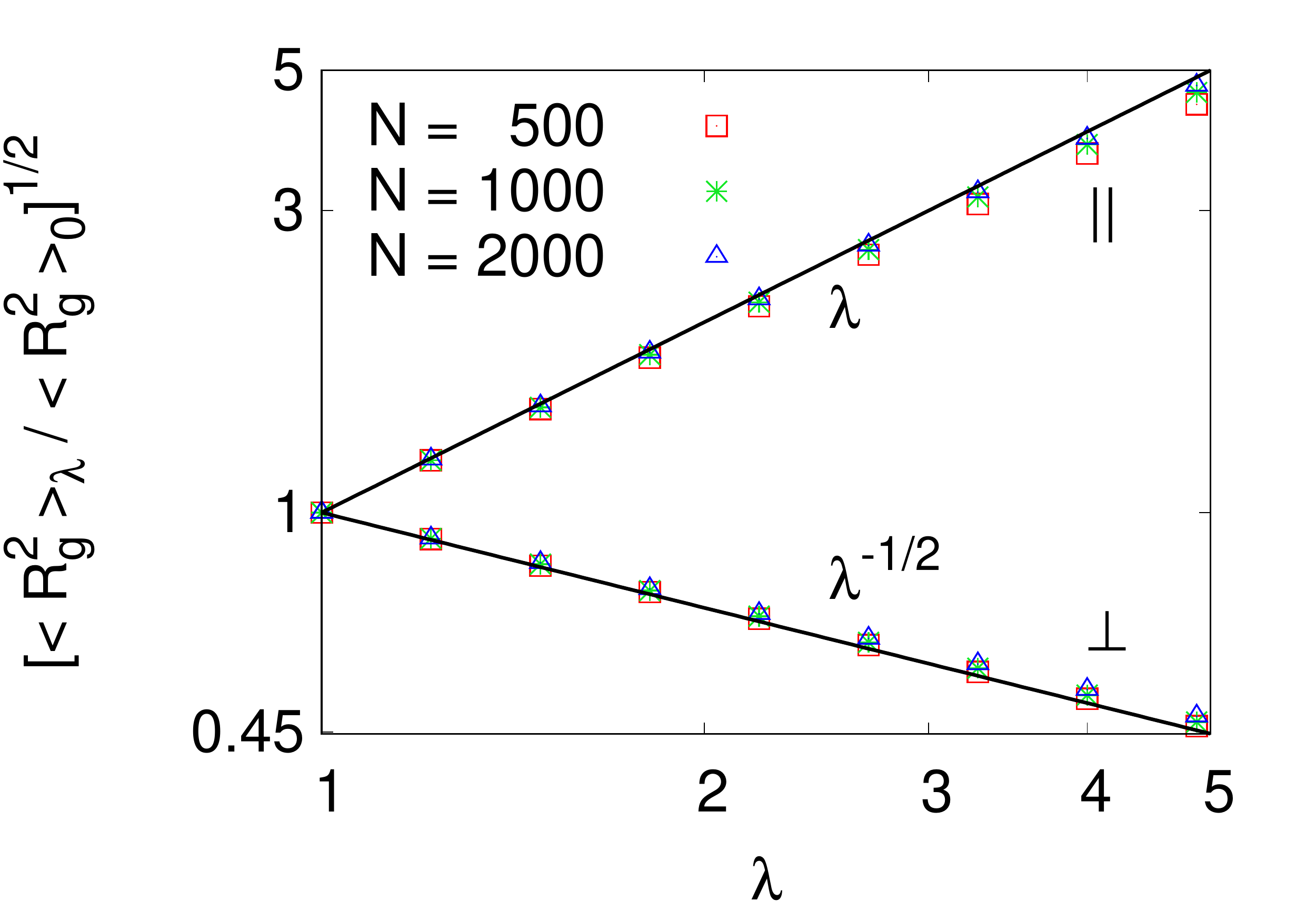}\hspace{0.2truecm}
(b)\includegraphics[width=0.32\textwidth,angle=270]{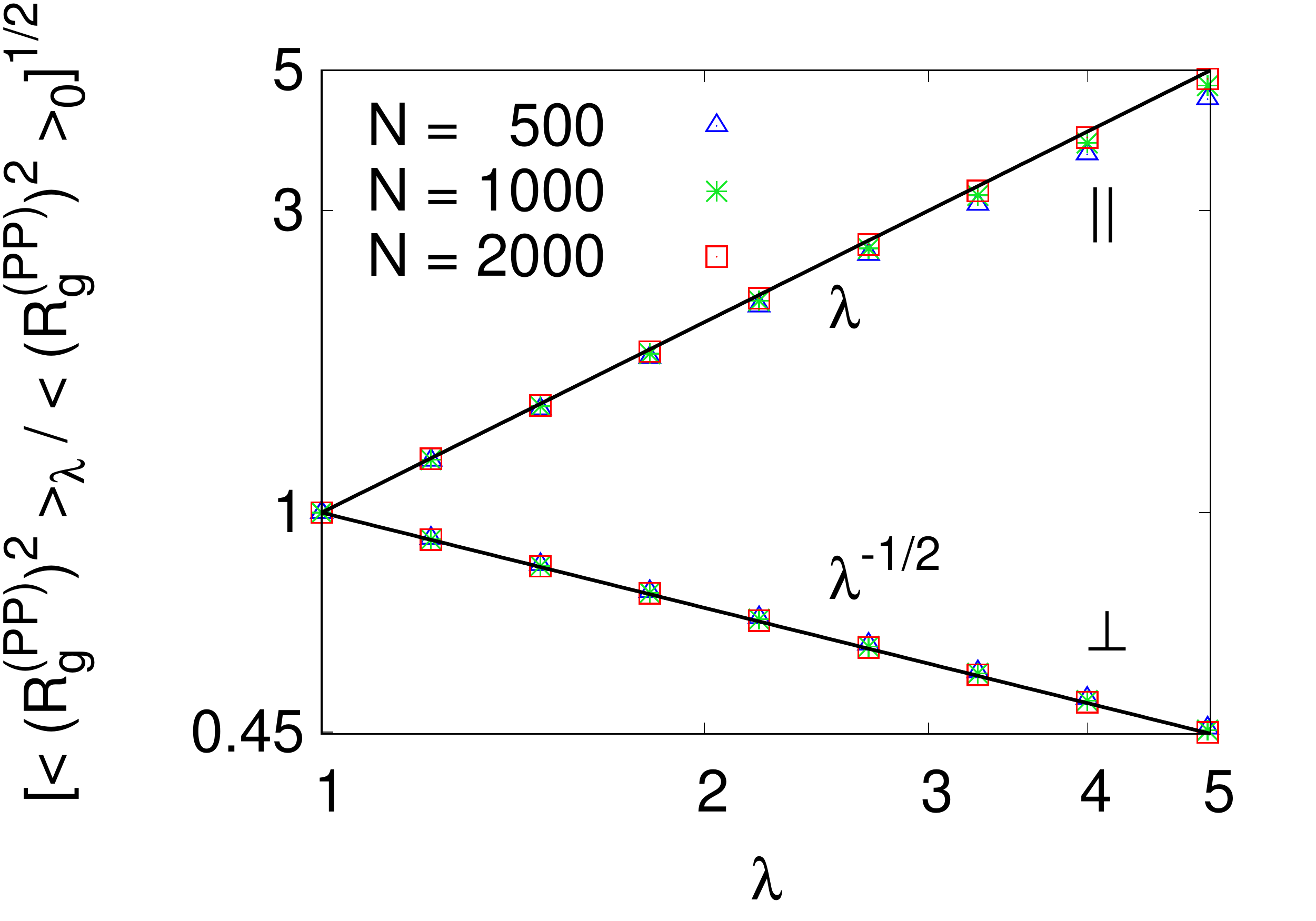}\\
(c)\includegraphics[width=0.32\textwidth,angle=270]{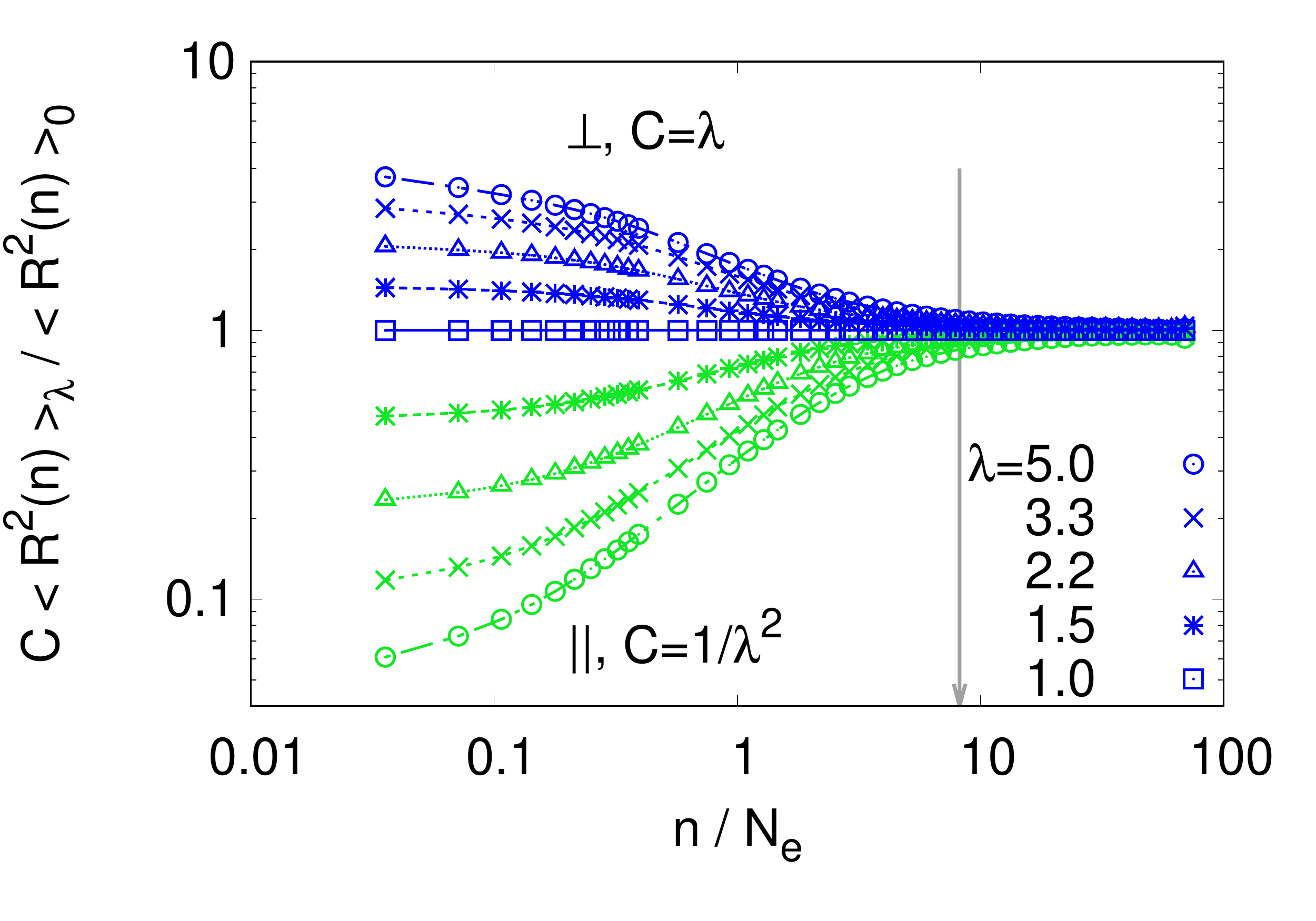}\hspace{0.2truecm}
(d)\includegraphics[width=0.32\textwidth,angle=270]{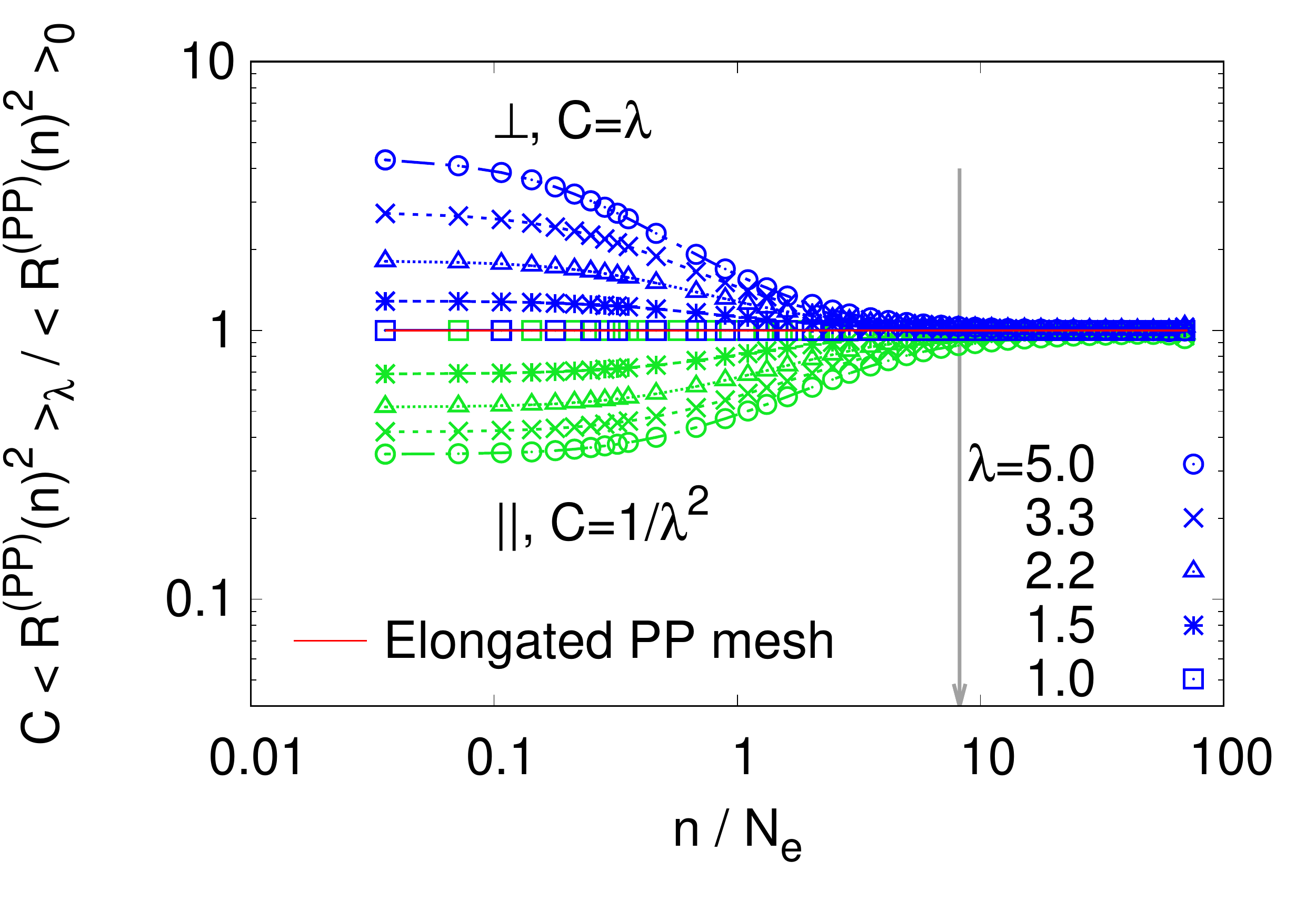}\\
\caption{Rescaled rms radius of gyration,
$[\langle R_g^2 \rangle_\lambda/ \langle R_g^2 \rangle_0]^{1/2}$ for the OPs of chains (a)
and $[\langle (R_g^{(PP)})^2 \rangle_\lambda/ \langle (R_g^{(PP)})^2 \rangle_0]^{1/2}$
for the PPs of chains (b), plotted versus the strain $\lambda$.
Rescaled mean square internal distance,
$C\langle R^2(n) \rangle_\lambda/\langle R^2(n) \rangle_0$ for the OPs of chains (c)
and $C\langle (R^{(PP)}(n))^2 \rangle_\lambda/\langle R^{(PP)}(n)^2 \rangle_0$ for the PPs of chains (d),
plotted versus the rescaled chemical distance $n/N_e$.
Two components in the directions perpendicular ($\perp$) and parallel ($||$) to the direction of stretching
are shown, as indicated.
Data are for elongated polymer melts of
chain sizes $N=500$, $1000$, and $2000$, as indicated in (a)(b), but only of $N=2000$ in (c)(d).
In (a)(b), the expected scaling laws for affine deformation are shown by straight lines.
In (c)(d), five strain values of $\lambda$ are chosen, as indicated, and $n=8.2N_e$ is
pointed out by an arrow (cf. text). Data for the elongated PP mesh are also included in (d) by
a red curve.}
\label{fig-Rgpp-el}
\end{center}
\end{figure*}

  We study three polymer systems each containing $n_c=1000$ 
chains of sizes $N=500$, $1000$, and $2000$, and the corresponding lengths of
the unperturbed simulation box being $L_0/\sigma=83.79$, $105.57$, $133.01$, respectively.
All estimates of physical quantities are taken as averages over all $n_c=1000$ chains 
except if a selected chain in a melt is explicitly mentioned and discussed. 
All $n_c=1000$ chains are labelled by $i=1$, $2$, $\ldots$, $1000$.
For the fully equilibrated 
melts chain conformations are characterized by
the mean square radii of gyration 
$\langle R_g^2(N) \rangle_0/\sigma^2$ of approximately  $221$, 
$438$, and $909$ for the OPs of chains, 
$\langle (R_g^{(PP)}(N))^2 \rangle_0/\sigma^2 \approx 209$, 
$424$, and $895$ for the PPs of chains. 
Correspondingly, the mean square end-to-end distances are
$\langle R_e^2 (N)\rangle_0/\sigma^2=\langle (R_e^{(PP)}(N))^2 \rangle_0/\sigma^2=1329$,
$2575$, and $5354$, for $N=500$, $1000$, and $2000$, respectively.
Results of the two components of the 
rms radius of gyration of the OPs (PPs) in the directions parallel $(||)$ and 
perpendicular $(\perp)$ to the stretching, rescaled to $1/3$ 
and $2/3$ of the estimates of $\langle R_g^2 \rangle_0^{1/2}$ ($\langle (R_g^{(PP)})^2 \rangle_0^{1/2}$)
, respectively, right after deformation
are shown in Figure~\ref{fig-Rgpp-el}.
Under uniaxial elongation, we see that the over-all conformations of the OPs as well as of the PPs deform affinely. Namely,
$\left[\langle R_{g,||}^2 \rangle_\lambda/\langle R_{g,||}^2 \rangle_0 \right ]^{1/2}=
\left[\langle (R_{g,||}^{(PP)})^2 \rangle_\lambda/\langle (R_{g,||}^{(PP)})^2 \rangle_0 \right ]^{1/2}=\lambda$
and $\left[\langle R_{g,\perp}^2 \rangle_\lambda/\langle R_{g,\perp}^2 \rangle_0 \right ]^{1/2}=
\left[\langle (R_{g,\perp}^{(PP)})^2 \rangle_\lambda/\langle (R_{g,\perp}^{(PP)})^2 \rangle_0 \right ]^{1/2}=\lambda^{-1/2}$.
The corresponding scaling laws also hold 
for $\langle R^2_{e,||} \rangle$ and $\langle R^2_{e,\perp} \rangle$, respectively (not shown).
In order to estimate the length scale at which the deformation of chains becomes affine, 
we also include the results of the rescaled internal
mean square distances of the OPs (PPs) right after deformation,
$C\langle R^2(n) \rangle_\lambda/\langle R^2(n) \rangle_0$
($C\langle R^{(PP)}(n)^2 \rangle_\lambda/\langle R^{(PP)}(n)^2 \rangle_0$),
in Figure~\ref{fig-Rgpp-el}. Here $n$ is the
chemical distance between two bonds along the OP (PP) of the same chain,
and $C$ is the scale parameter.   
According to the scaling law for affine deformation,  
$C=\lambda$ and $C=1/\lambda^2$ for the two components in the directions
perpendicular and parallel to the stretching, respectively.
Our data demonstrate that $n\approx 8.2N_e$ 
related to the chosen strain rate $\dot{\varepsilon}$
represents the characteristic length scale along the chains and above which both OPs and PPs of chains deform affinely. 
However, chains obviously cannot fully relax on short length scales, since 
the connectivity and the constraints lead to deviations from this picture. 
This qualitatively compares to the expectation based on the strain rate chosen,
which allows for local but not global relaxation during the deformation. 
For comparison, we also include data for the affinely (instantly) elongated original primitive
path mesh of the unperturbed melt named 
the elongated PP mesh thereafter in Figure~\ref{fig-Rgpp-el}d. 
Our results of the mean square internal distances show
that PPs of chains in the PP mesh generated from the corresponding elongated 
OPs of chains in a melt through the PPA~\cite{Everaers2004,Sukumaran2005}
display a deformation pattern very similar to the one of the OPs. 

\begin{figure*}[thb!]
\begin{center}
\includegraphics[width=1.00\textwidth,angle=0]{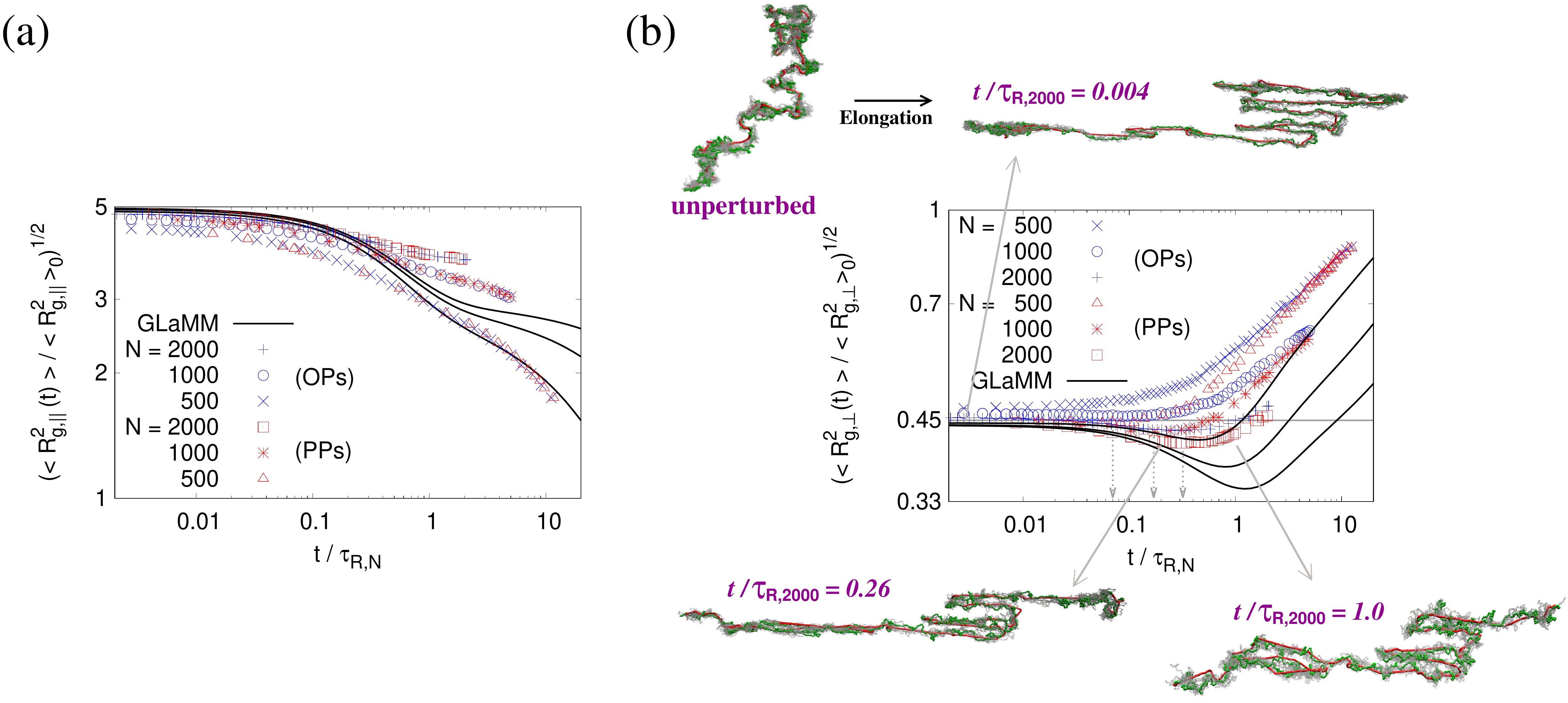}
\caption{Two components of the rescaled rms radius of gyration for
the PPs of chains in the directions
parallel ($||$) (a) and perpendicular ($\perp$) (b) to the stretching,
$[ \langle (R_g^{(PP)})^2(t) \rangle/ \langle (R_g^{(PP)})^2 \rangle_0 ]^{1/2}$,
plotted versus subsequent relaxation times, $t/\tau_{R,N}$, at
$\lambda \approx 5.0$ on a log-log scale. Data are for $n_c=1000$ primitive
paths of chains of sizes $N=500$, $1000$, and $2000$, as indicated.
Data for the OPs of chains are also shown for comparison.
In (b), four typical configuration snapshots of
chain $i=300$ of $N=2000$ are included for better illustration, as indicated.
At a chosen relaxation time $t$, the PP of the selected chain is shown by a
curve in red, the OP is shown by beads in green, and the other $6$ OPs of the same chain
at $t\pm1.32\tau_e$, $t\pm 0.88\tau_e$, and $t\pm 0.44 \tau_e$
with $\tau_e \approx 2266\tau \approx 0.0002\tau_{R,N=2000}$ are also shown by beads in gray.
The minimum occur at $t/\tau_{R,N}\approx 0.07$, $0.17$, and $0.32$ for the PPs of $N=500$,
$1000$, and $2000$, respectively, are pointed out by arrows.
Theoretical predictions from the GLaMM model~\cite{Graham2003,Hsu2018b} for
$Z=72$, $36$, and $18$ are also shown by black curves from top to bottom in (a), and bottom to top in (b)
for comparison.}
\label{fig-Rgpp-rel}
\end{center}
\end{figure*}

\subsection{Conformational change of single chains in deformed melts}

In the non-linear viscoelastic regime, the tube model predicts for the initial relaxation right after deformation an over-damped initial retraction process of the individual chains in both directions parallel and
perpendicular to the stretching direction~\cite{Doi1986,Graham2003}.
Such an immediate chain retraction mechanism has been observed for the time evolution of the rescaled two components of radius of gyration
for the OPs of chains~\cite{Hsu2018a,Hsu2018b} of sizes $N=1000$ and $2000$.
One should also expect a similar behavior for the end-to-end distance of the OPs of chains,
but with larger fluctuation due to the end effect. 
Therefore, we still only focus on the chain conformations 
described by the radius of gyration here.
For the PPs the same holds as shown in Figure~\ref{fig-Rgpp-rel}.
Typical snapshots of a selected chain of
size $N=2000$ in the melt before and after deformation and after different
relaxation times $t/\tau_{R,2000}=0.004$, $0.26$, and $1.0$
are also presented in Figure~\ref{fig-Rgpp-rel}b for better illustration of the conformational
variations. 
In both parallel and perpendicular directions to the stretching direction,
$\langle R_{g,||}^{(PP)}(t)^2 \rangle$ and $\langle R_{g,\perp}^{(PP)}(t)^2 \rangle$
decrease initially with increasing time showing the signature of chain retraction.
$\left[\langle (R_{g,\perp}^{(PP)})^2 \rangle_\lambda/\langle (R_{g,\perp}^{(PP)})^2 \rangle_0 \rangle \right ]^{1/2}$
reaches a deeper minimum compared to the OPs while
parallel to stretching, data for the PPs and OPs, respectively, coincide.
The minima of $(\langle R_{g,\perp}^{(PP)}(t)^2\rangle/\langle R_{g,\perp}^{(PP)}(t)^2\rangle_0)^{1/2}$ 
for PPs also occur slightly later than for OPs~\cite{Hsu2018a}, 
but the difference for the occurrence times becomes negligible within fluctuation as 
the chain size increases
(at $t/\tau_{R,N} \approx 0.30(4)$ (OPs), $0.32(4)$ (PPs) for $N=2000$,
at $t/\tau_{R,N} \approx 0.09(5)$ (OPs), $0.17(4)$ (PPs) for $N=1000$, and at no time (OPs),
$0.07(4)$ (PPs) for $N=500$). 
The minimum becomes deeper with increasing N.
However, the duration of this global retraction process is still
below the predicted longest time scale~\cite{Doi1986, Graham2003}, i.e. the Rouse time of the whole chains
in unperturbed melts, $\tau_{R,N}$. This indicates a rather strong contribution from tension along the PPs.
Setting the two parameters $c_\nu=0.1$ and $R_s=2.0$ in the GLaMM model~\cite{Graham2003}
(see the supplementary material in Ref.~\citenum{Hsu2018b}),
results of the time evolution of the radius gyration perpendicular and parallel to the stretching direction have been
obtained by solving the constitutive equation iteratively in our previous work~\cite{Hsu2018b}.
Obviously, our results for both the OPs and PPs qualitatively capture the signature 
of the initial chain retraction mechanism~\cite{Doi1986, Graham2003} after
a large step elongation while quantitatively, the GLaMM model predicts a significantly
stronger signature of retraction for the same chain size $N$, i.e., the same
number of entanglements $Z=N/N_e$.
Furthermore, the rate of retraction
$(\langle R_{g,||}^{(PP)}(t)^2\rangle/\langle R_{g,||}^{(PP)}(t)^2\rangle_0)^{1/2}$ 
becomes smaller with time for $N=1000$ and $2000$, and eventually 
$(\langle R_{g,||}^{(PP)}(t)^2\rangle/\langle R_{g,||}^{(PP)}(t)^2\rangle_0)^{1/2}$
indicates a turn to an intermediate plateau 
(which is more pronounced for $N=2000$ than for $N=1000$)
showing a substantially delayed conformational relaxation well above and significantly 
earlier than the regime predicted by the GLaMM model.
This relaxation retardation of deformed chains, not accounted for in current  theoretical models, has been attributed to an inhomogeneous 
distribution of entanglement points along the PPs~\cite{Hsu2018a}, and will be discussed later. 
A similar delay has also been observed in the context of rheological experiments of very long, highly entangled polymer chains by several other authors~\cite{Archer1995, Archer2002,Venerus2006}.

\begin{figure*}[th!]
\begin{center}
(a)\includegraphics[width=0.32\textwidth,angle=270]{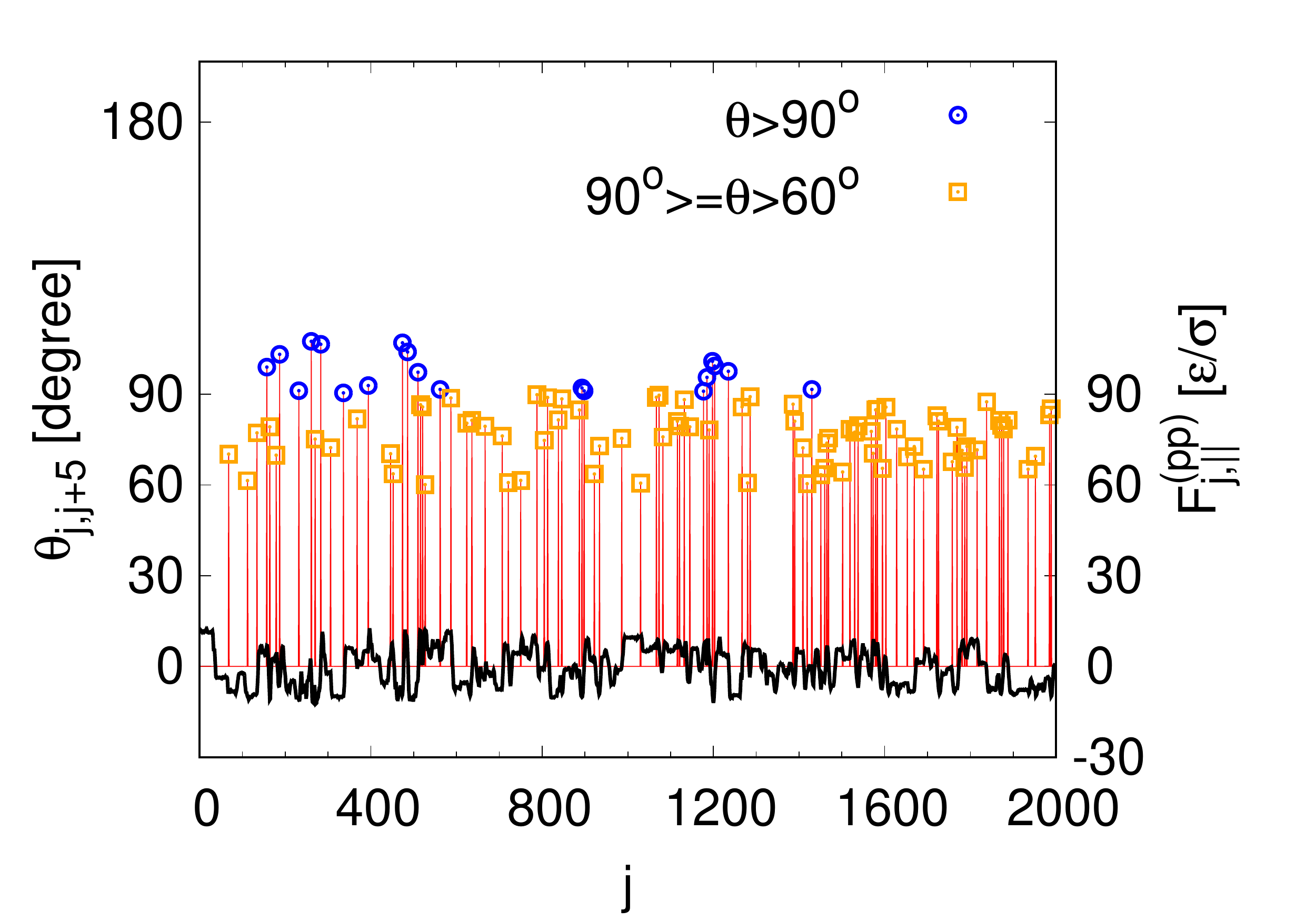}\hspace{0.2truecm}
(b)\includegraphics[width=0.32\textwidth,angle=270]{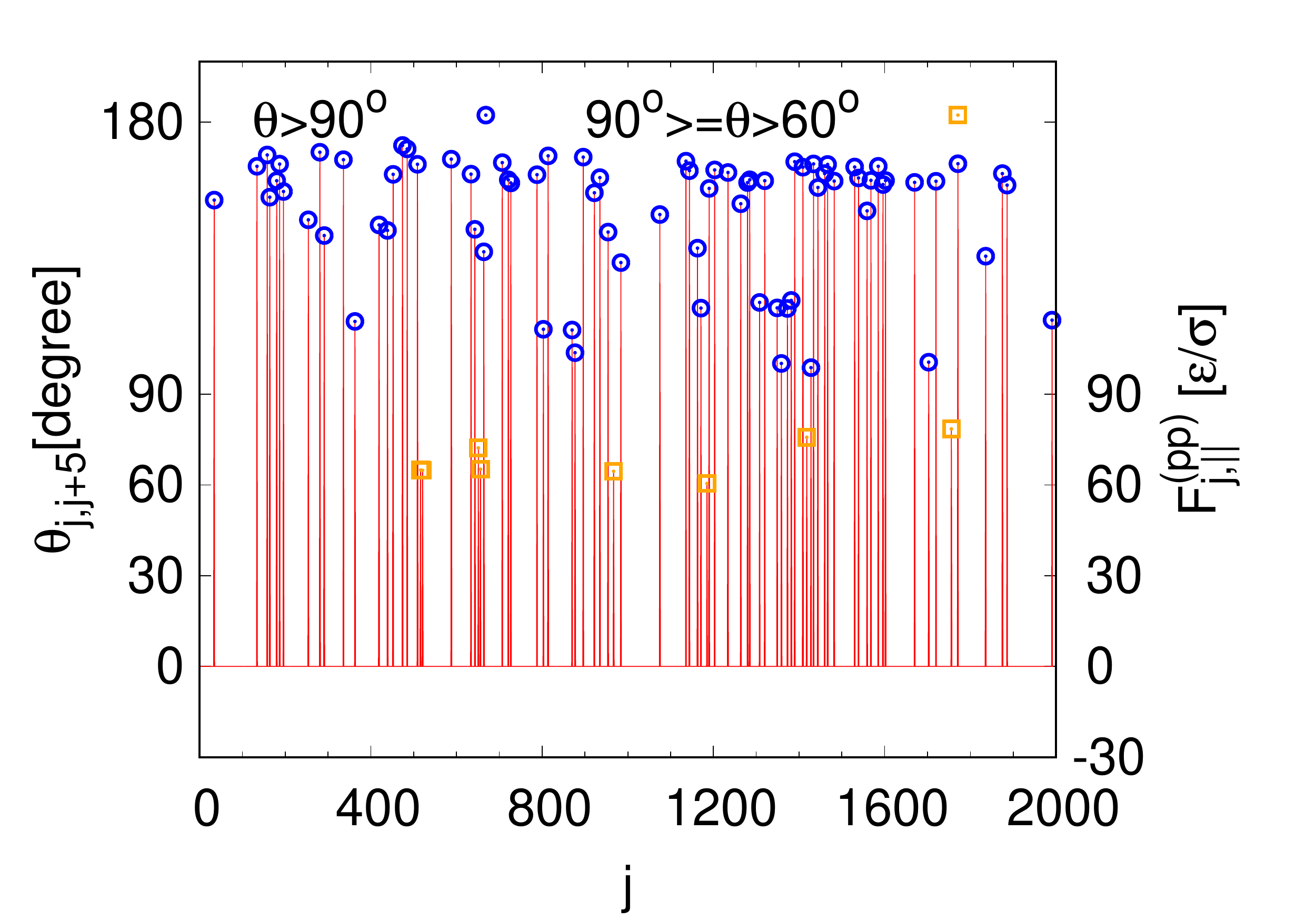}\\
(c)\includegraphics[width=0.32\textwidth,angle=270]{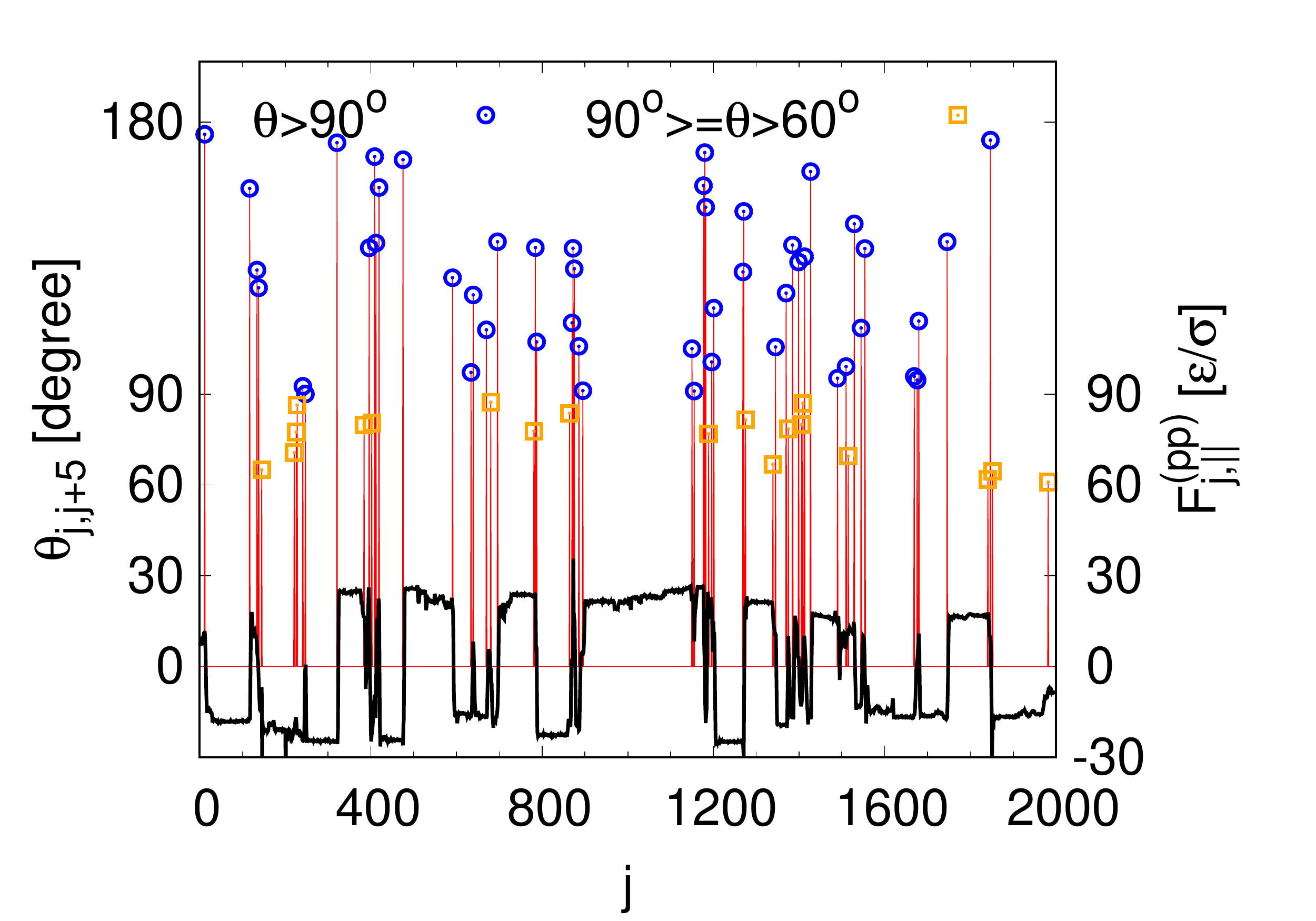}\hspace{0.2truecm}
(d)\includegraphics[width=0.32\textwidth,angle=270]{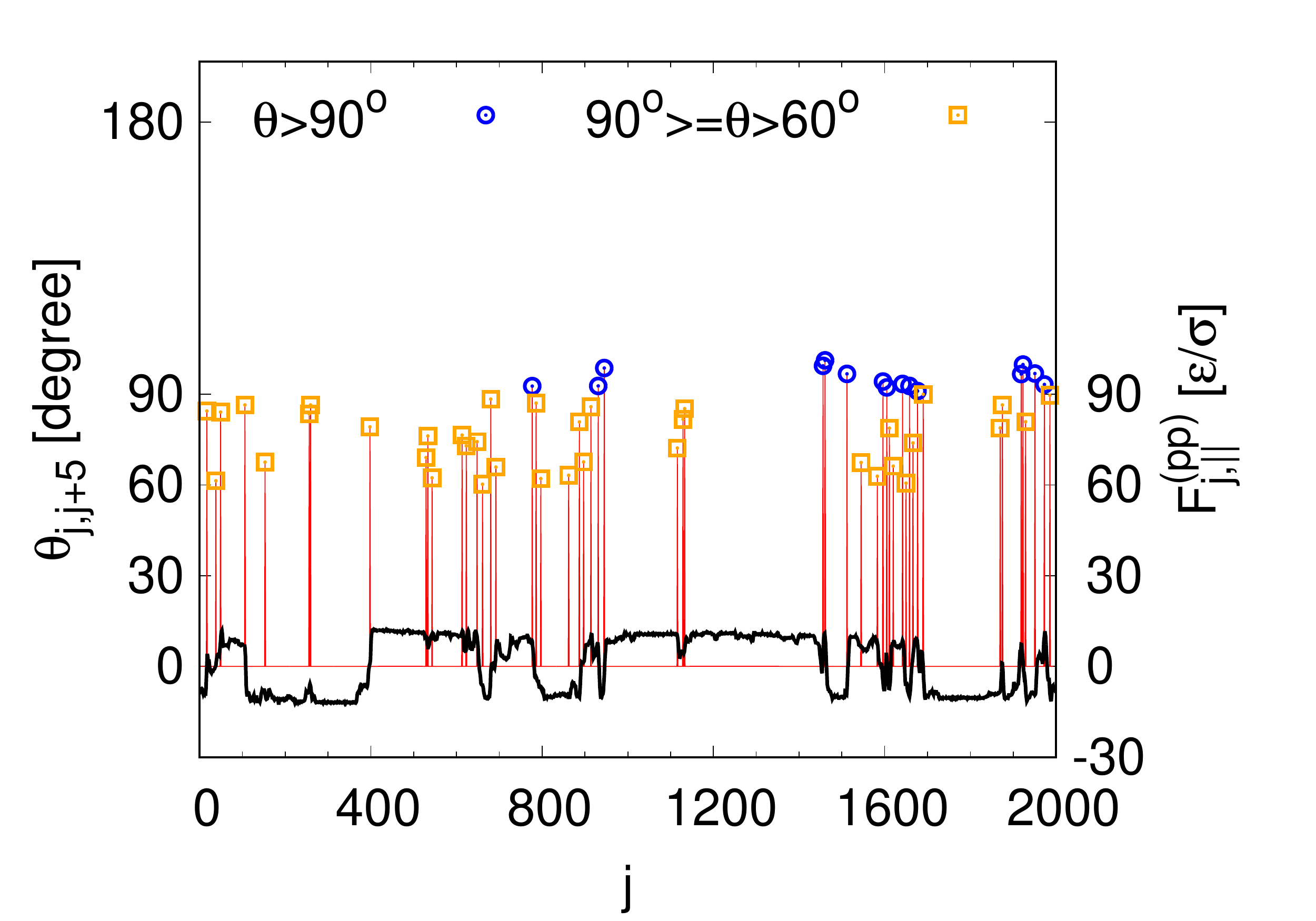}
(e)\includegraphics[width=0.32\textwidth,angle=270]{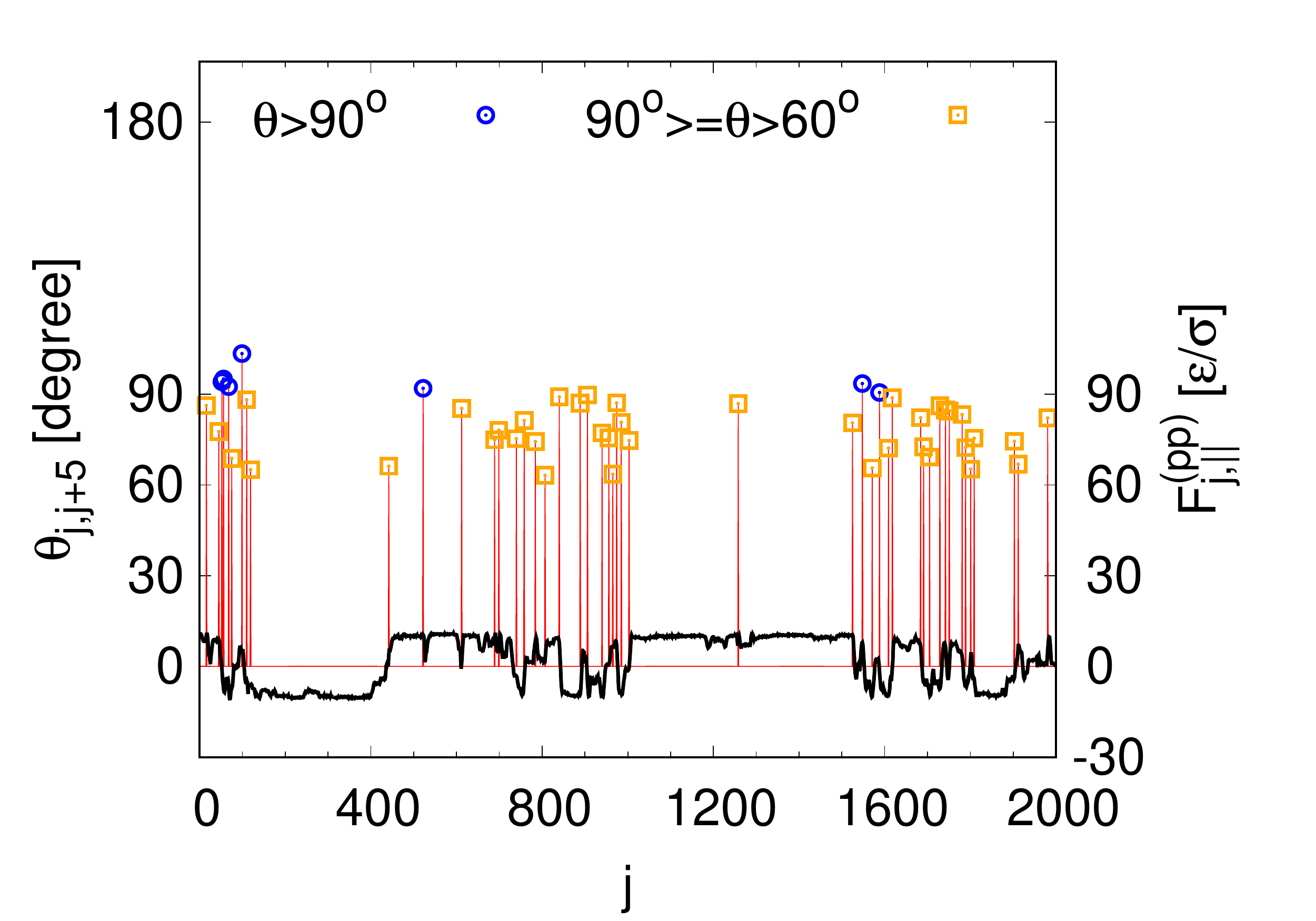}\hspace{0.2truecm}
(f)\includegraphics[width=0.32\textwidth,angle=270]{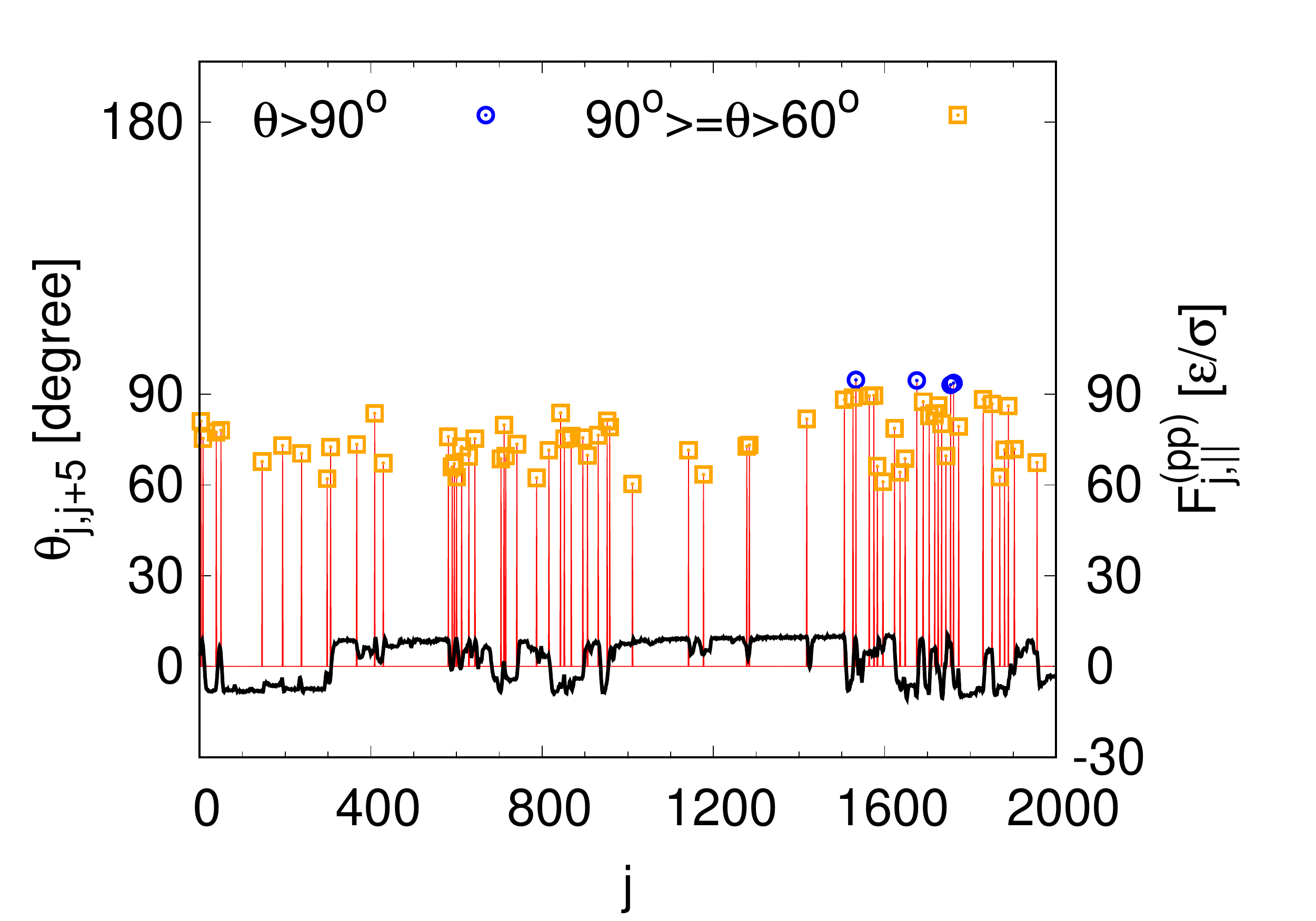}
\caption{Comparison of the bond angle $\theta_{j,j+5}$ (curvature) to the force
in the direction parallel to the stretching direction,
$F_{j,||}^{(PP)}$ (black curve), along the PP
for chain $i=2$ of $N=2000$ before (a) and after (c) stretching to the
strain of $\lambda \approx 5.0$, and
at the rescaled subsequent relaxation time $t/\tau_{R,N}=0.5$ (d), $1.0$ (e), and $2.0$ (f).
In (b) data are for the instantly elongated original PP mesh at $\lambda \approx 5.0$, but
the force pattern is not shown due to large fluctuation.
The number of kinks $N_{\rm kink}^{(i=2)}=99$ (a), $83$ (b), $68$ (c), $52$ (d), $47$ (e), and $66$ (f).}
\label{fig-fcosbb-N2000}
\end{center}
\end{figure*}

\subsection{Transient entanglement effects}

\begin{figure*}[t!]
\begin{center}
(a)\includegraphics[width=0.32\textwidth,angle=270]{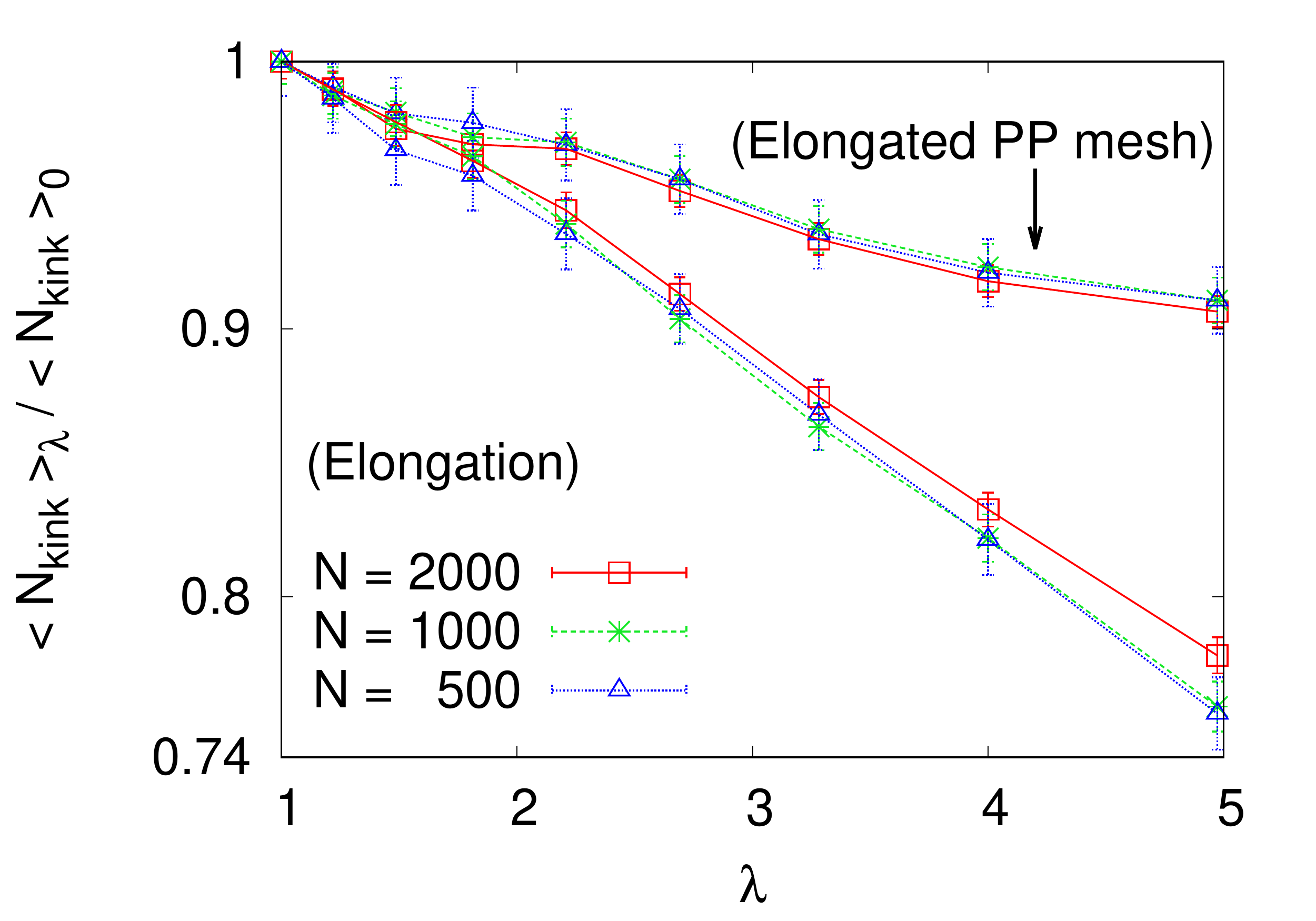}\hspace{0.2truecm}
(b)\includegraphics[width=0.32\textwidth,angle=270]{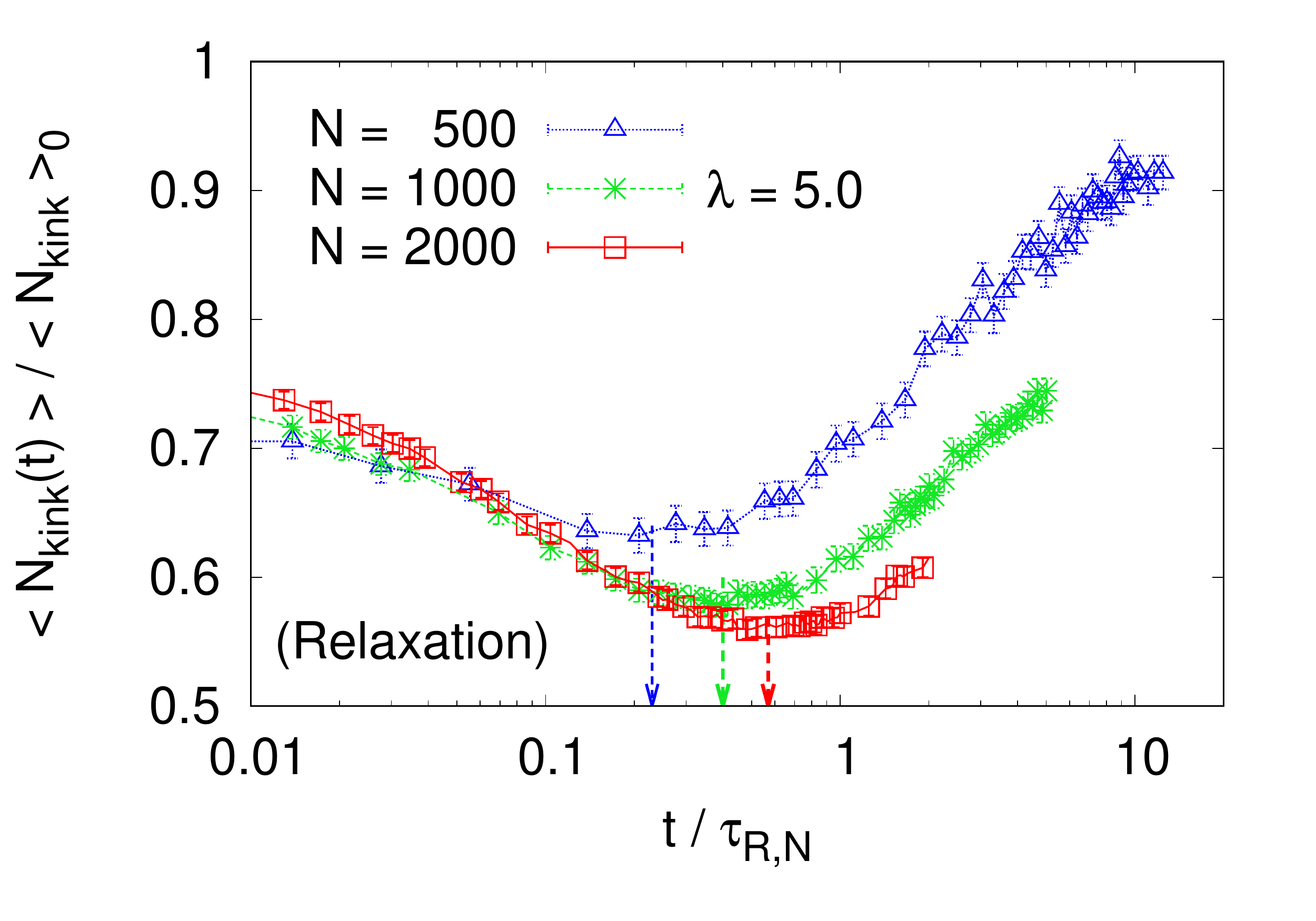}
\caption{Rescaled average number of kinks, $<N_{\rm kink}>_\lambda/<N_{\rm kink}>_0$,
plotted versus strain of $\lambda$ (a), and
$<N_{\rm kink}(t) >/<N_{\rm kink}>_0$ plotted versus the subsequent relaxation
time $t/\tau_{R,N}$, at $\lambda \approx 5.0$ (b) on a log-log scale.
Here $<N_{\rm kink}>_0\approx 22.67$, $46.88$,
and $90.74$ are the average number of kinks for unperturbed polymer melts of
chain sizes $N=500$, $1000$, and $2000$, respectively.
In (a), the upper set of data for the instantly elongated PP mesh 
is included for comparison.
In (b), the minima occur at $t/\tau_{R,N} \approx 0.23$,
$0.40$, and $0.57$ for $N=500$, $1000$, and $2000$, respectively,
are pointed out by arrows.}
\label{fig-Nkink}
\end{center}
\end{figure*}

In the following we focus on understanding the relaxation of effective entanglement points along the PPs.
Figure~\ref{fig-fcosbb-N2000} shows the comparison between the curvature characterized by the local bond angle maxima $\theta_{j,j+5} \ge 60^o$ (significant kinks) and the tension force ${\bf F}_{j,||}^{(PP)}=-{\bf \bigtriangledown} U_{\rm FENE}^{(PP)} \cdot \hat{e}_x$  in the direction parallel to the stretching direction along the PP of chain $i=2$ of size $N=2000$ (a typical case).
Before the system is elongated, significant kinks are roughly equally distributed along the PP of the chain. 
The actual number of kinks is $N_{\rm kink}^{(i=2)}=99$, while it varies between $40$ and $100$ for the whole system. 
The sign or magnitude switches of the tension force pattern, and the kinks are closely correlated (see Figure~\ref{fig-fcosbb-N2000}a).
To set the stage for comparison, we also take the original PP mesh of the unperturbed melt and deform this affinely up to $\lambda \approx 5$. The distribution of kinks along the PP of chain $i=2$ in this case remains very similar (see Figure~\ref{fig-fcosbb-N2000}b) to the original one, however some kinks become sharper, some less sharp, and the number of kinks is slightly smaller, $N_{\rm kink}^{(i=2)}=83$.
Differently from that, kinks along the corresponding PP of chain $i=2$ in the elongated melt not only become sharper but also the number of kinks is reduced, $N_{\rm kink}^{(i=2)}=68$ (see Figure~\ref{fig-fcosbb-N2000}c).
At the same time, the correlation between curvature and force pattern 
becomes even more pronounced.
The linear correlation between the kinks of high curvature and sign
switches of the tension force along the PP has been shown in the supplementary information of Ref.~\citenum{Hsu2018a}.
Our results indicate that the entanglement points, i.e. significant kinks, 
already in the very beginning do
not follow the affine deformation, if compared to the 
results for the elongated PP mesh (see Figure~\ref{fig-Nkink}),
while the average conformation of chains does, cf. Figure~\ref{fig-Rgpp-el}.
This inhomogeneous distribution even becomes more pronounced upon relaxation of the deformed system up to $0.5\tau_{R,N=2000}$ ($N_{\rm kink}^{(i=2)}=52$)  or even up to $1.0\tau_{R,N=2000}$ ($N_{\rm kink}^{(i=2)}=47$).
Obviously, the length of the regions without force sign change is increasing instead of decreasing and kinks become less sharp. 
At $t/\tau_{R,N}=2.0$, the inhomogeneous pattern still persists although the number of kinks ($N_{\rm kink}^{(i=2)}=66$) again increases.
It should be kept in mind that these processes are subject to (up to chain end effects) the (approximate) conservation of topological constraints, but $N_{\rm kink}$ can vary. 
This implies that entanglement points initially do not redistribute along the chain backbone as the chain tries to retract within a still globally affinely deformed tube.
Qualitatively all chains display very similar patterns.
Taking the clustering of entanglement points, i.e. the most confining topological constraints, one expects more conformational freedom between these regions.
This is reminiscent of knotted polymers, where entropic forces tend to pull knots tight~\cite{Grosberg2016,Narsimhan2016} (e.g. jamming knots).
In this ``jammed'' regime, the knot's diffusivity decays exponentially above a critical tension force such that monomers can only move slowly due to high monomeric friction~\cite{Narsimhan2016}.
This is consistent with our finding that the state of the system is stabilized by the significant clustering of original kinks along the PP. 

The theoretical predictions of the GLaMM model are based on the assumption of a constant Kuhn length of the PP.
The authors assume that for deformation rates as in the case of the present study the melt structure on the scale of the tube diameter $d_T$ is only weakly perturbed from that of an equilibrium melt, $d_T \approx d_T^{(0)}$ ($d_T^{(0)}$ being the tube diameter in an equilibrium melt) since entanglements should be considered as mutual, delocalized topological interactions acting on a length scale of $d_T$. 
To facilitate this either the number of entanglement points would have to increase while the Kuhn length of the PP is not changing or the other way around. Another possibility would be that a significant amount of topological constraints does not lead to kinks, e.g. like in a mesh where many chains are somewhat aligned.
Nevertheless, this is not the case here.
In Figure~\ref{fig-Nkink}a, we see that differently from such a naive extension of the original tube model,
the average number of kinks for elongated melts, $\langle N_{\rm kink} \rangle_\lambda$, neither increases nor remains constant. 
This might be partially due to the reason that the cross-section of the tube perpendicular to the tube axis is not always perfect circle, but can be rather
elliptical under deformation (see Figure~\ref{fig-Rgpp-el}b).
Comparing to the estimates of $\langle N_{\rm kink} \rangle_\lambda$
along all PPs in elongated PP meshes, 
we see that both sets of data are quantitatively the same within error bars under small perturbation while the deviation between these two sets of data becomes stronger with the increase of strain $\lambda$.
Thus topological constraints in polymer melts under large deformation do not follow affine deformation, but polymer melts under small deformation do (see Figure.~\ref{fig-Rgpp-el}d).
The results of the affinely elongated PP mesh also indicate that a significant amount of topological constraints does not lead to kinks.
Apparently, our results indicate a limitation of the GLaMM model, namely the assumption of the unperturbed Kuhn length and the homogeneous distribution of kinks breaks down.
During subsequent relaxation, the characteristic behavior of $\langle N_{\rm kink} (t) \rangle$ shown in Figure.~\ref{fig-Nkink}b is very similar to the retraction of the overall chain (see Figure~\ref{fig-Rgpp-rel}b).
$\langle N_{\rm kink}(t) \rangle$ initially decreases, reaches a minimum at $t/\tau_{R,N} \approx 0.23(3)$, $0.40(3)$, and $0.57(12)$ for $N=500$, $1000$, and $2000$, respectively, and then slowly begins to increase to eventually approach $\langle N_{\rm kink} \rangle_0$.
Taking the data for $N=500$ this final relaxation seems to occur on the time scale of the disentanglement time $\tau_d\cong \tau_{R,N} (N/N_e)^x$, $x=1$ in the original reptation scheme~\cite{deGennes1979,Doi1986} and $x\cong1.4$ experimentally~\cite{Ferry1980}.

\begin{figure*}[t!]
\begin{center}
(a)\includegraphics[width=0.32\textwidth,angle=270]{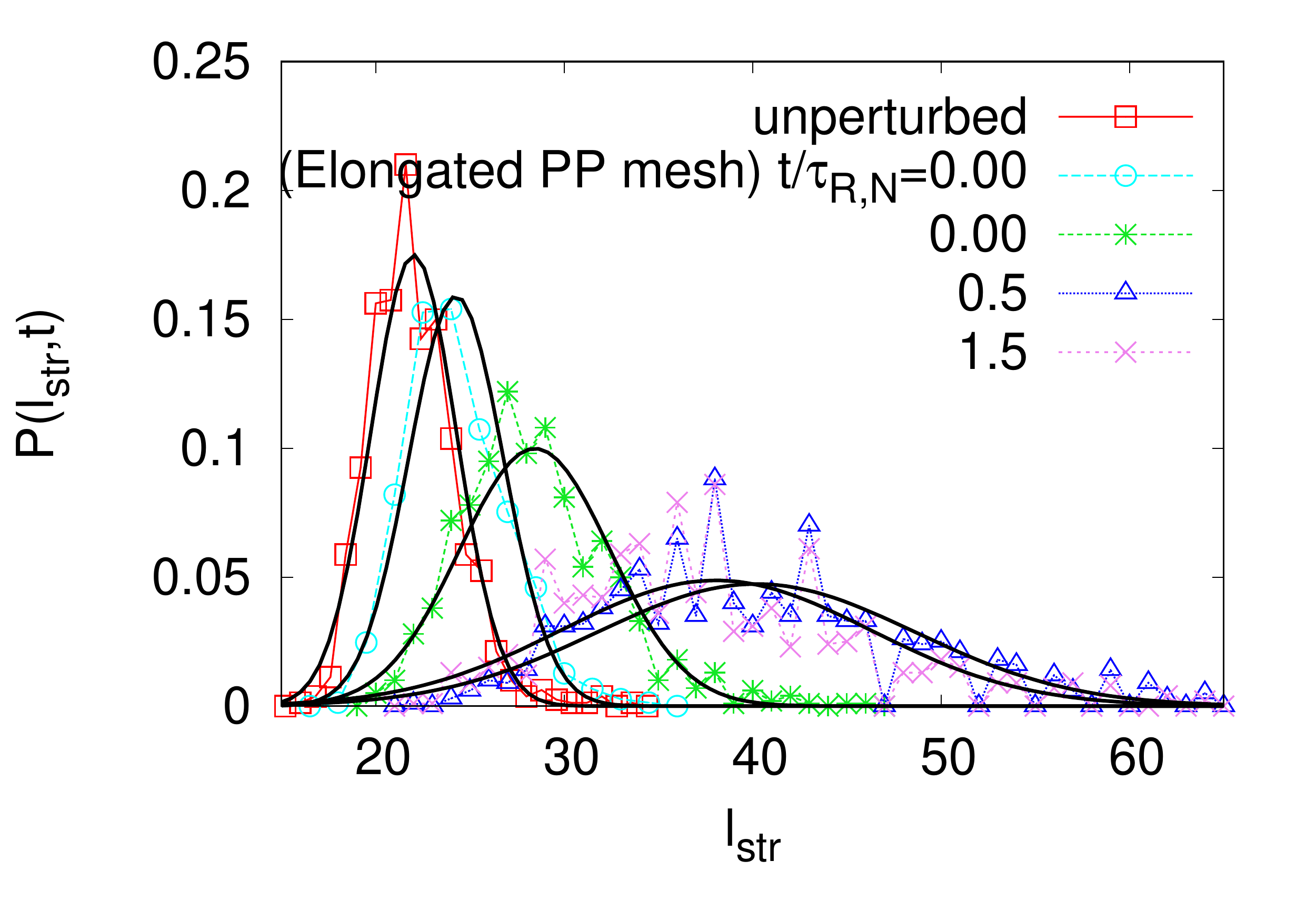}\hspace{0.2truecm}
(b)\includegraphics[width=0.32\textwidth,angle=270]{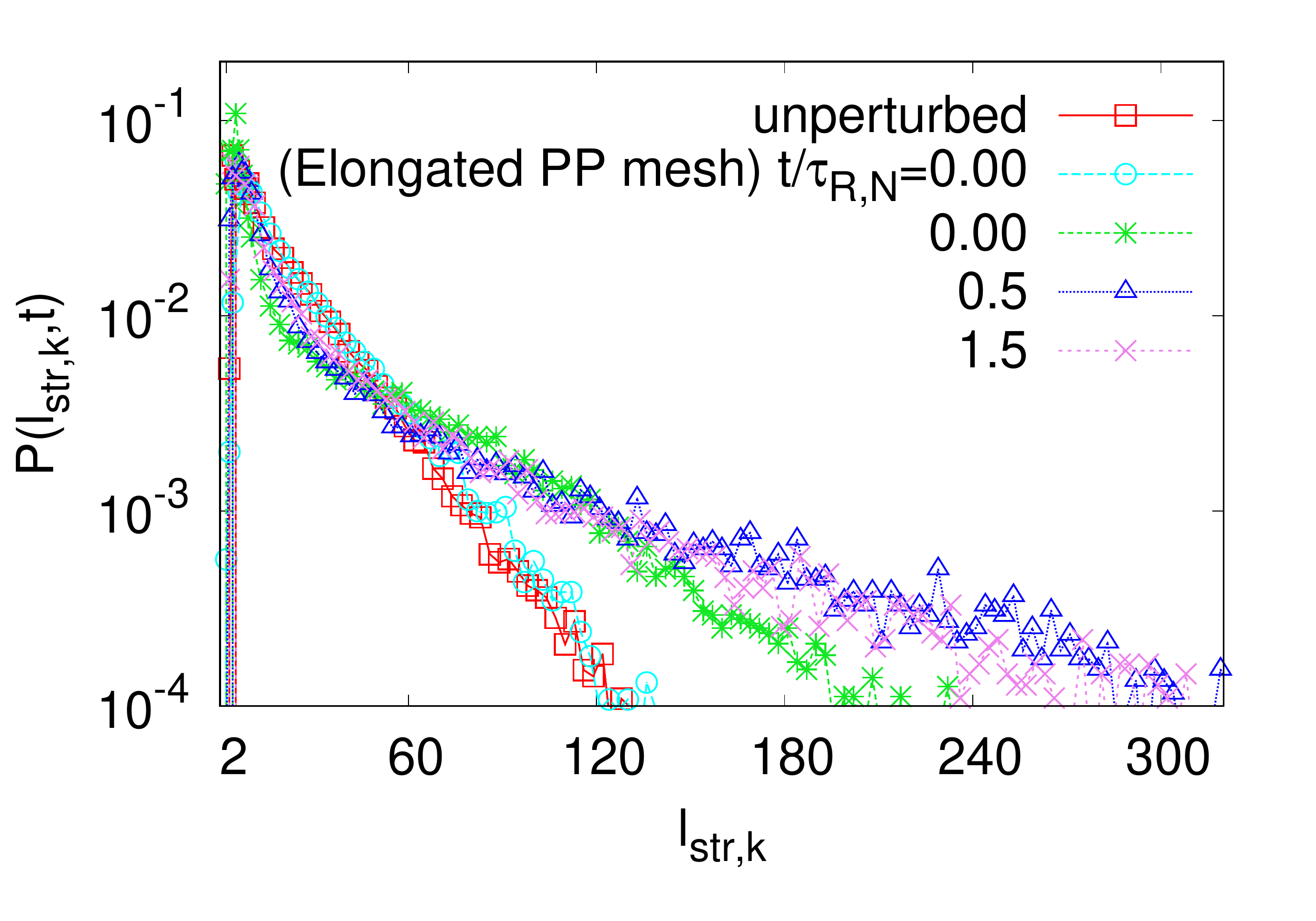}
\caption{The probability distributions $P(l_{\rm str},t)$ plotted as a function of number of the average of effective entanglement lengths $l_{\rm str}$ per
chain (a), and of the distribution of individual effective entanglement lengths, $P(l_{{\rm str},k},t)$, within all chains (b).
Data are for polymer melts containing $n_c=1000$ chains of chain size $N=2000$
before and right after deformation to the strain of $\lambda \approx 5.0$,
and at subsequent relaxation times
$t/\tau_{R,N}=0.5$ and $1.5$ with fixed $\lambda$. Results obtained for the elongated PP mesh
are also shown for comparison.
The curves of shifted Gaussian distributions are also shown in (a) for comparison.}
\label{fig-plstr}
\end{center}
\end{figure*}

\begin{figure*}[t!]
\begin{center}
\includegraphics[width=1.00\textwidth,angle=0]{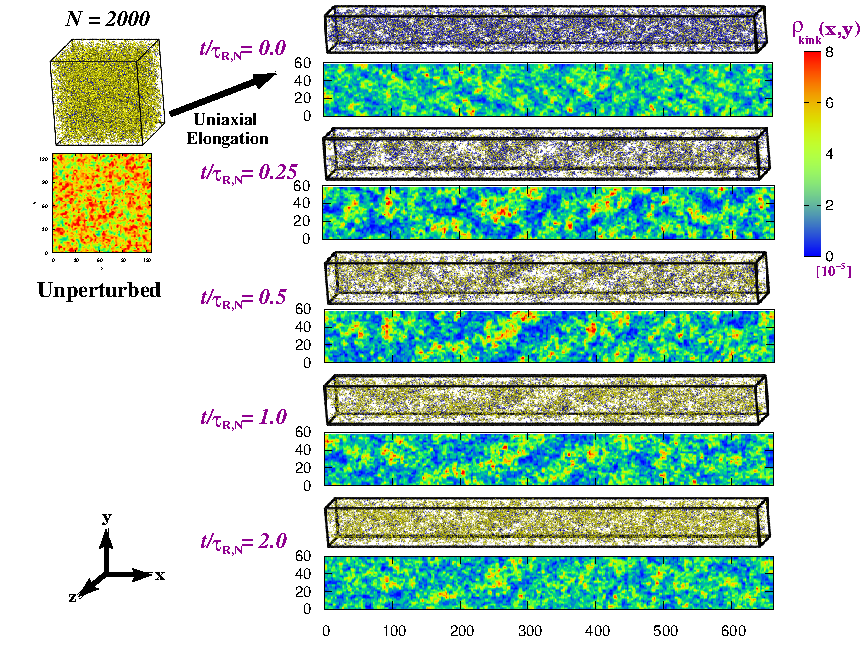}
\caption{Snapshots of entanglement points (significant kinks identified
in the way mentioned in the text) before and after the whole
polymer melt containing $n_c=1000$ chains of chain size $N=2000$ is
elongated to the strain of $\lambda \approx 5.0$, and at several selected relaxation times
$t/\tau_{R,N}$, as indicated.
The corresponding probability densities of entanglement points
projected onto x-y plane, $\rho_{\rm kink}(x,y)$, are also shown for comparison.
In the snapshots, the entanglement points correspond to the sharper kinks having
the local maximum $\theta_{j,j+5}>90^o$ are shown by blue balls, and the kinks
having $90^o \ge \theta_{j,j+5} >60^o$ are shown by yellow balls.
The linear dimensions of the simulation box are $L_x\approx 661.43\sigma$ and
$L_y=L_z \approx 59.64 \sigma$. The interval of the intensity of
$\rho_{\rm kink}(x,y)$ is set to $(0,8 \times 10^{-5})$ for better illustration.}
\label{fig-Ents-N2000}
\end{center}
\end{figure*}

PP strands between two neighboring entanglement points are almost straight lines (see Figure~\ref{fig-curvature}).
Thus a PP consists of straight segments of fluctuating length, and the average length of straight segments gives the length of an ``effective entanglement length'', which in the case of unperturbed melts is close to $N_e$.
The number of bonds along such an almost straight segment of the PP between the
$k$th and $(k-1)$th entanglement points of a chain is denoted by $l_{{\rm str},k}$,and the average length of the straight segments along the PP of each chain is given by $l_{\rm str}=\sum_k l_{{\rm str},k} /(N_{\rm kink}+1)$ with  $\sum_k l_{{\rm str},k}=N-1$, where the two ends of the chain are treated as kinks. 
Thus, results of $\langle l_{{\rm str}} \rangle$ are expected to be proportional to  $1/\langle N_{\rm kink} \rangle$. 
The distributions of $l_{\rm str}$ and $l_{{\rm str},k}$, $P(l_{\rm str},t)$ and $P(l_{{\rm str},k},t)$, respectively right after deformation, and upon relaxation to $t/\tau_{R,N}=1.5$ for $N=2000$ are presented in Figure~\ref{fig-plstr}. 
For comparison, results for unperturbed polymer melts and for elongated polymer meshes are also included in Figure~\ref{fig-plstr}.
There the inverse of the estimated slope in Figure~\ref{fig-plstr}b roughly corresponds to 
$\langle l_{{\rm str},k} \rangle \approx 22$ which is close to $N_{e,PPA}^{(0)}=28$.
We see that the profiles of $P(l_{\rm str},t)$ are quite well described by shifted Gaussian distributions. 
The peak first shifts to significantly larger values of $l_{\rm str}$, while only after $t/\tau_{R,N}>0.5$ the peak of $P(l_{\rm str},t)$ slowly begins to move towards smaller values of $l_{\rm str}$, still being far away from the unperturbed case, even for $t \approx 1.5 \tau_{R,N}$. 
Furthermore the width of the distribution $(\langle l_{\rm str}^2 \rangle -\langle l_{\rm str}\rangle^2)$ seems to become even wider with time up to $t \approx 0.5\tau_{R,N}$. 
Within the chains the distribution of the lengths of straight PP segments $P(l_{{\rm str},k},t)$, Figure~\ref{fig-plstr}b, shows long tail distributions of $l_{{\rm str},k}$.
Right after deformation, the tail of $P(l_{{\rm str},k},t)$ becomes significantly broader than for an unperturbed polymer melt. 
At subsequent relaxation times, instead of moving back to the distribution for unperturbed polymer melt, we see that the tail initially becomes even significantly longer up to $t\approx 0.5\tau_{R,N}$, before it very slowly turns toward the distribution for the unperturbed melt as shown in Figure~\ref{fig-fcosbb-N2000}, which it eventually has to reach. 
The apparent slope for $l_{{\rm str},k}>120$ would indicate an intermediate effective $N_e$ of about $100$ based on a small fraction of the total straight PP segments, which should not be confused with the entanglement length based on the theoretical considerations for the PPA~\cite{Everaers2004,Sukumaran2005}, discussed later.
For the elongated PP mesh at $\lambda \approx 5.0$, we see only a slight shift to larger values of $l_{\rm str}$ for $P(l_{\rm str},t)$ compared to that of the unperturbed polymer melt since $\langle N_{\rm kink} \rangle$ is only about  $10\%$ smaller than $\langle N_{\rm kink} \rangle_0$ right after deformation (Figure~\ref{fig-Nkink}a).
However, such effects are too weak to analyze them quantitatively, as they are within the error bars of the profiles of $P(l_{{\rm str},k},t)$.

Our results clearly show that the number of kinks
along all PPs in polymer melts under large deformation does not follow affine deformation and
the kinks along the individual PPs of chains distribute unequally right after deformation.
The delayed relaxation observed in the profiles of $\langle N_{\rm kink}(t) \rangle$, $P(l_{\rm str},t)$, and
$P(l_{{\rm str},k},t)$ indicate that topological constraints play an essential role 
in the relaxation retardation of deformed chains.

\subsubsection{Entanglement point distribution in space}

\begin{figure*}[thb!]
\begin{center}
(a)\includegraphics[width=0.32\textwidth,angle=270]{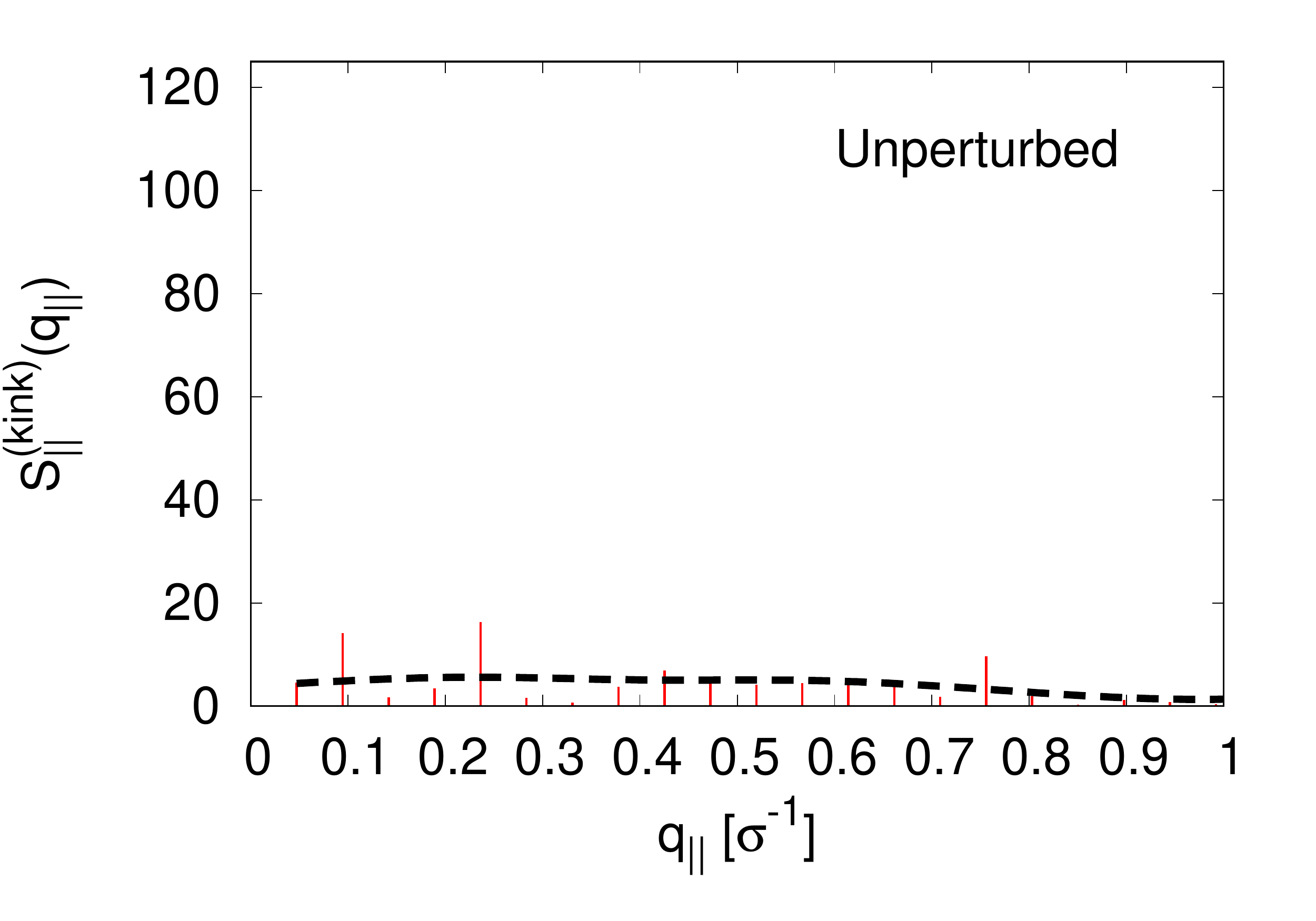}\hspace{0.2truecm}
(b)\includegraphics[width=0.32\textwidth,angle=270]{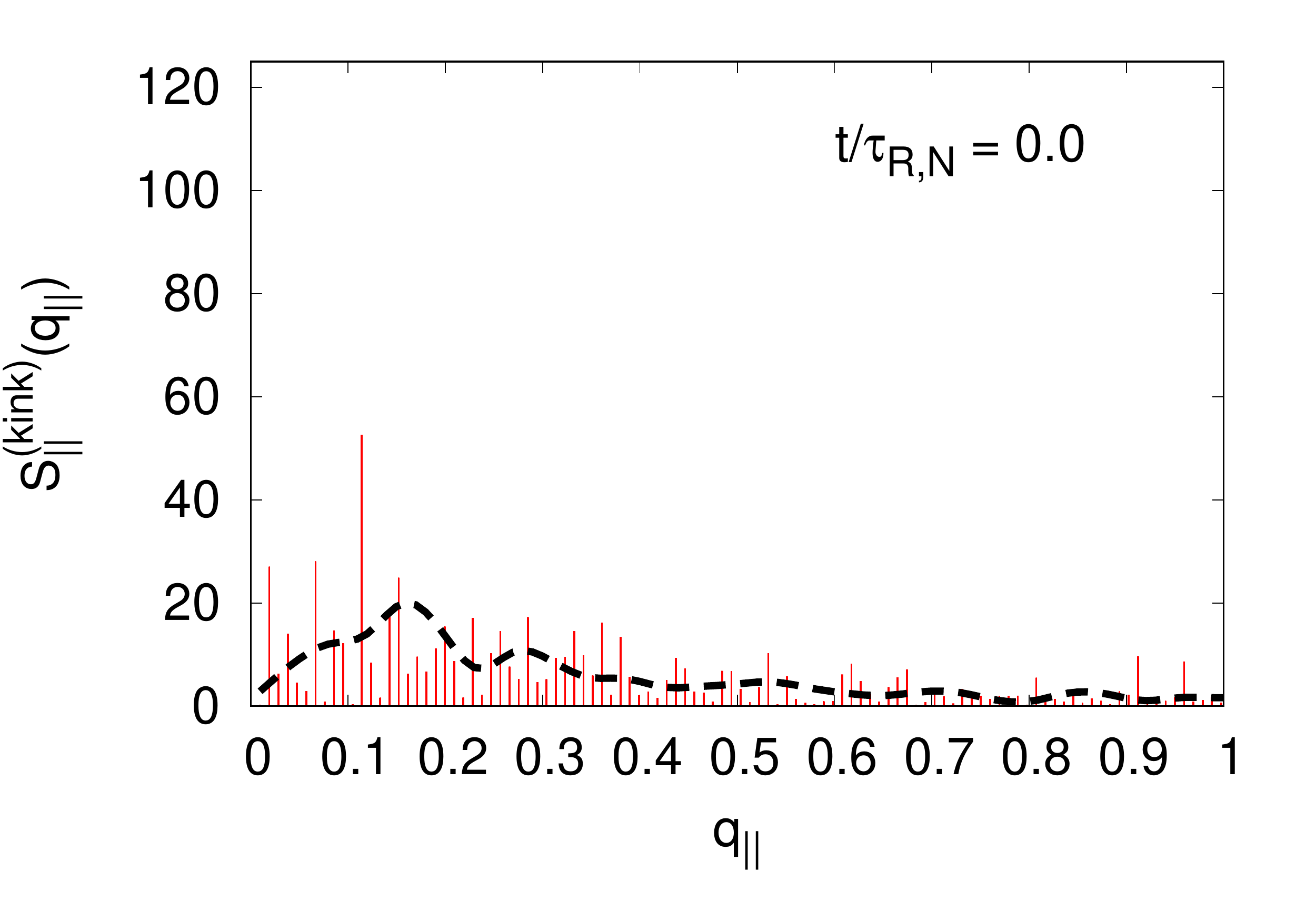}\\
(c)\includegraphics[width=0.32\textwidth,angle=270]{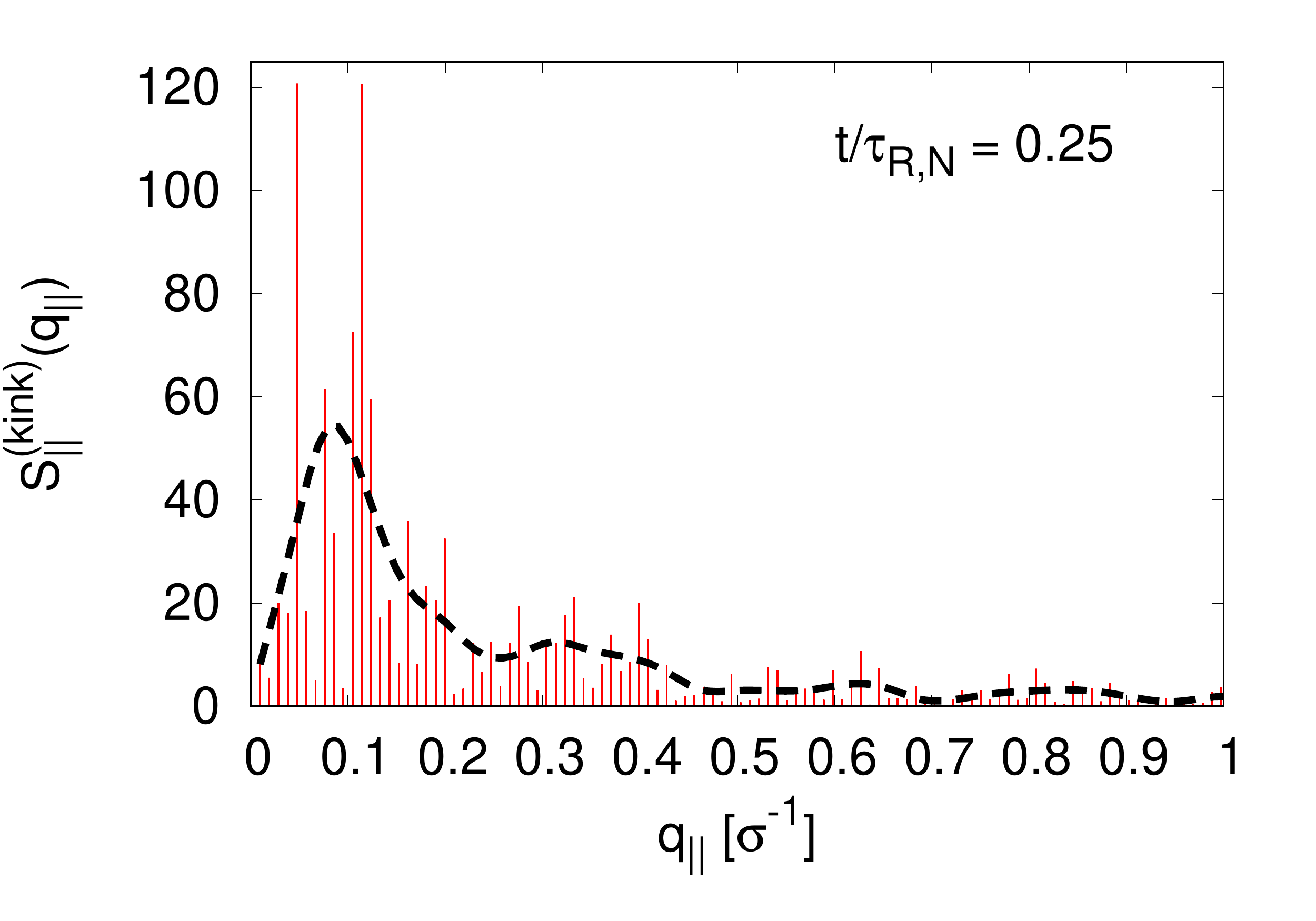}\hspace{0.2truecm}
(d)\includegraphics[width=0.32\textwidth,angle=270]{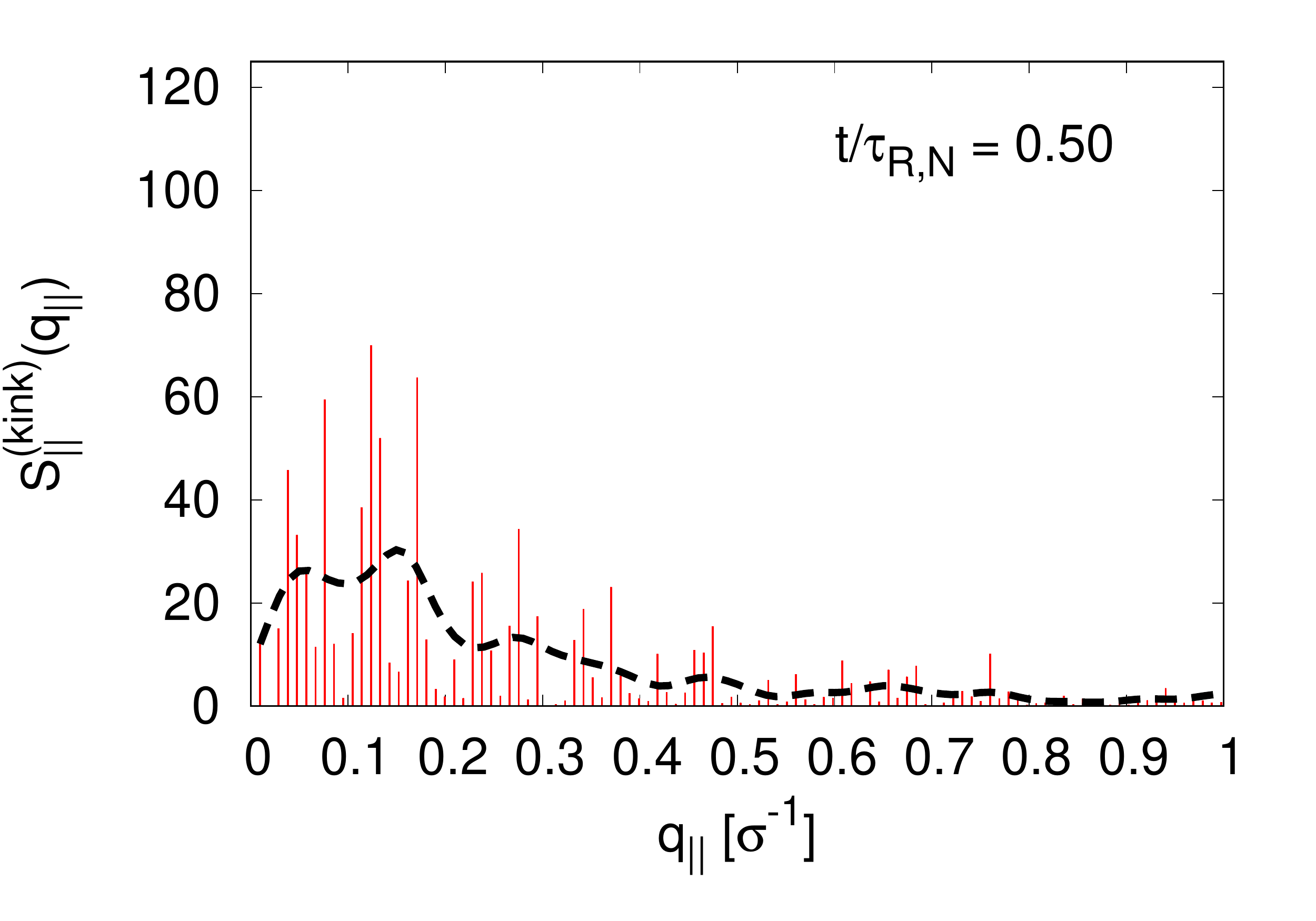}
(e)\includegraphics[width=0.32\textwidth,angle=270]{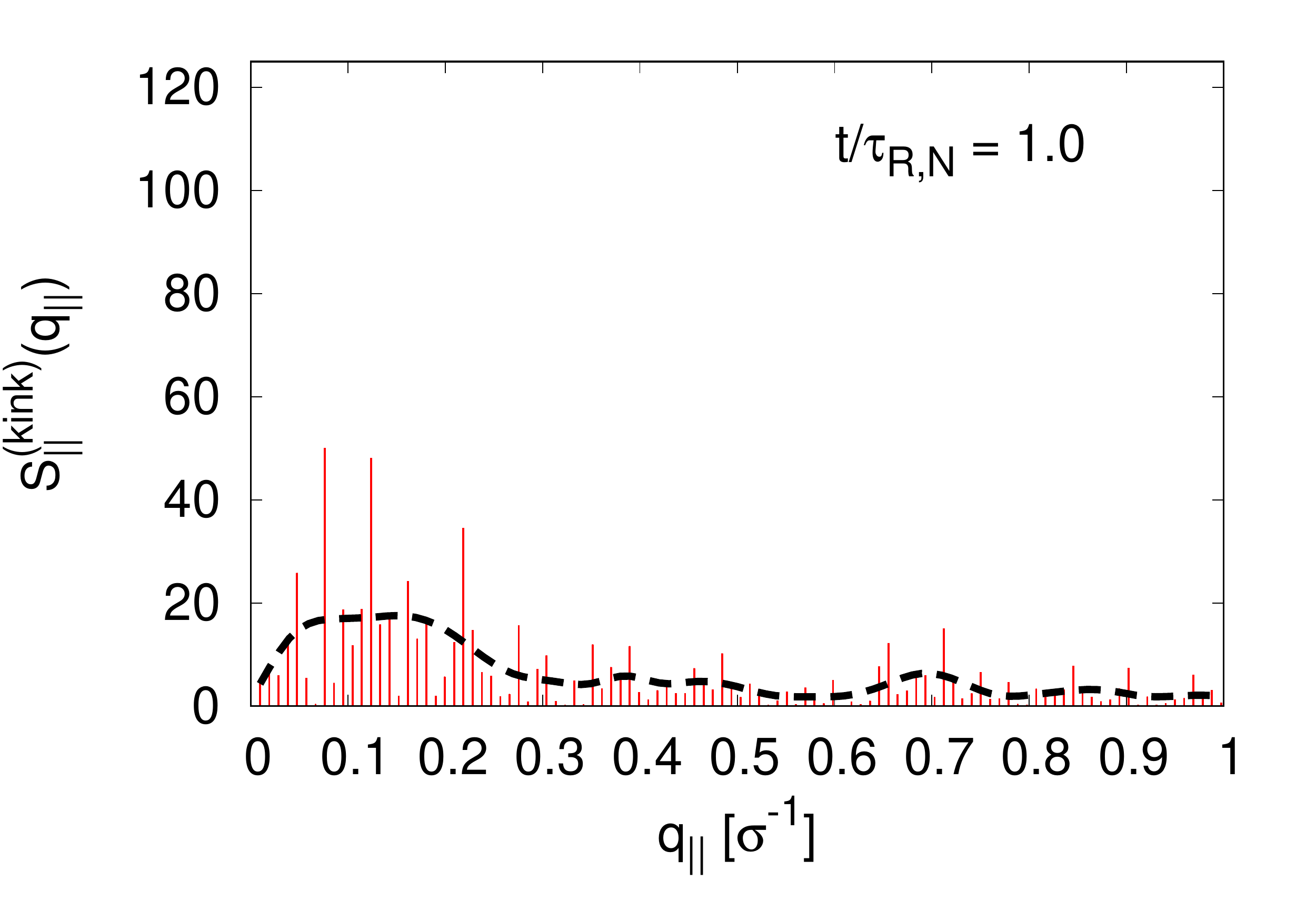}\hspace{0.2truecm}
(f)\includegraphics[width=0.32\textwidth,angle=270]{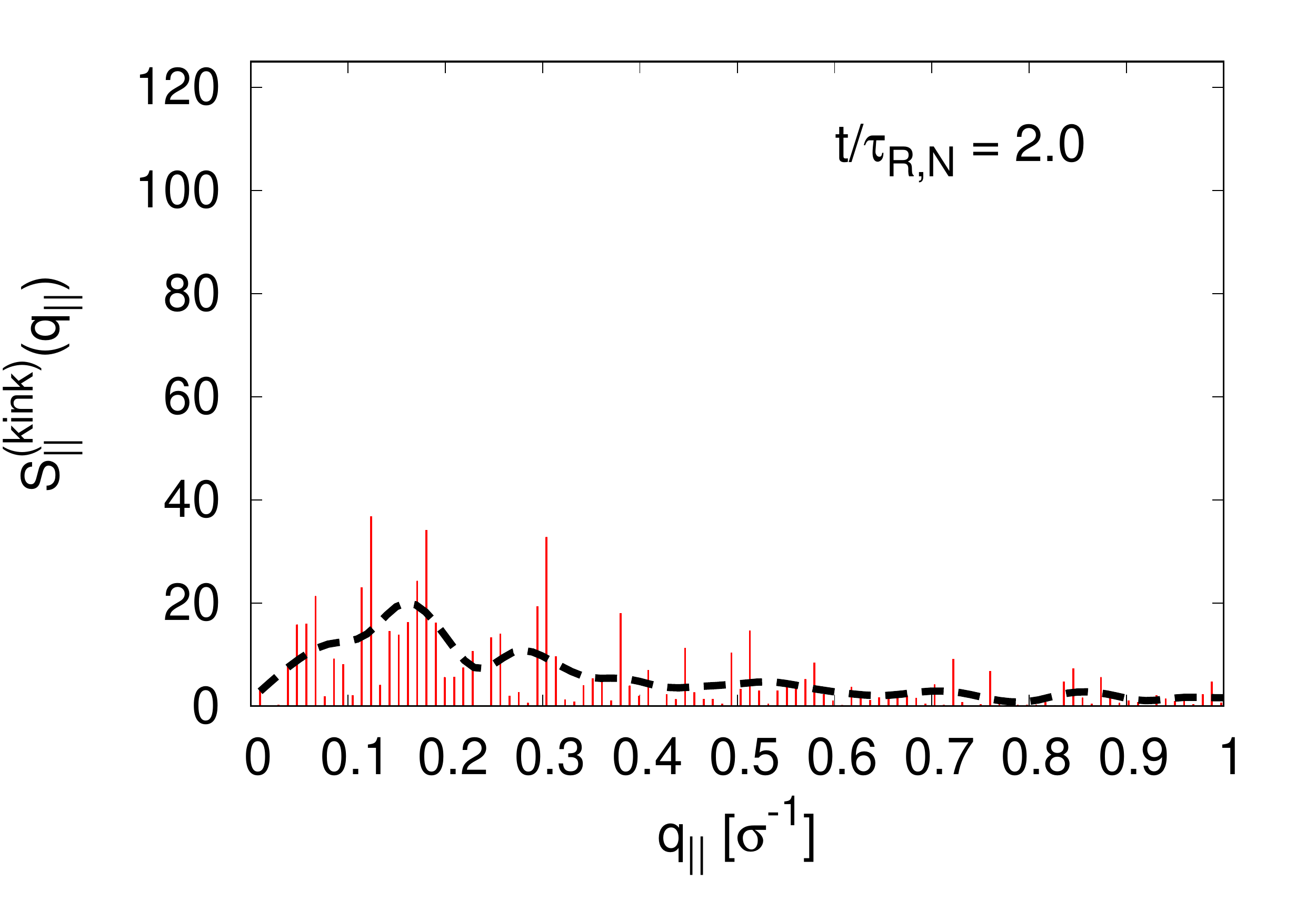}
\caption{The one dimensional collective structure factor $S^{\rm (kink)}(q_{||})$
of entanglement points in a melt shown in Figure~\ref{fig-Ents-N2000},
plotted versus $q_{||}$, before (a) and after deformation (b),
and at several selected subsequent relaxation times $t/\tau_{R,N}=0.25$ (c), $0.50$ (d), $1.0$ (e) $2.0$ (f).
Here $q_{||}=(2\pi/L_x)m_x$ with $m_x=1,2,3,\ldots$. In (a), $L_x=133.01\sigma$, and
In (b)-(f), $L_x=661.43\sigma$.
The oscillation of $S^{\rm (kink)}(q_{||})$ due to the finite-size effect is smeared out
by taking the average value of $S^{\rm (kink)}(q_{||})$ over eight values of $q_{||}$
(black dashed curves).}
\label{fig-sqkink}
\end{center}
\end{figure*}

So far we have been focusing on conformations of individual chains and their respective primitive paths. We now turn to the distribution of topological constraints, as represented by entanglement points, in space.  
Figure~\ref{fig-Ents-N2000} shows snapshots of such distributions as a function of time.
Quantitatively, the distribution can be described by the corresponding density of entanglement points projected onto the $x$-$y$ plane, $\rho_{\rm kink}(x,y)$. 
$\rho_{\rm kink}(x,y)$ is estimated by simply counting the number of entanglement points located at $(x,y)$ and normalized such that  $\int_0^{L_x} \int_0^{L_y}  \rho_{\rm kink}(x,y)dxdy =1$.
For this the grid spacing is set to $2.0\sigma$. 
For a large disorder system, structural inhomogeneities presented in such projection 
would be smeared out. Thus one can view Figure~\ref{fig-Ents-N2000} as a rephrase native slice 
of a large system. 
Before and right after deformation, the entanglement points distribute homogeneously while the kinks become sharper. For the unperturbed melt ($\lambda=1.0$), 
$N_{\rm kink}=88128$ for $60^o<\theta_{j,j+5}<90^o$ and  
$N_{\rm kink}=7617$ for $\theta_{j,j+5}>90^o$ while 
$N_{\rm kink}=7920$ for $60^o<\theta_{j,j+5}<90^o$ and
$N_{\rm kink}=74344$ for $\theta_{j,j+5}>90^o$ for the polymer melt right after deformation ($\lambda\approx 5.0$).
Upon relaxation, we observe a clustering of entanglement points and the clustering pattern becomes more distinct and seems to reach a maximum around $t=0.5\tau_{R,N}$.
Even up to $t=2.0\tau_{R,N}$, these clusters persist although the kinks along deformed chains become less sharp ($N_{\rm kink}=46917$ for $60^o<\theta_{j,j+5}<90^o$ and  $N_{\rm kink}=8888$ for $\theta_{j,j+5}>90^o$).
Since the larger clusters of entanglement points are rather fuzzy, and since there are still many isolated entanglement points, it is difficult to identify a characteristic length scale which leads to the instability in the homogeneous distribution of entanglement points.
Based on the original tube concept one would expect that the onset of the separation into these jammed areas should be of the order of at most a few tube diameters (times the elongation amplitude), since this is the only length originating from the topological constraints.
 The one dimensional scattering function $S^{\rm (kink)}(q_{||})$ of entanglement points in the melt gives an impression of the correlation, where  $q_{||}$ is the component of the vector ${\bf q}$ in the direction parallel to the stretching.
The discretization in $q$-space along the stretching direction is given by $q_{||}=2\pi m_x/L_x$ with $m_x=0,1,2,\ldots$, where $L_x=L_0$ and $L_x \approx 5.0L_0$ before and after deformation, respectively. 
Results of $S^{\rm (kink)}(q_{||})$ plotted versus $q_{||}$ are shown in Figure~\ref{fig-sqkink}.
Here we only focus on the regime $q_{||} \le 1.0 \sigma^{-1}$.
At first glance, the distributions of entanglement points seem to be more structured during initial relaxation after polymer melts are deformed comparing to the structure for the unperturbed polymer melt.
This is consistent with the observed clustering of entanglement points shown in
Figure~\ref{fig-Ents-N2000}. 
If we only focus on clusters with high density of kinks, the characteristic distance $d_{\rm cluster}$ can be roughly determined by  $d_{\rm cluster}=2\pi/q_{||}^{\rm max}$, i.e., at $q_{||}=q_{||}^{\rm max}$, $S^{\rm (kink)}(q_{||})$ reaches a maximum along the stretching direction.
For $t=0.25\tau_{R,N}$, we see one broad maximum around $q_{||}^{\rm max} \approx 0.08\sigma^{-1}$ which gives $d_{\rm cluster} \approx 78\sigma \sim 16 d_T^{(0)}$, consistent with the above argument. 
At subsequent relaxation times, this broad maximum slowly moves to a larger value of $q$ and becomes even broader.
This clustering structure remains even up to $t=2.0\tau_{R,N}$, and $d_{\rm cluster} \approx 40\sigma \sim 8 d_T^{(0)}$ ($q_{||}^{\rm max} \approx 0.16\sigma^{-1}$).
Thus, the clustering of the entanglement points is related to the relaxation retardation of the deformed polymer melt. 

\subsection{Mobility of monomers in non-equilibrium states}

\begin{figure*}[t!]
\begin{center}
(a)\includegraphics[width=0.32\textwidth,angle=270]{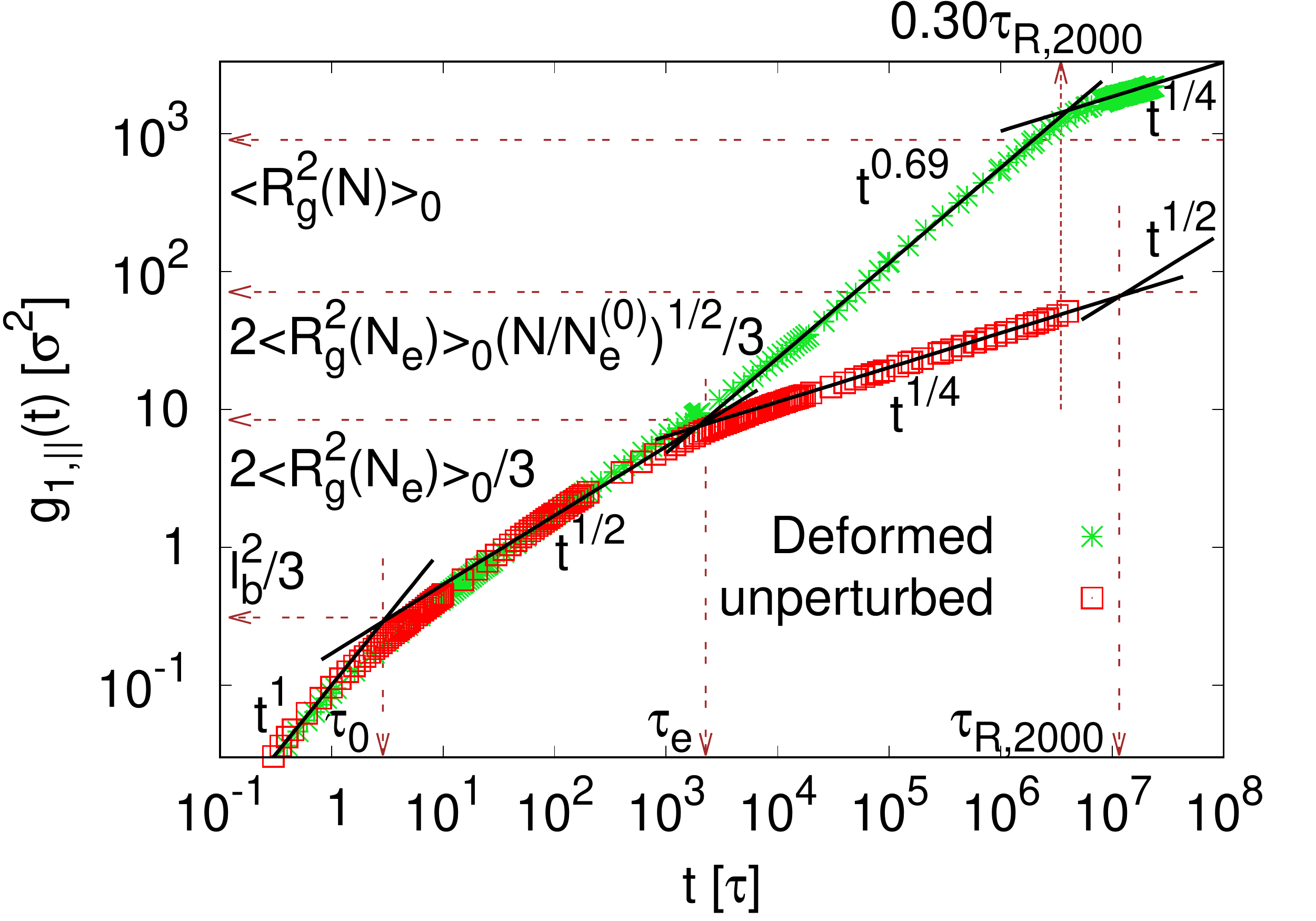}\hspace{0.2truecm}
(b)\includegraphics[width=0.32\textwidth,angle=270]{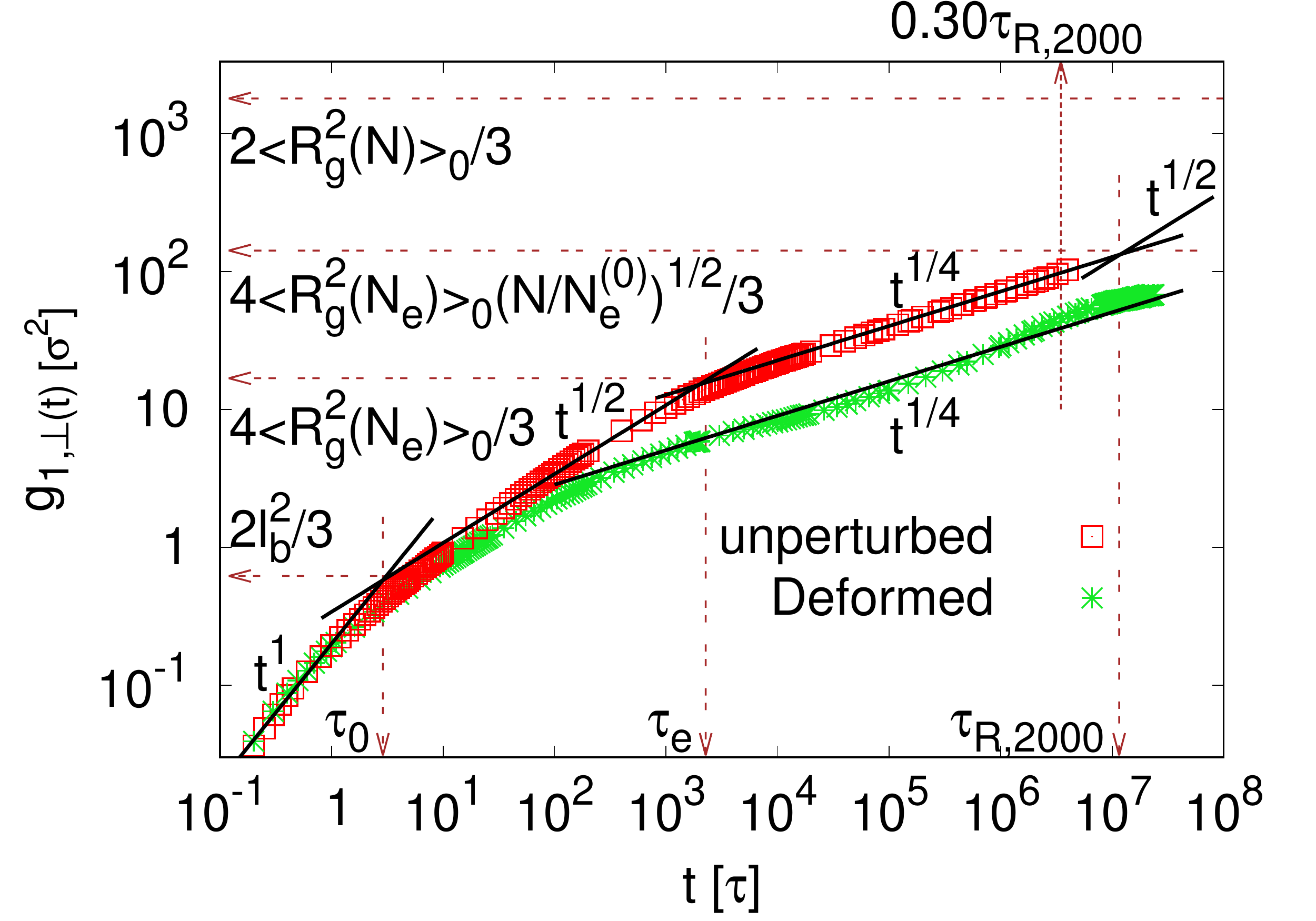}\\
(c)\includegraphics[width=0.32\textwidth,angle=270]{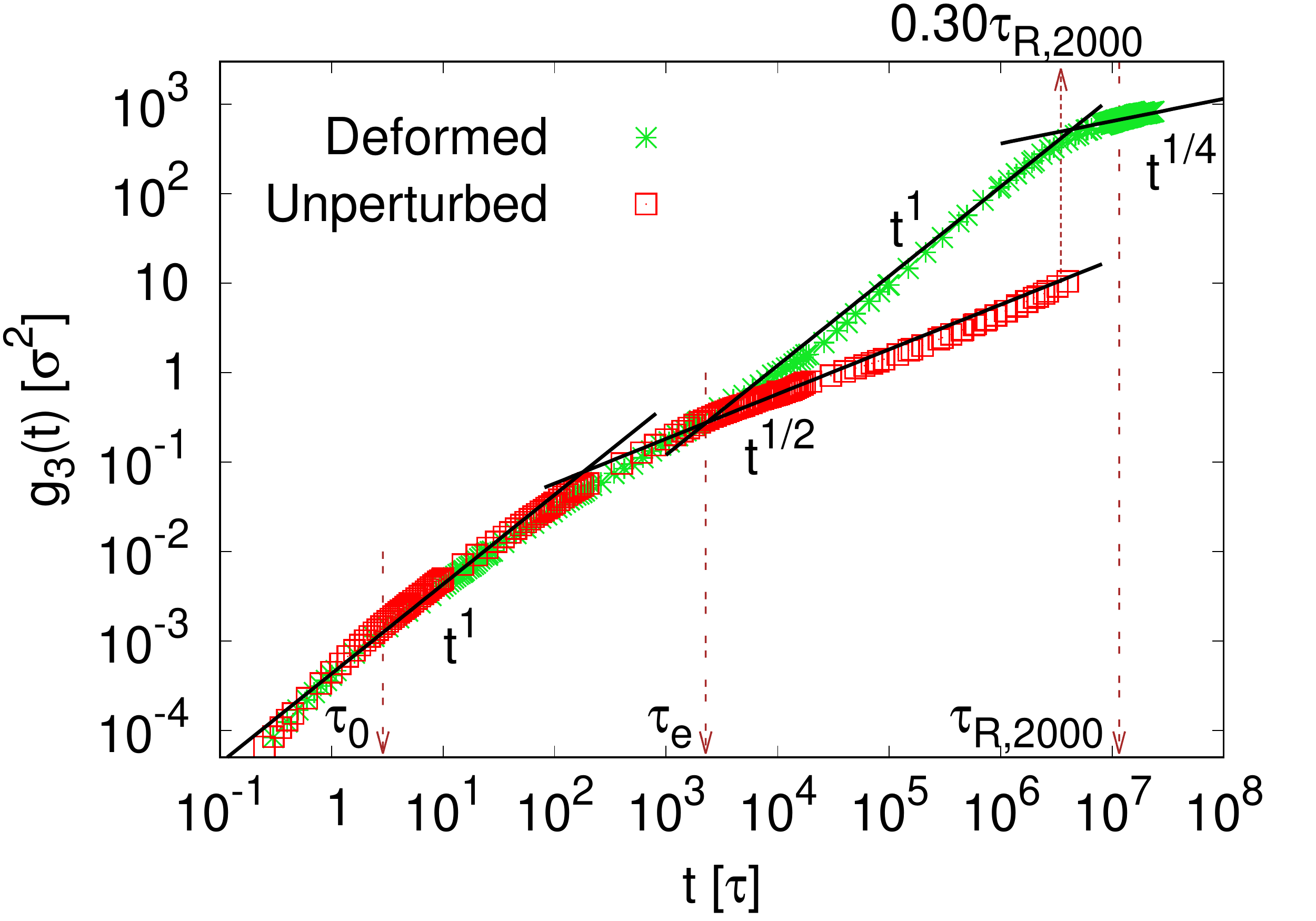}\hspace{0.2truecm}
(d)\includegraphics[width=0.32\textwidth,angle=270]{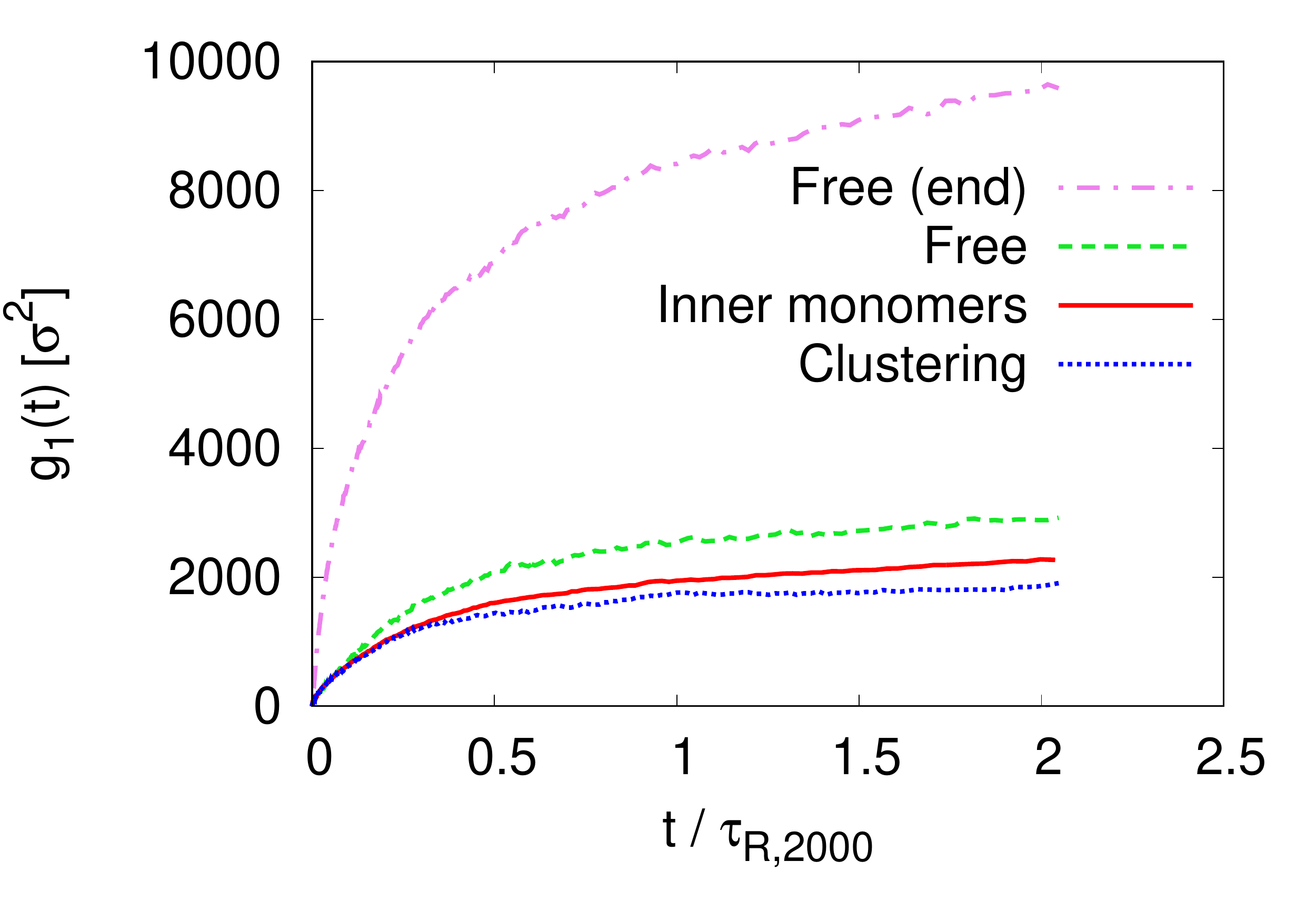}\\
\caption{Mean square displacement of inner monomers in the directions parallel,
$g_{1,||}(t)$ (a), and perpendicular, $g_{1,\perp}(t)$ (b), to the stretching
direction, and of center of mass, $g_3(t)$, plotted versus the relaxation time $t$.
(d) $g_1(t)=g_{1,||}(t)+g_{1,\perp}(t)$ of all inner monomers from (a)(b),
$20$ monomers in one of the free and one of the clustering regions for all chains (cf. text), plotted
versus the rescaled relaxation time $t/\tau_{R,2000}$.
Data are for the polymer melt consisting of $n_c=1000$ chains of chain size $N=2000$.
In (a)-(c), the crossover scaling laws between different regimes and the corresponding
characteristic length and time scales for fully equilibrated polymer melts are shown, as indicated.
For the deformed polymer melt, the exponents of the power laws at different regimes are also
shown for comparison.
In (d), $g_1(t)$ of $20$ monomers in the free region at the end for all chains is also shown for comparison.}
\label{fig-g13}
\end{center}
\end{figure*}

Polymer chain dynamics is usually characterized by different regimes of the mean square displacement (MSD) of monomers. 
Reptation theory~\cite{deGennes1979,Doi1986} predicts the crossover between these regimes at the characteristic time $\tau_0$ for local fluctuations, the entanglement time $\tau_e \propto N_e^2$, the Rouse time $\tau_{R,N} \propto N^2$, and the disentanglement time $\tau_{d,N} \propto N^{3.4}$ (when contour length fluctuations, constraint release, and correlation hole effects are taken into account~\cite{Doi1983, Milner1998, McLeish2002, Likhtman2002}).
The crossover scaling predictions of the MSD of monomers, $g_1(t)$, of monomer with respect to the corresponding center of mass, $g_2(t)$, and of the center of mass $g_3(t)$ have also been verified by our large time scales MD simulations for highly entangled and fully equilibrated polymer melts~\cite{Hsu2016}, i.e., the unperturbed systems studied here.

In Figure~\ref{fig-g13}, we compare the motions of inner $(N/2+1)$ monomers (i.e., eliminating the strong fluctuations caused by chain ends~\cite{Kremer1990,Kremer1992,Hsu2016}) parallel and perpendicular to the stretching direction, $g_{1,||}(t)$ and $g_{1,\perp}(t)$, respectively, and of the center of mass, $g_3(t)$, in the polymer melt of size $N=2000$ after deformation to that in the fully equilibrated polymer melt. 
Note that we here do not average over starting times as we begin to measure $g_1(t)$ and $g_3(t)$ right after deformation.
We see that along the stretching direction, $g_{1,||}(t)$ for the deformed polymer melt initially follow the same scaling behavior up to the original tube diameter $d_T^{(0)}$ as that for the unperturbed polymer melts ($g_{1,||}(t) \sim t^1$ for $t<\tau_0$, and $g_{1,||}(t) \sim t^{1/2}$ for $\tau_0 < t < \tau_e=\tau_0(N^{(0)}_{PPA})^2$)
For $t>\tau_e$, monomers move faster, $g_{1,||}(t) \sim t^{0.69}$, until reaching $t \approx 0.30\tau_{R,2000}$, i.e., the duration of the chain retraction process. 
Within this time frame, the change of $g_{1,||}(t)$, $\Delta g_{1,||}(t) \approx {\cal O}(10^3)$ is of the same order as the change of $\langle R_{g,||}^2 (t) \rangle$, 
while the center of mass displays unperturbed diffusion $g_3(t) \sim t$ up to $0.30\tau_{R,2000}$, but
moves a little more slowly around $t=\tau_e$.
For $t>0.3\tau_{R,2000}$, monomers motion is slowed down again due entanglement effects and the corresponding relaxation retardation, similar as in a tube-like regime created by the surrounding chains, $g_{1,||}(t) \sim t^{1/4}$.  
However, the effect of entanglements varies in the process of equilibration. 
A strong relaxation retardation is also observed for $g_3(t)$ that $g_3(t) \sim t^{1/4}$. 
Perpendicular to the stretching, the situation is somewhat different. 
Apparently, monomers move more slowly that a gradual deviation from $g_{1,\perp} \sim t^{1/2}$ to 
$g_{1,\perp} \sim t^{1/4}$ develops as $t$ approaches $\tau_e$. 
After deformation, chains are somewhat aligned
along the stretching direction such that monomers presumably have less freedom to move in the crowded space 
along the perpendicular direction.

To test our finding that the deformed chain conformations are stabilized state by the clustering/jamming of entanglement points, we estimate the MSD of groups of $20$ inner monomers in the clustering (jammed) region, and $20$ monomers in the free region for all $n_c=1000$ chains. Here the clustering and free regions are identified according to the 
curvature the PPs of chains at $t=\tau_{R,N}$. For example, in the case of $i=2$ (see Figure~\ref{fig-fcosbb-N2000}e),
monomers $j=960$ to $j=979$ in the clustering region, and monomers $j=1480$ to $j=1499$ in the free regime 
are considered.
This is compared to MSD of inner $(N/2+1)$ monomer during the relaxation process.
Results of $g_1(t)$ plotted as a function of $t/\tau_{R,2000}$ are shown in Figure~\ref{fig-g13}d. 
We see that monomers in the constrained regime move much slower comparing to those in the free regime.
Indeed, the MSD of inner monomers is dominated by the motion of inner monomers in the constrained regime.
In figure~\ref{fig-g13}d, we have also included the MSD of $20$ monomers in the free regime near one end of all chains. 
Monomers apparently move much faster as we have expected due to the weak topological constraints at the end and the retraction process.

\subsection{Intrinsic properties of primitive paths and the tube picture} 

  Based on the tube picture~\cite{deGennes1979,Doi1980,Doi1986}, one would expect that entanglement effects appear at $t\approx \tau_e$ and monomers are restricted to move along the contour of an imaginary tube of diameter $d_T$  created by surrounding chains. 
The primitive paths represent the backbone of the tube that can be constructed following this picture. 
In this subsection we investigate the time dependent intrinsic properties of PPs created by PPA~\cite{Everaers2004,Sukumaran2005} for polymer melts under strain and link them to the tube concept.

\begin{figure*}[t!]
\begin{center}
(a)\includegraphics[width=0.32\textwidth,angle=270]{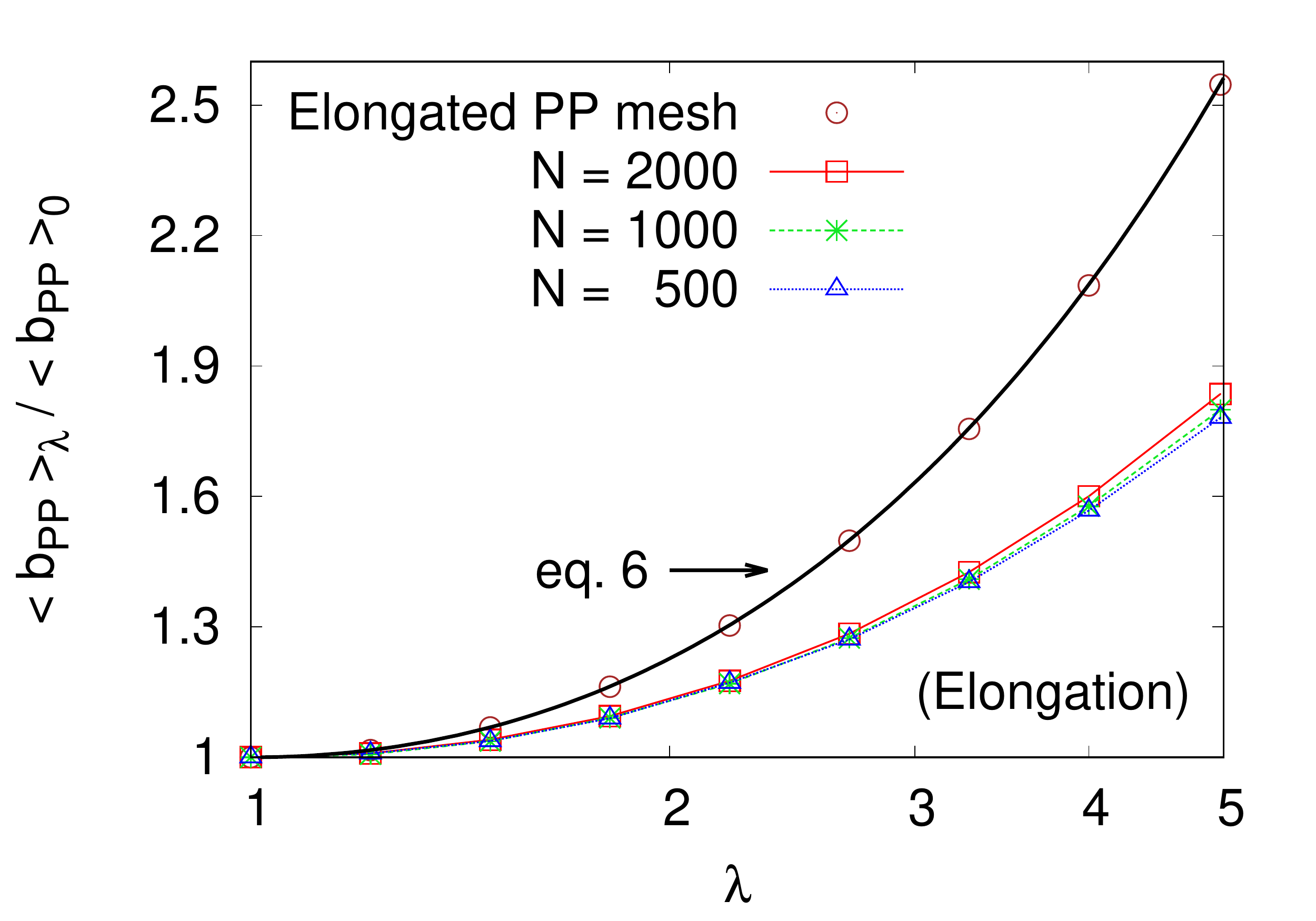}\hspace{0.2truecm}
(b)\includegraphics[width=0.32\textwidth,angle=270]{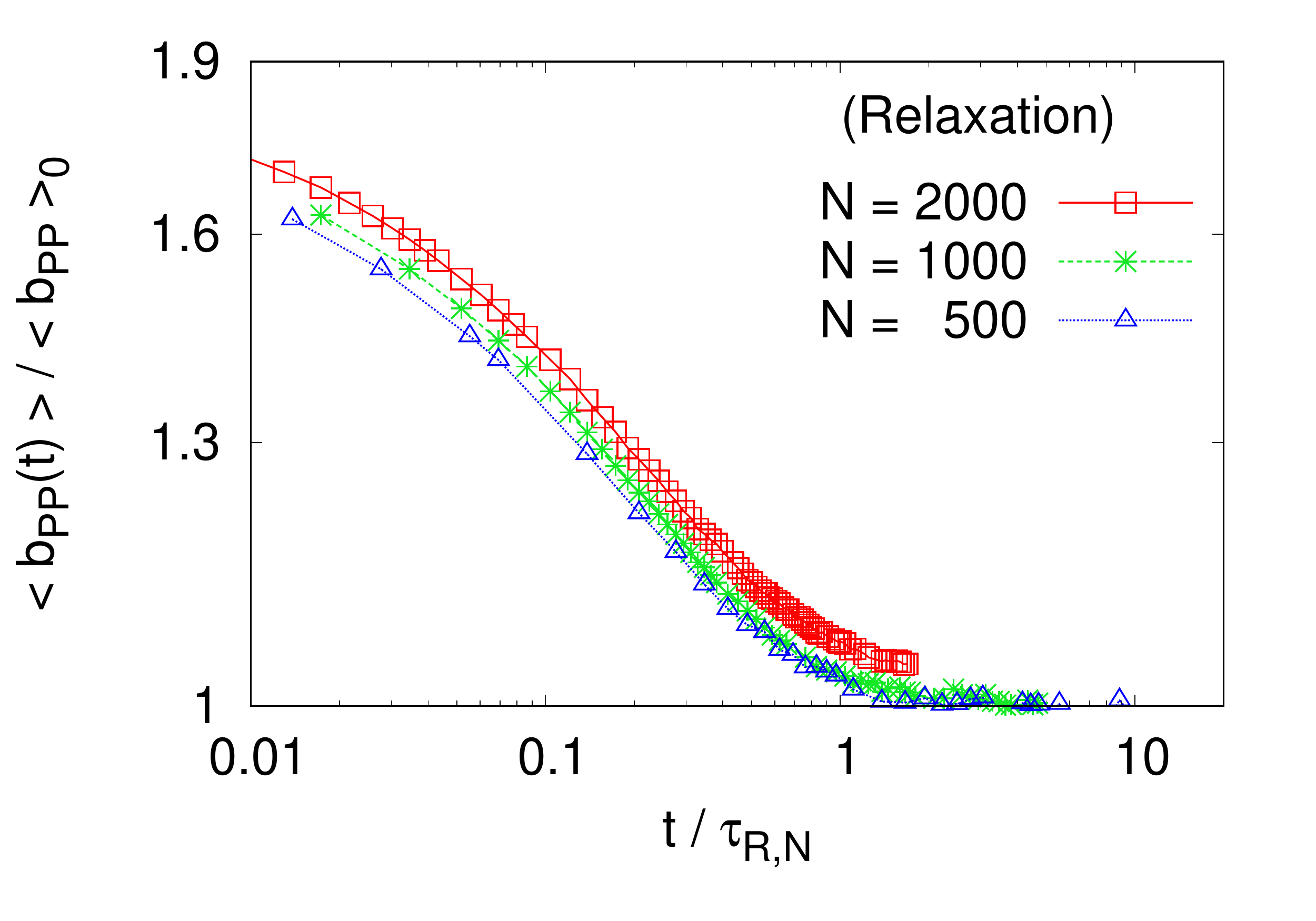}
(c)\includegraphics[width=0.32\textwidth,angle=270]{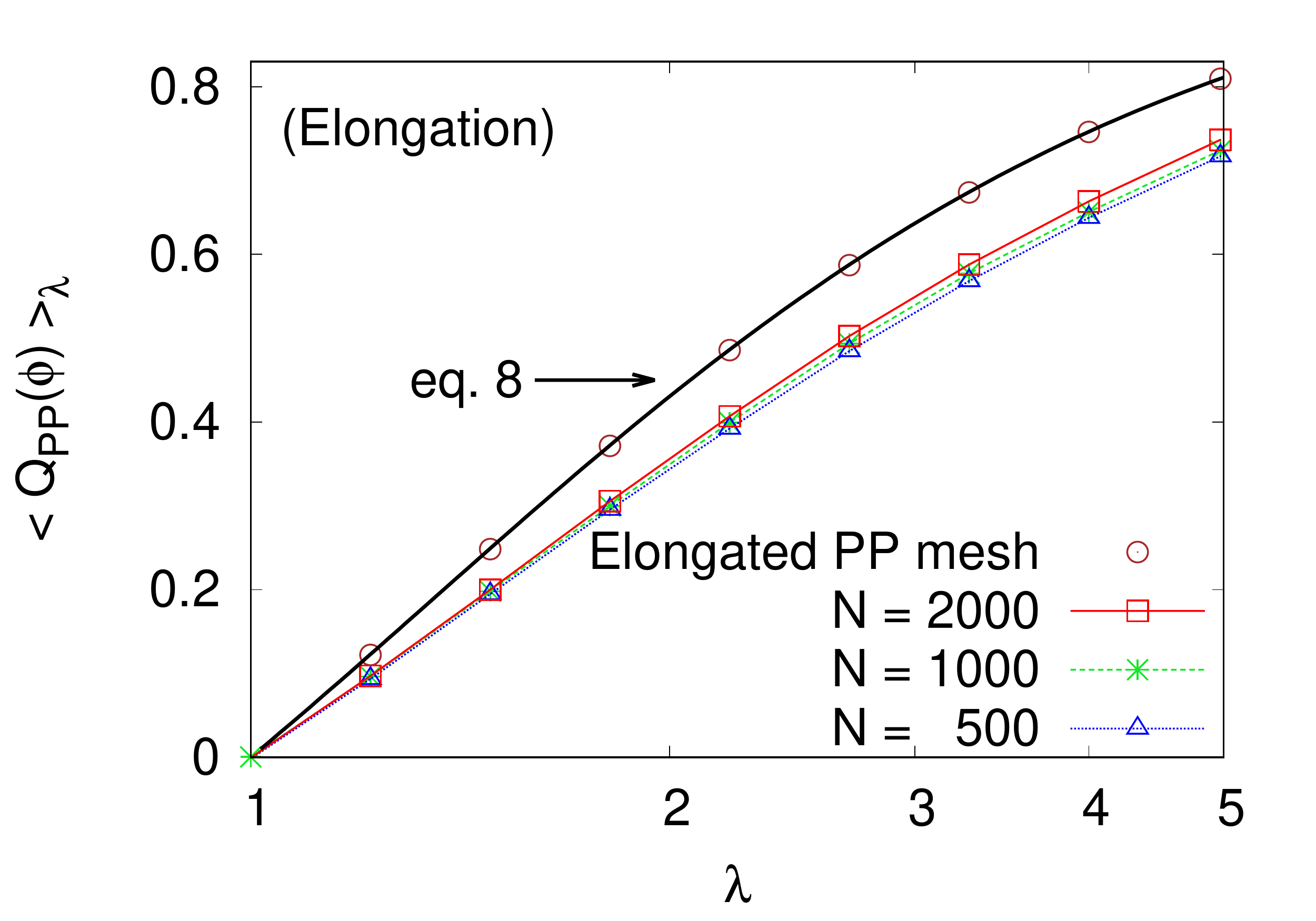}\hspace{0.2truecm}
(d)\includegraphics[width=0.32\textwidth,angle=270]{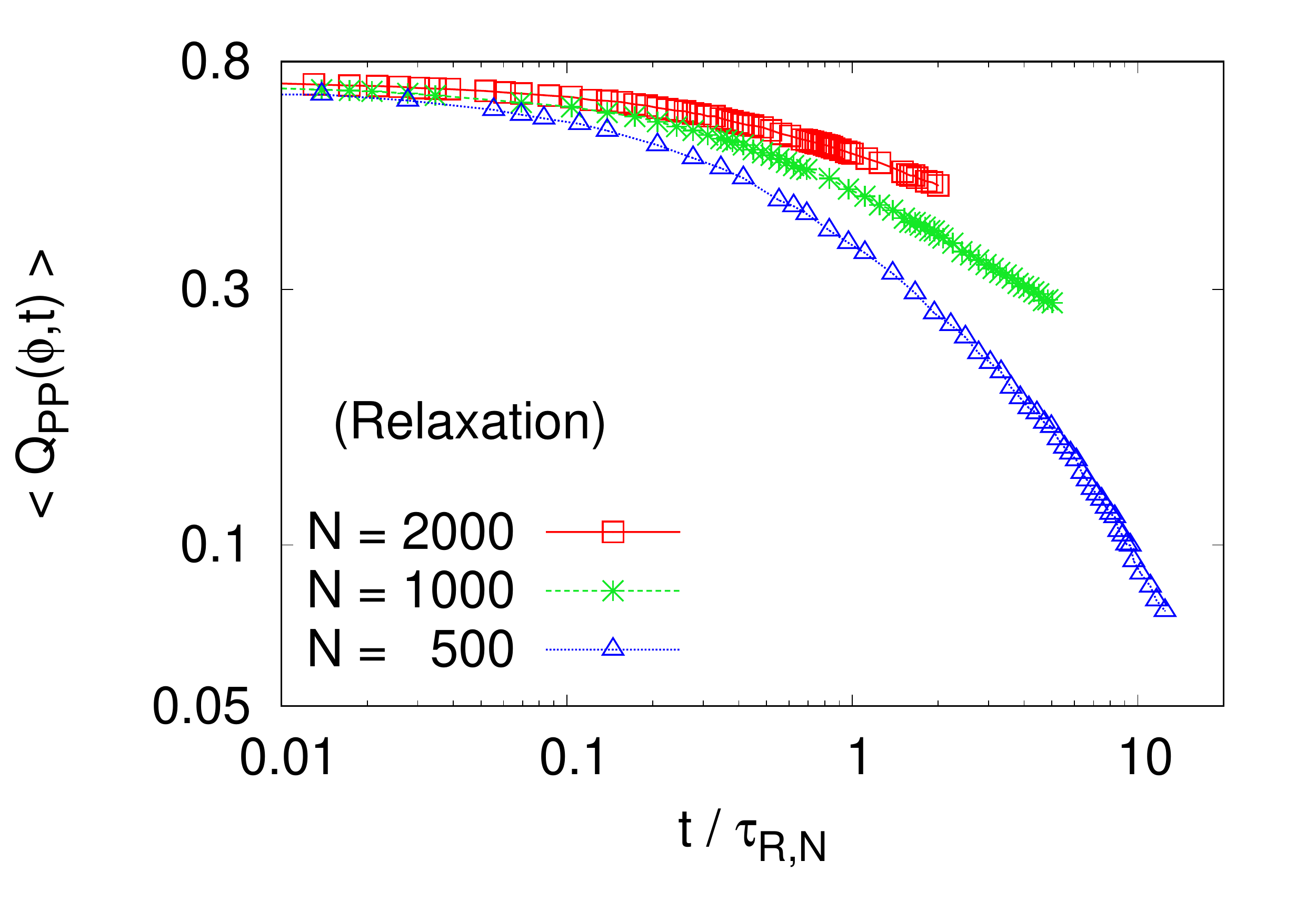}
\caption{Average bond length $\langle b_{PP} \rangle_\lambda$ (a) and orientational
order parameter $\langle Q_{PP}(\phi) \rangle_\lambda$ (c), plotted versus $\lambda$ for
elongated polymer melts. Average bond length $\langle b_{PP}(t) \rangle$ (b)
and orientational order parameter $\langle Q_{PP}(\phi,t) \rangle$ (d), plotted
versus the subsequent rescaled relaxation time $t/\tau_{R,N}$ after elongation.
In (a)(b), data are rescaled to $\langle b_{PP} \rangle_0 \approx 0.31\sigma$ for
the PPs of unperturbed chains in melts.
In (a)(c), data for elongated PP mesh are also shown for comparison.
In (a), the theoretical prediction \{eq.~\ref{eq-bpp-sph}\} for isotropic PPs
is also shown for comparison.}
\label{fig-bpp}
\end{center}
\end{figure*}

The average contour length of PPs is defined by $L_{PP}=(N-1) \langle b_{PP} \rangle$ with the average bond length of PPs
\begin{equation}
\langle b_{PP} \rangle=\frac{1}{n_c(N-1)} \sum_{i=1}^{n_c} \sum_{j=1}^{N-1}
\mid {\bf r}_{i,j+1}^{(PP)} - {\bf r}_{i,j}^{(PP)} \mid
\end{equation}
where ${\bf r}_{i,j}^{(PP)}$ is the coordinate of the $j$th monomer of the PP of chain $i$.
If we assume the distribution of the original bonds to be isotropic, taking the integral of the deformed bonds over the surface of unit sphere, normalized by $4\pi$, the average bond length as a function of the strain $\lambda$ is given by
\begin{widetext}
\begin{eqnarray}
\langle b_{PP} \rangle_{\lambda}
&=&
 \frac{1}{4\pi} \int_0^{2\pi} d\theta \int_0^\pi d \phi \sin \phi \mid {\bf E} \cdot {\bf u} \mid \nonumber \langle b_{PP} \rangle_0\\
&=& \left [
\frac{\lambda}{2}+\frac{1}{4\lambda h(\lambda)^{1/2}} \ln 
\left | \frac{h(\lambda)+\lambda \sqrt{h(\lambda)}}
{-h(\lambda)+\lambda \sqrt{h(\lambda)}} \right | \right] \langle b_{PP} \rangle_0 
\label{eq-bpp-sph}
\end{eqnarray}
\end{widetext}
with
\begin{equation}
     h(\lambda)=\lambda^2-1/\lambda
\end{equation}
where ${\bf E}$ is a diagonal deformation tensor with elements $\{\lambda, \lambda^{-1/2}, \lambda^{-1/2} \}$,
${\bf u}$ is a unit vector in the spherical coordinate, 
${\bf u}=(\cos \phi, \sin \phi \cos \theta, \sin \phi \sin \theta)$, $\phi$ is the polar angle 
between ${\bf u}$ and the $x$-axis, and $\theta$ is the azimuthal angle. 
One gets $\langle b_{PP} \rangle_{\lambda} \approx 2.56 \langle b_{PP} \rangle_0$ at $\lambda = 5.0$. 
The deviations between different approximations discussed in our previous work~\cite{Hsu2018a} are a consequence of the fact that there is a lower cutoff, given by $N_e$.
In general one should expect that $\langle b_{PP} \rangle_{\lambda}/\langle b_{PP}\rangle_0$ for the affinely elongated PP mesh follow  eq.~\ref{eq-bpp-sph}, which indeed is seen in Figure~\ref{fig-bpp}a. 
At the same time the deviation between the elongated original PP mesh and the PP meshes of deformed polymer melts becomes more pronounced with increasing strain $\lambda$.
At $\lambda =5.0$, $\langle b_{PP} \rangle_{\lambda} \approx 1.84 \langle b_{PP} \rangle_0$ for the deformed polymer melt, which is about $28\%$ below that of the deformed original PP mesh. 
At the subsequent relaxation after stretching, we see that in Figure~\ref{fig-bpp}b, $\langle b_{PP}(t) \rangle$ reaches $\langle b_{PP} \rangle_0$ for $N=500$ and $N=1000$ for $t>\tau_{R,N}$ while for $N=2000$, $\langle b_{PP}(t) \rangle$ seems to settle at a slightly larger value than $\langle b_{PP} \rangle_0$.

Choosing the $x$-axis (along the stretching direction) as reference,
the orientational order parameter of bond vectors along PPs is defined by $Q_{PP}(\phi)=(3 \cos^2 \phi -1)/2$
where $\phi$ is the angle between the bond vector ${\bf b}_{PP}$ and the $x$-axis.
Assuming the distribution of the original bonds to be isotropic, $\langle Q_{PP}(\phi) \rangle_\lambda$ for the PP meshes of polymer melts under elongation is given by
\begin{widetext}
\begin{eqnarray}
   \langle Q_{PP}(\phi) \rangle_\lambda &=&
\left [\frac{3}{2\pi}\int_0^{2 \pi} d \theta \int_0^{\pi} d \phi \sin \phi
\left(\frac{{\bf E} \cdot {\bf u}}{\mid {\bf E} 
\cdot {\bf u} \mid} \cdot \hat{e}_x \right)^2 -1 \right]/2
\nonumber \\
&=& \left [ \frac{3\lambda^2}{h(\lambda)} \left (
1-\sqrt{\frac{1}{\lambda h(\lambda)}} \tan^{-1}(\sqrt{\lambda h(\lambda)}) \right )-1 \right]/2 \,.
\label{eq-Q-sph}
\end{eqnarray}
\end{widetext}

\noindent 
Under elongation Figure~\ref{fig-bpp}c shows that the orientation distribution of bonds changes from isotropic, $\langle Q_{PP}(\phi) \rangle_\lambda=0.0$ for $\lambda=1.0$, to anisotropic, $\langle Q_{PP}(\phi) \rangle_\lambda \approx 0.8$ for $\lambda=5.0$, where most bond vectors align along the stretching direction. 
The subsequent relaxation of bond orientation (Figure~\ref{fig-bpp}d) is apparently significantly delayed compared to that of the bond length. 
Furthermore the relaxation rate decreases as the chain size $N$ increases, 
Despite quantitative discrepancy for $N=2000$ both $\langle b_{PP}(t) \rangle$ and $\langle Q_{PP}(\phi,t) \rangle$ show relaxation retardation starting at $t \approx \tau_{R,N}$.

\begin{figure*}[t!]
\begin{center}
(a)\includegraphics[width=0.32\textwidth,angle=270]{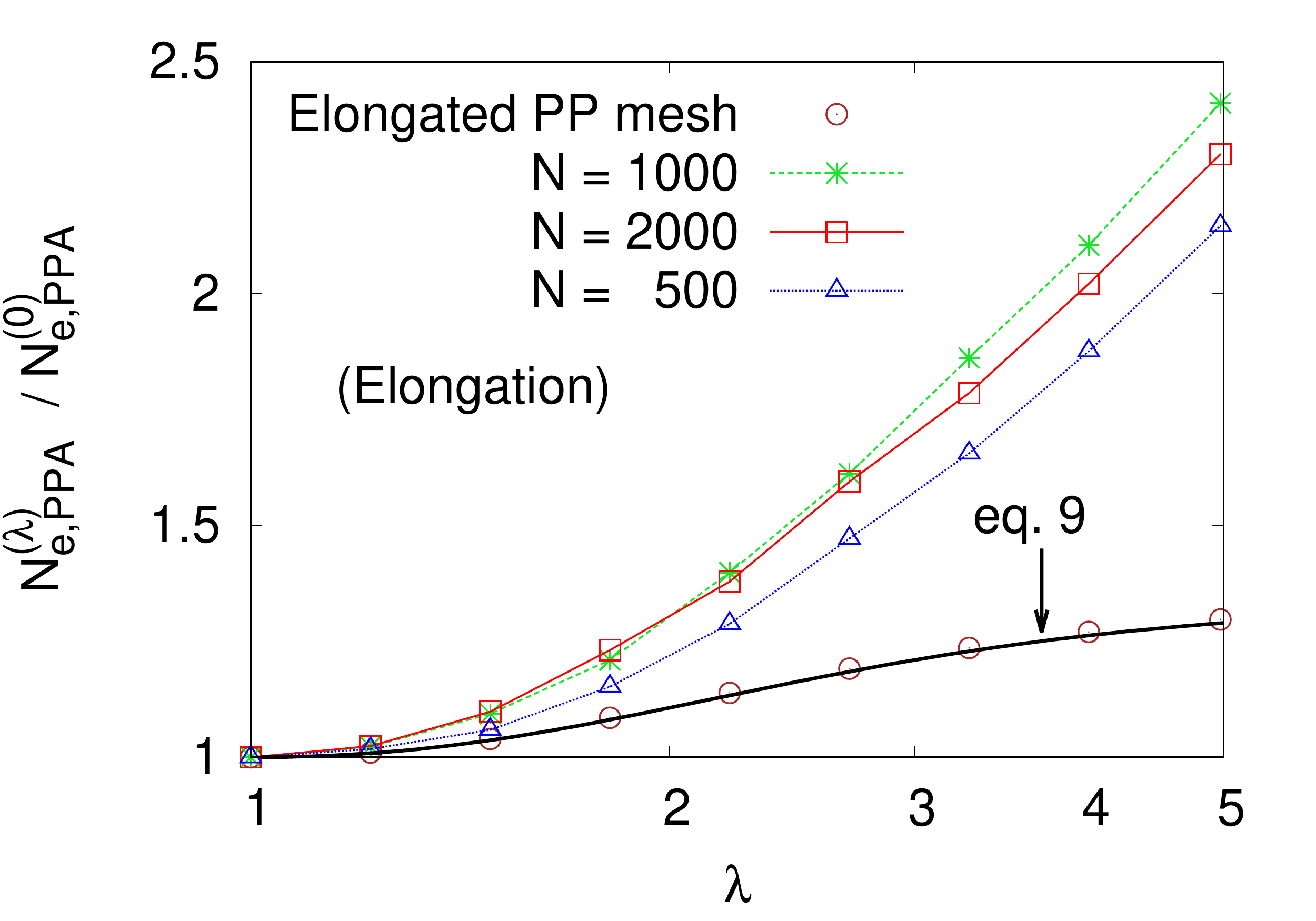}\hspace{0.2truecm}
(b)\includegraphics[width=0.32\textwidth,angle=270]{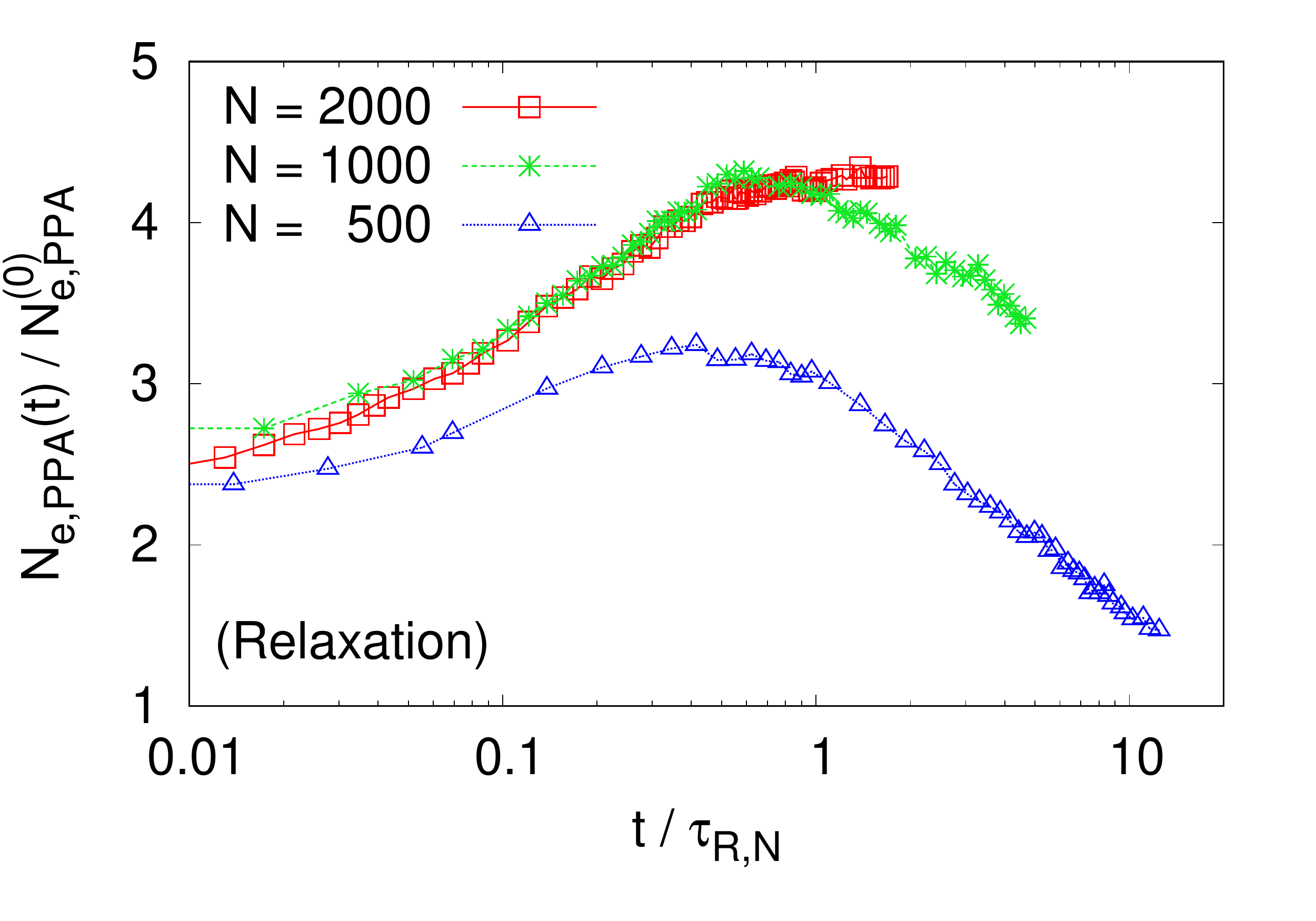}
\caption{Rescaled entanglement length estimated according to Gaussian chain assumption,
$N_{\rm e,PPA}^{(\lambda)}/N_{\rm e,PPA}^{(0)}$ plotted as a function of $\lambda$ (a), and
$N_{\rm e,PPA}(t)/N_{\rm e,PPA}^{(0)}$ plotted as a function of $t/\tau_{R,N}$ (b).
Here $N_{e,PPA}^{(0)} \approx 28$ for unperturbed polymer melts.
In (a), data for elongated PP mesh and
the theoretical prediction \{eq.~\ref{eq-Ne}\} for affine deformation are
also shown for comparison.}
\label{fig-Ne}
\end{center}
\end{figure*}

\begin{figure*}[t!]
\begin{center}
(a)\includegraphics[width=0.30\textwidth,angle=270]{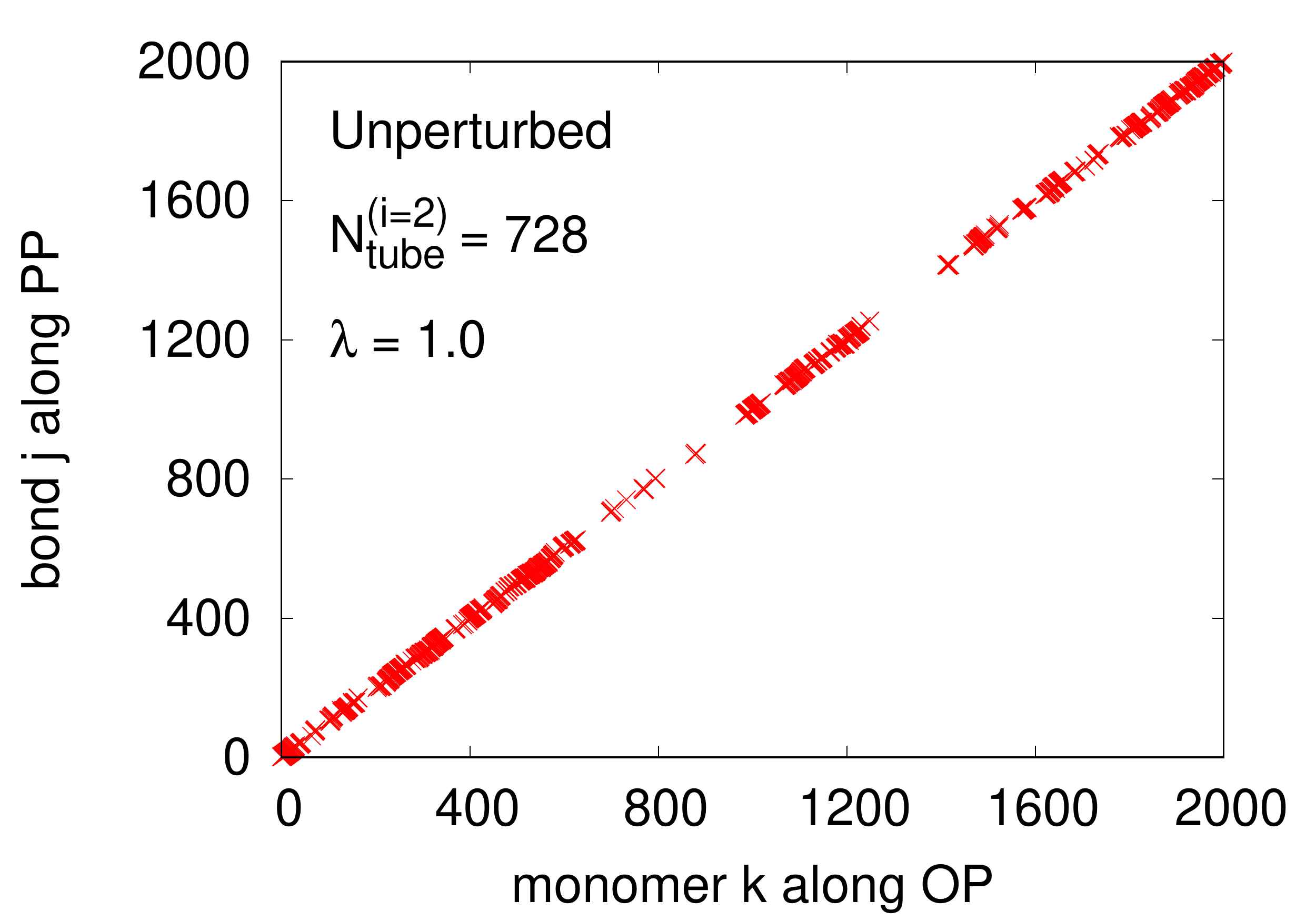} \hspace{0.4cm}
(b)\includegraphics[width=0.30\textwidth,angle=270]{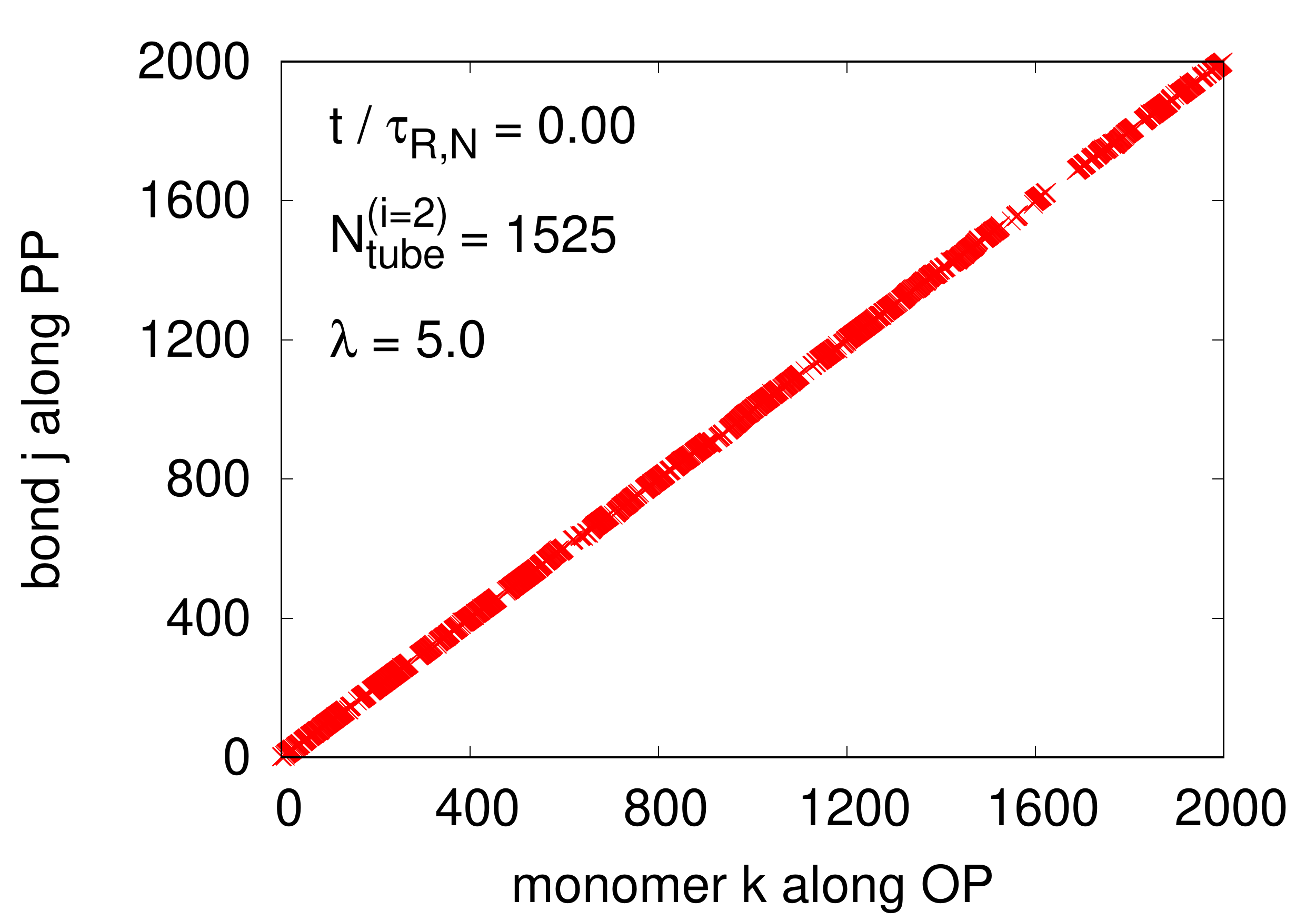}\\
(c)\includegraphics[width=0.30\textwidth,angle=270]{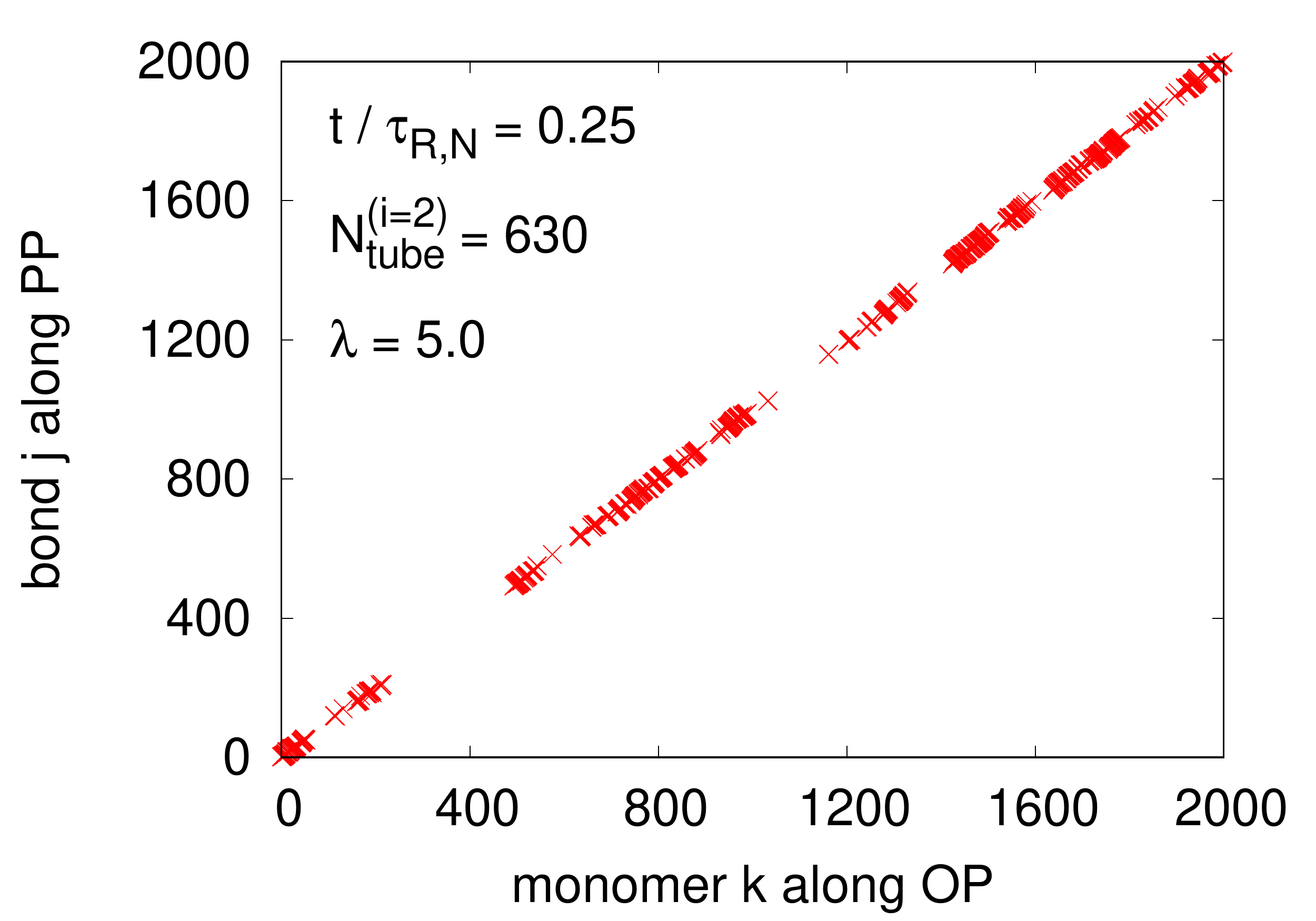} \hspace{0.4cm}
(d)\includegraphics[width=0.30\textwidth,angle=270]{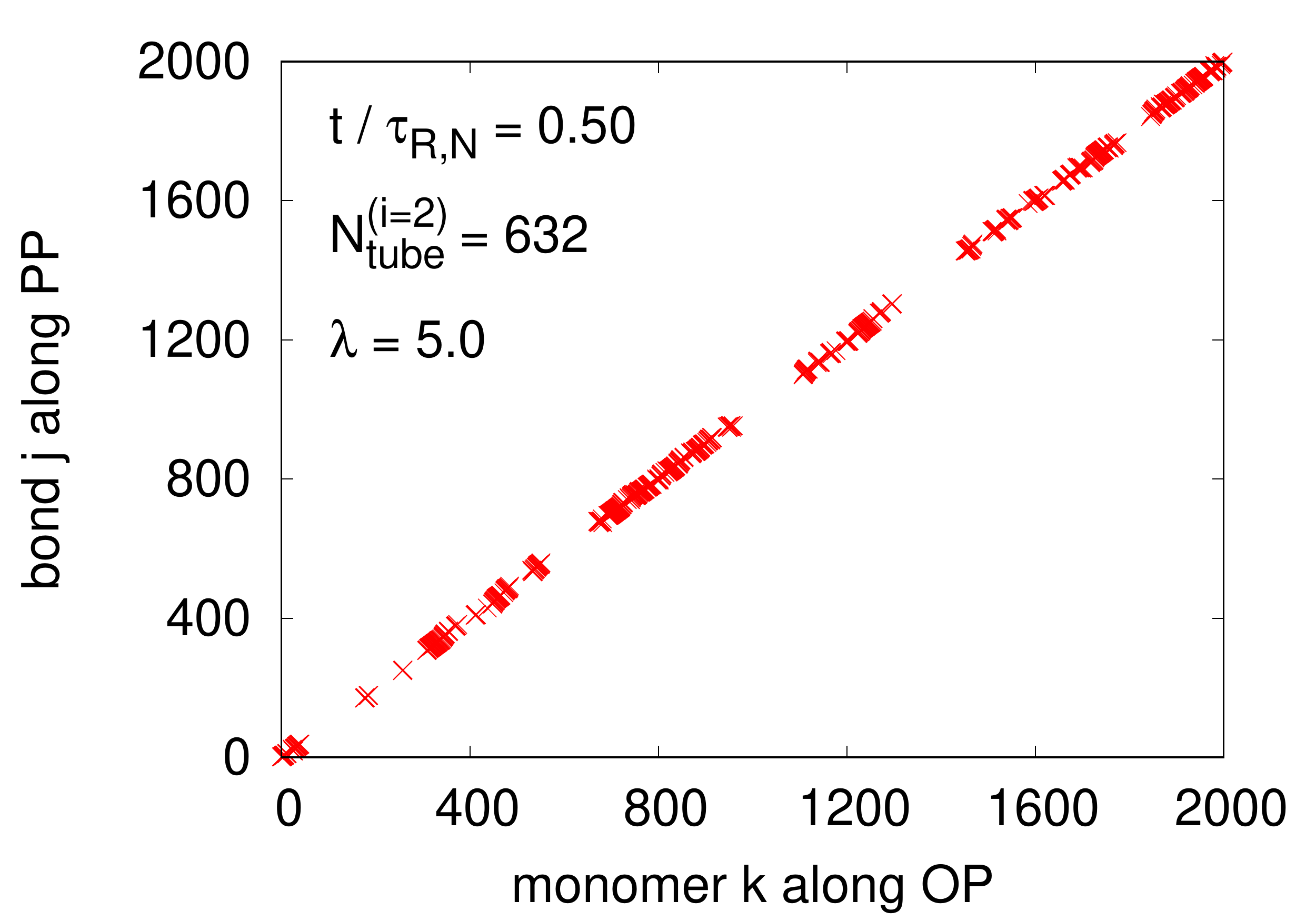}\\
(e)\includegraphics[width=0.30\textwidth,angle=270]{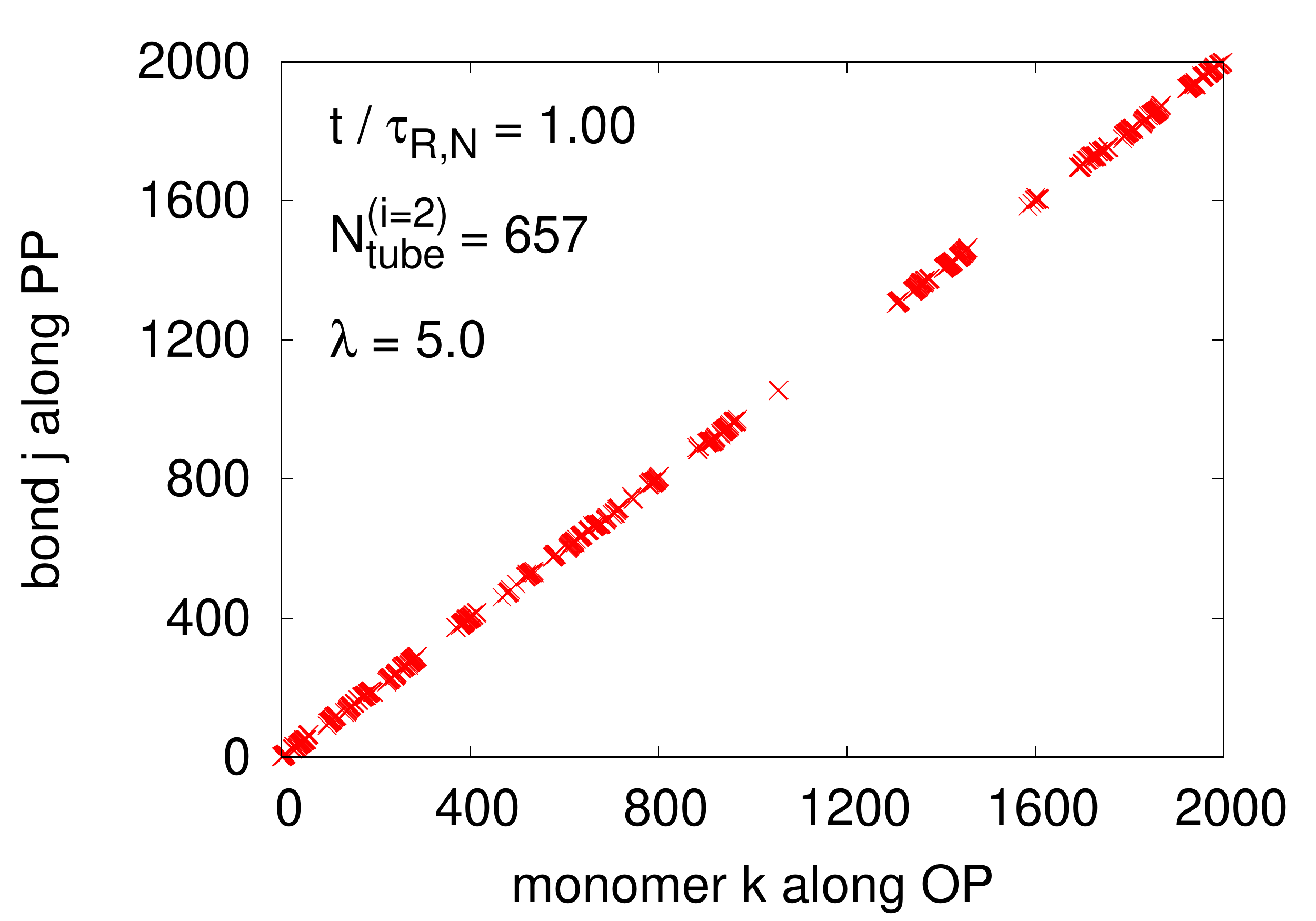} \hspace{0.4cm}
(f)\includegraphics[width=0.30\textwidth,angle=270]{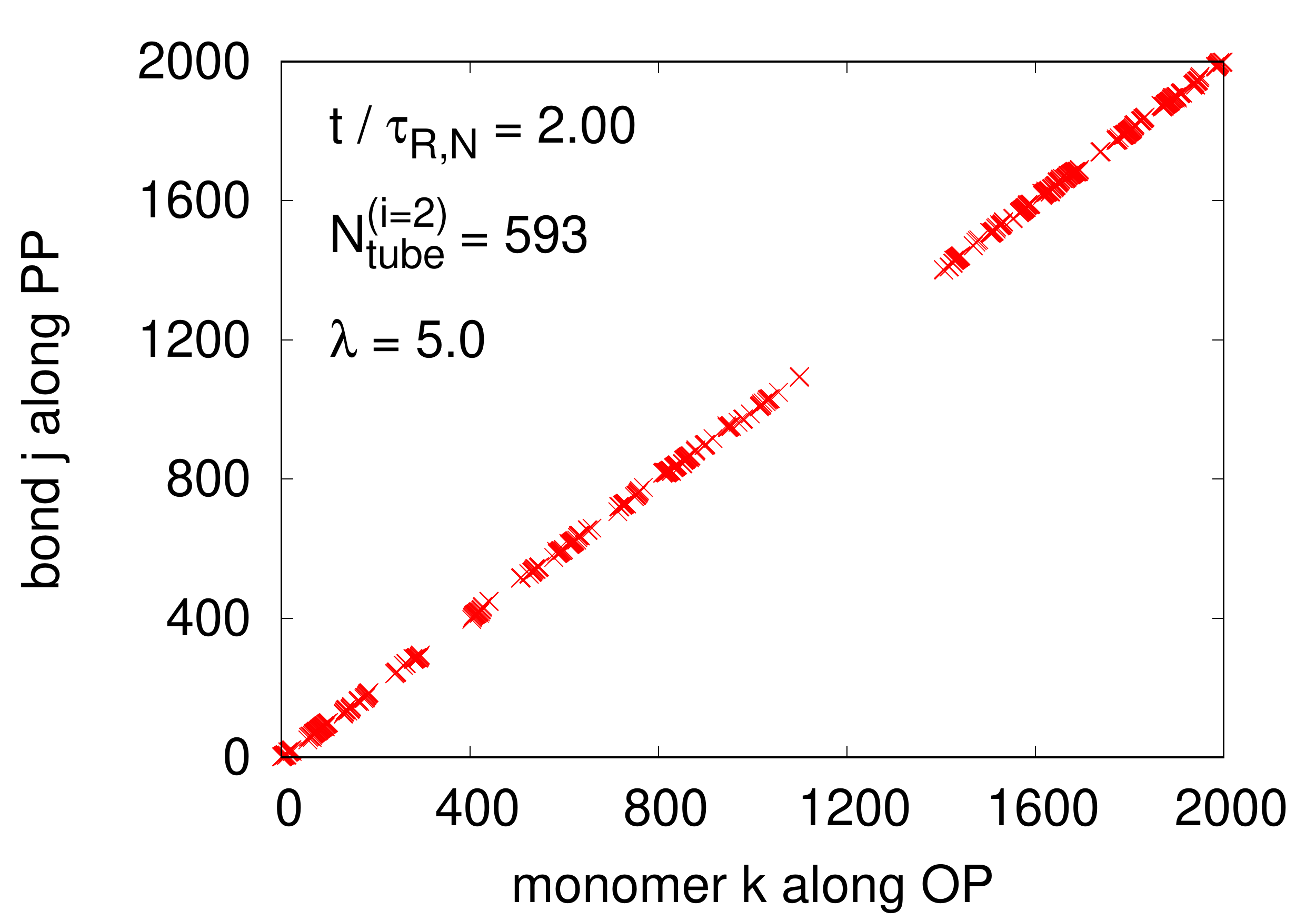}\\
\caption{Monomers along the original path of the same selected chain $i=2$ of size $N=2000$ as shown
in Figure~\ref{fig-fcosbb-N2000} located in the tube-like regime along the primitive path at six
different states: (a) unperturbed ($\lambda=1.0$), (b) after deformation ($\lambda \approx 5.0$),
(c)(d)(e)(f) at the rescaled relaxation times $t=0.25$, $0.50$, $1.0$, and $2.0$, respectively.
The estimates of $N^{(i=2)}_{\rm tube}$ are also shown for comparison.
Note that only small fluctuations are observed within $t \pm \tau_e$ at each time $t$.}
\label{fig-Ntube}
\end{center}
\end{figure*}

\begin{figure*}[t!]
\begin{center}
(a)\includegraphics[width=0.30\textwidth,angle=270]{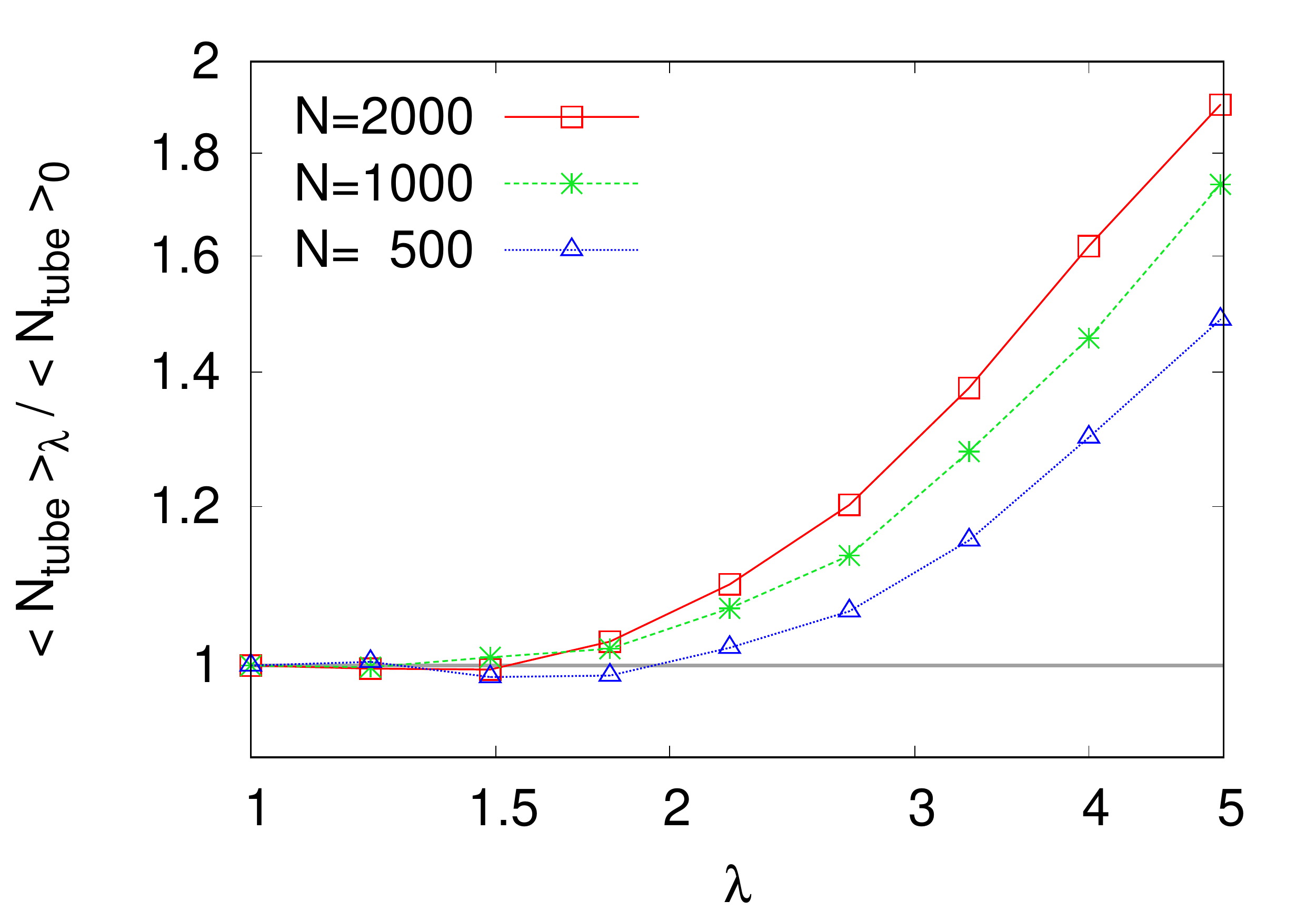} \hspace{0.4cm}
(b)\includegraphics[width=0.30\textwidth,angle=270]{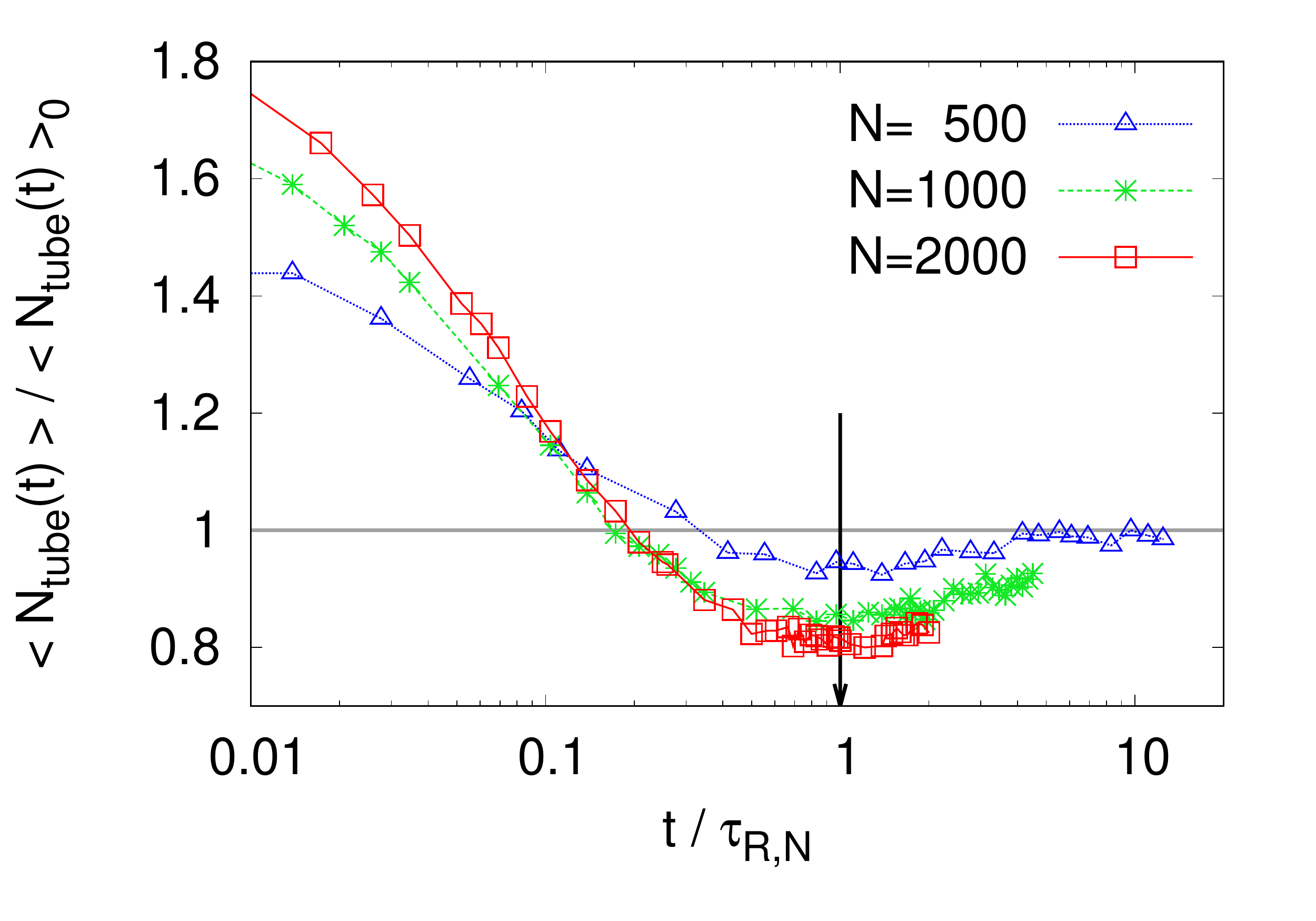}\\
(c)\includegraphics[width=0.30\textwidth,angle=270]{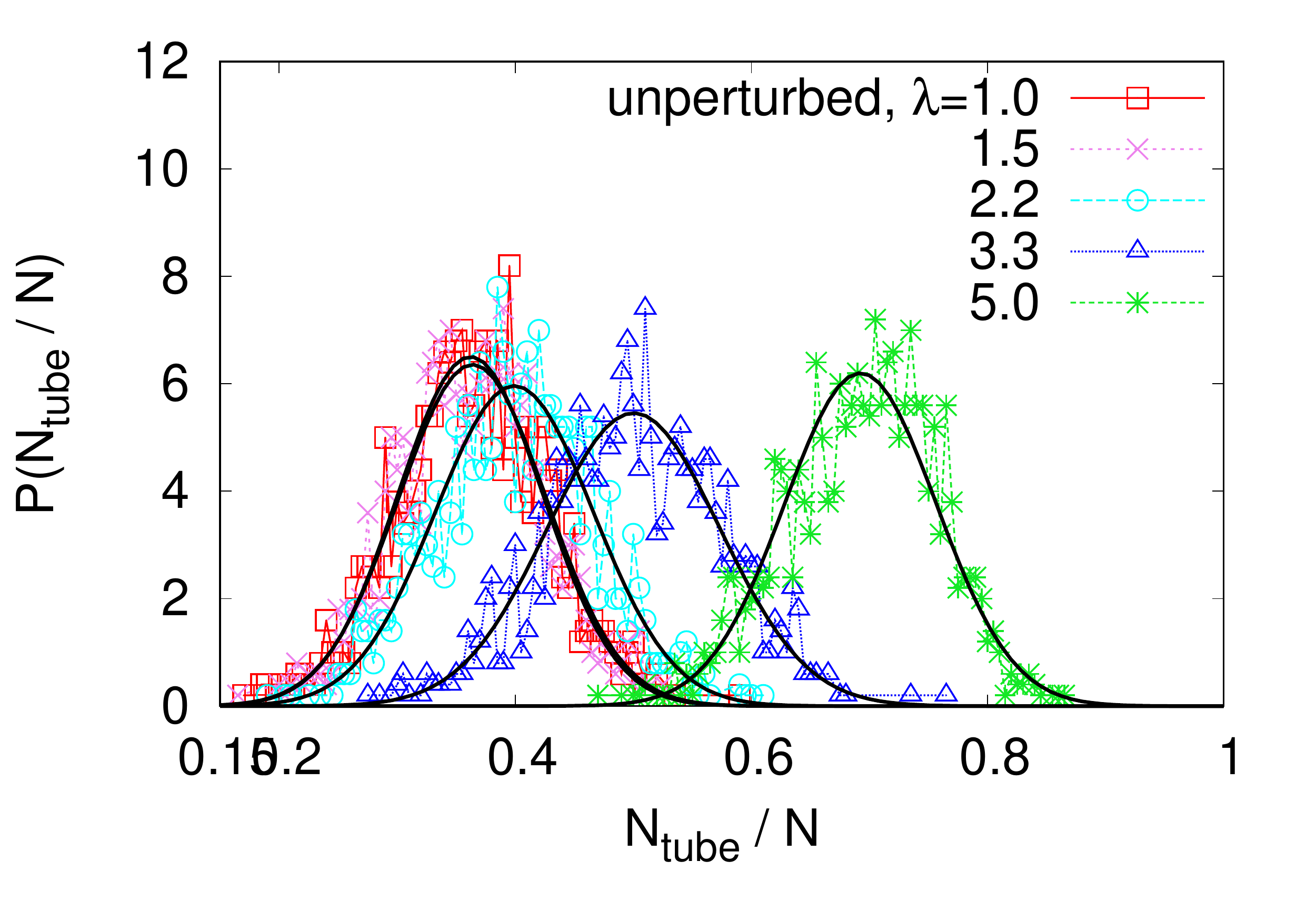} \hspace{0.4cm}
(d)\includegraphics[width=0.30\textwidth,angle=270]{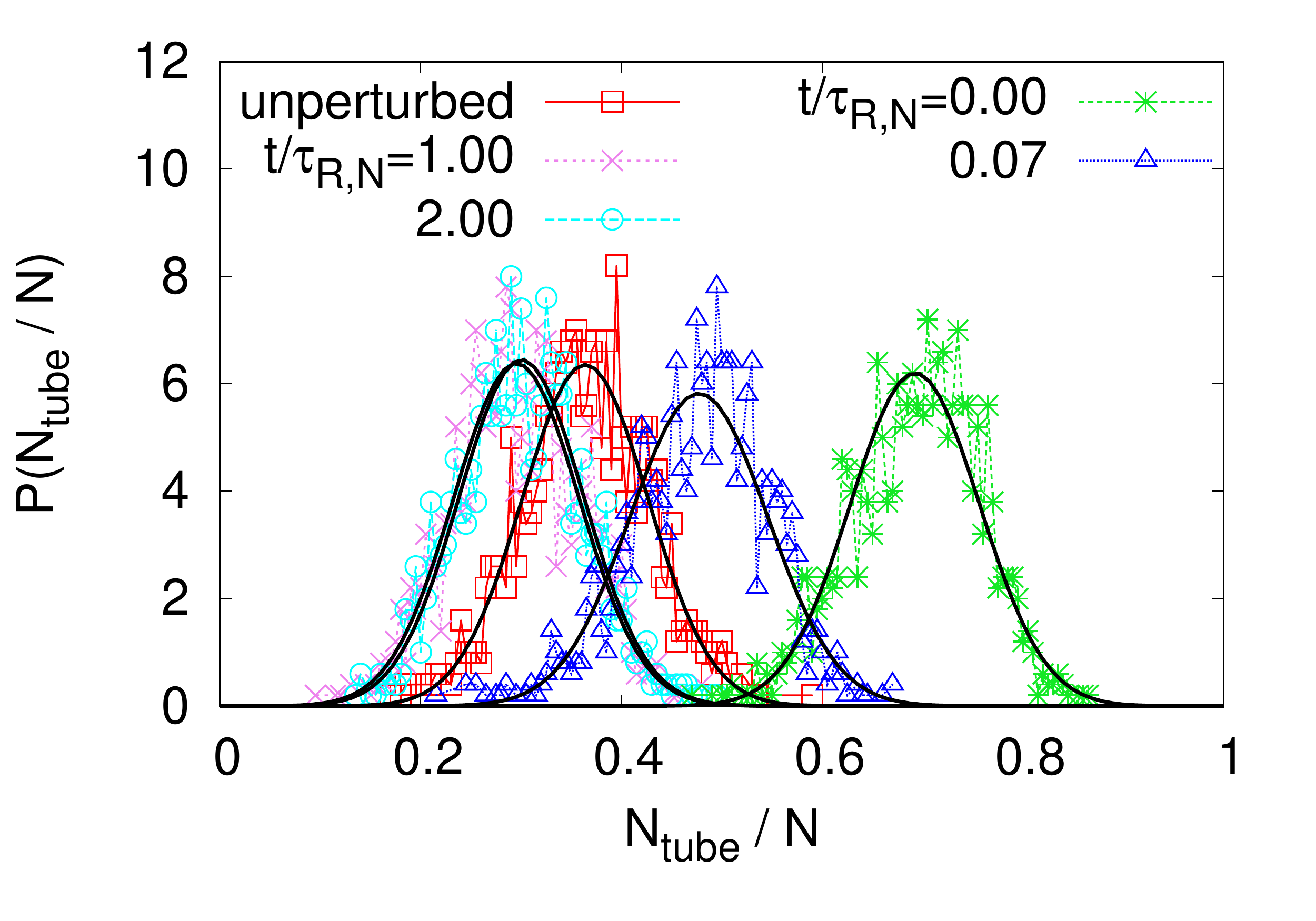}\\
\caption{Average number of monomers in the tube-like regime,
$\langle N_{\rm tube} \rangle_\lambda$ (a), and
$\langle N_{\rm tube}(t) \rangle$ (b) rescaled to $\langle N_{\rm tube} \rangle_0 \approx 200.50$,
$385.50$, and $727.55$ for unperturbed polymer melts of chain sizes $N=500$, $1000$, and $2000$,
respectively.
Normalized histogram of number of monomers in the tube-like regime
along the primitive path, $P(N_{\rm tube}/N)$, for deformed polymer melts
 at several chosen strain values of $\lambda$ (c) and at several selected rescaled relaxation
times $t/\tau_{R,N}$ at $\lambda \approx 5.0$.
In (c)(d), data are only for $N=2000$.} 
\label{fig-pNtube}
\end{center}
\end{figure*}

Finally, we turn to the entanglement length $N_{e,PPA}$ as estimated from the PPA for unperturbed and strongly deformed polymer melts, cf. Figures~\ref{fig-Ne}-\ref{fig-pNtube}.
For strongly deformed polymer melts, the estimate of $N_{e,PPA}$ becomes less obvious, since the basic original concept assumes a Gaussian chain conformation. 
The Gaussian chain assumption still holds for each component individually, however, is different in different directions since the contours of PPs globally deform affinely. 
Keeping this in mind, we nevertheless apply the standard formula $\langle R_e^2 \rangle =\langle (R_e^{(PP)})^2 \rangle= L_{PP}\ell_K^{(PP)}=(N-1)N_{e,PPA}\langle b_{PP} \rangle^2$, $\ell_K=N_{e,PPA}\langle b_{PP} \rangle$ being the Kuhn length of the PPs of chains~\cite{Everaers2004,Sukumaran2005}.
For the overall affine deformation of the chains along the three orthogonal directions (see Figure~\ref{fig-Rgpp-el}, one would expect $\langle R_e^2 \rangle_{\lambda}=\lambda^2 \langle R_{e,x}^2 \rangle_0+\langle R^2_{e,y} + R^2_{e,z} \rangle_0/\lambda = (\lambda^2 + 2/\lambda)\langle R_e^2 \rangle_0/3$.
Following eq.~\ref{eq-bpp-sph}, we obtain 
\begin{widetext}
\begin{equation}
\frac{N_{e,PPA}^{(\lambda)}}{N_{e,PPA}^{(0)}} = \frac{(\lambda^2 +2/\lambda)}{3}
\left \{\frac{\lambda}{2}+\frac{1}{4\lambda h(\lambda)^{1/2}} \ln 
\left | \frac{h(\lambda)+\lambda \sqrt{h(\lambda)}}
{-h(\lambda)+\lambda \sqrt{h(\lambda)}} \right | \right \}^{-2} \,.
\label{eq-Ne}
\end{equation}
\end{widetext}\\
\noindent
Figure~\ref{fig-Ne}a shows that $N_{e,PPA}^{(\lambda)}$ for the elongated PP mesh follows affine deformation up to $\lambda \approx 5.0$ while for the PPs of the deformed polymer melts it does not. 
This is consistent with the estimates of ``effective entanglement length'' $\langle l_{\rm str} \rangle_\lambda \propto 1/\langle N_{\rm kink} \rangle_\lambda $ (Figure~\ref{fig-Nkink}a).
At subsequent relaxation time $t \approx 0.5 \tau_{R,N}$, $N_{e,PPA} (t)$ reaches a maximum for $N=500$, and $1000$ while for $N=2000$, it seems to reach a plateau value without any further decay within the time window studied here,
and the plateau value is very close to the effective entanglement length ($100$) extracted from the 
long tail probability distribution at $t=\tau_{R,N}$. 
This is another direct piece of evidence for a significantly delayed relaxation.

Along the original path the monomers can fluctuate in space, confined by fuzzy tube boundaries.
If the average monomer positions and thus the tube are deformed affinely, 
one would expect that the average number of monomers located inside the tube
of tube diameter $d_T=d_T^{(0)}/\sqrt{\lambda}$ , $\langle N_{\rm tube} \rangle$ is approximately kept as a constant. 
The GLaMM model even assumes that $d_T=d_T^{(0)}$ remains unchanged, which suggests that even more monomers would be located inside the tube, since the contour length of the tube becomes larger.
From our results shown in previous sections, we find that this cannot be expected here.
Still, following the assumption of the GLaMM model and fixing the tube diameter to that of the fully equilibrated melt, namely to~\cite{Hsu2016}  
$d_T^{(0)}\approx \sqrt{2}\langle R_g^2(N_e) \rangle_0^{1/2} \approx 5.02 \sigma \propto \sigma \sqrt{N_{e,PPA}^{(0)}}$, we follow the variation of $N_{\rm tube}^{(i)}$ for each chain $i$ and $\langle N_{\rm tube} \rangle$ upon deformation and subsequent relaxation.
The way of counting $N_{\rm tube}^{(i)}$ for each chain $i$ is as follows:
To determine whether monomer $k$ is located within its actual reptation tube we assume the tube to be constructed of piecewise cylinders with diameter $d_T^{(0)}$ and length $d_T^{(0)}$.
Monomer $k$ belongs to the tube if the shortest distance to any PP bond $j$ with $k-\delta \le j \le k+\delta$ is less or equal to $d_T^{(0)}/2$ with $\delta=8$, being determined by $d_T^{(0)}$ ($\delta \langle b_{PP} \rangle_0 \sim d_T^{(0)}/2$). Note that the precise number of $N_{\rm tube}$ depends on details of the algorithm and the chosen value for $\delta$, but results of $\langle N_{\rm tube} \rangle$ should be qualitatively the same. 
To provide some insight into the relation between the location of kinks and the fluctuation of monomers of OP along the PP for the same chain, we choose the same chain $i=2$ of size $N=2000$ as it is selected in Figure~\ref{fig-fcosbb-N2000} for comparison.
Indeed, the monomer $k$ of the OP belonging to the bond $j$ along the PP of chain $i=2$ in a melt presented in Figure~\ref{fig-Ntube} show approximate agreement with to the data before deformation, and at several selected subsequent relaxation times, but not right after deformation. 
This makes sense since monomers along the OP have less freedom moving away from the tube-like regime along its corresponding PP as the chain is strongly stretched, and thus $N_{\rm tube}^{(i)}$ increases. 
Once the deformed chain starts to relax, $N_{\rm tube}^{(i)}$ immediately decreases due to the dramatic changes of the entangled surrounding, and then monomers of OP between two neighboring kinks can start to fluctuate.

$\langle N_{\rm tube} \rangle$, cf. Figure~\ref{fig-pNtube}ab, fits to the overall scheme observed so far. 
In the linear regime ($\lambda < 1.5$), 
$\langle N_{\rm tube} \rangle_\lambda \approx \langle N_{\rm tube} \rangle_0$.
In the non-linear regime ($\lambda>1.5$), upon deformation $\langle N_{\rm tube} \rangle_\lambda$ increases with a rate larger with larger $N$. 
Right after deformation, at the subsequent initial relaxation time, $\langle N_{\rm tube}(t) \rangle$ decreases, if normalized by the Rouse time. 
Eventually, it even drops below the equilibrium value, slows down for $t/\tau_{R,N}>0.2$, and 
reaches a minimum around the Rouse time. 
However, the minimum is quite shallow and shows the delayed relaxation around the Rouse time.
The relaxation back to the unperturbed tube occupancy at least takes several Rouse times of the chains, just as for other tube related quantities studied here. 
Choosing $d_T=d_T^{(0)}(N^{(\lambda \approx 5.0)}_{e,PPA}/N^{(0)}_{e,PPA})^{1/2}\approx 7.61\sigma \approx 1.5d_T^{(0)}$, the profiles of $\langle N_{\rm tube}(t)\rangle / \langle N_{\rm tube} \rangle_0$
having the minimum around $t \approx \tau_{R,N}$ within fluctuation are similar as the results 
shown in Figure~\ref{fig-pNtube}b, 
while quantitatively, the estimates of $\langle N_{\rm tube}(t)\rangle$ are somewhat larger.
Again this agrees with the formation and growth of topologically highly congested areas along the chains. 
Regions of low density of entanglement points, where configurations of monomers along the OPs can fluctuate significantly in space, seem to stabilize regions with high density of entanglement points.
The probability distributions $P(N_{\rm tube}/N)$ as a function of $(N_{\rm tube}/N)$ for deformed polymer chains of size $N=2000$ in a melt at several selected strain values of $\lambda$ and at several selected rescaled relaxation times $t/\tau_{R,N}$ after deformation are shown in Figure~\ref{fig-pNtube}c,d.
We see that the $P(N_{\rm tube})$ is simply a shifted Gaussian distribution in terms of the mean value $\langle N_{\rm tube} \rangle$, and the standard deviation $\sigma(N_{\rm tube})=\sqrt{\langle N^2_{\rm tube} \rangle -\langle N_{\rm tube} \rangle^2}$.
The distribution $P(N_{\rm tube}/N)$ remains the same for $\lambda=1.0$ and $\lambda \approx 1.5$, and then the profile of $P(N_{\rm tube}/N)$ shifts to a larger value of $N_{\rm tube}/N$ as $\lambda$ increases.
After deformation, the profile shifts to smaller value of $N_{\rm tube}/N$ with increasing $t$, and even moves to the left-hand side of the profile for unperturbed chains similar as the behavior of $\langle N_{\rm tube}(t) \rangle$ shown in Fig.~\ref{fig-pNtube}b.

We see that the intrinsic properties of PPs, analyzed according to the tube concept 
provide profound insights into the relaxation paths of entangled chains in deformed polymer melts
although the ``effective entanglement length'' for deformed chains in a melt can no longer 
be extracted from the theoretical considerations for the PPA~\cite{Everaers2004,Sukumaran2005}.

\subsection{Stress relaxation}

 After having discussed details of individual and collective conformational relaxation we turn to the related stresses in the systems, a quantity which would be experimentally accessible more easily. 
Since the entanglement structure of melts is closely connected to their viscous and elastic properties we expect that the previously described primitive path relaxation also leads to characteristic signals in the viscoelasticity of the strained melts. 
The viscoelasticity of polymer melts is normally characterized by the time dependent stress relaxation modulus $G(t)$. 
For fully equilibrated, entangled polymer melts, the short time dynamics of chains is described by the Rouse model.
This leads to $G(t) \sim t^{-1/2}$ for $t<\tau_e$, while for $\tau_e <t \ll \tau_{d,N}$  ($\tau_{d,N}=\tau_{R,N}(N/N_e)^{1.4}$ being the disentanglement time of chains of size $N$) where chains are assumed to move in a tube-like regime due to entanglements, $G(t)$ reaches a plateau value $G_N^0=(4/5)(\rho k_BT/N_e)$, which depends on the entanglement length or the molecular weight between entanglements as predicted by the Doi-Edwards tube model~\cite{Doi1980,Doi1986}. 
In the linear viscoelastic regime, $G(t)$ and its approach to $G_N^0$ for intermediate time scales are well understood. 
In contrast our understanding of the stress relaxation scenario for strongly deformed polymer melts in the non-linear viscoelastic regime and the relationship between the strain rate and stress relaxation is significantly less developed.

\begin{figure*}[t!]
\begin{center}
(a)\includegraphics[width=0.32\textwidth,angle=270]{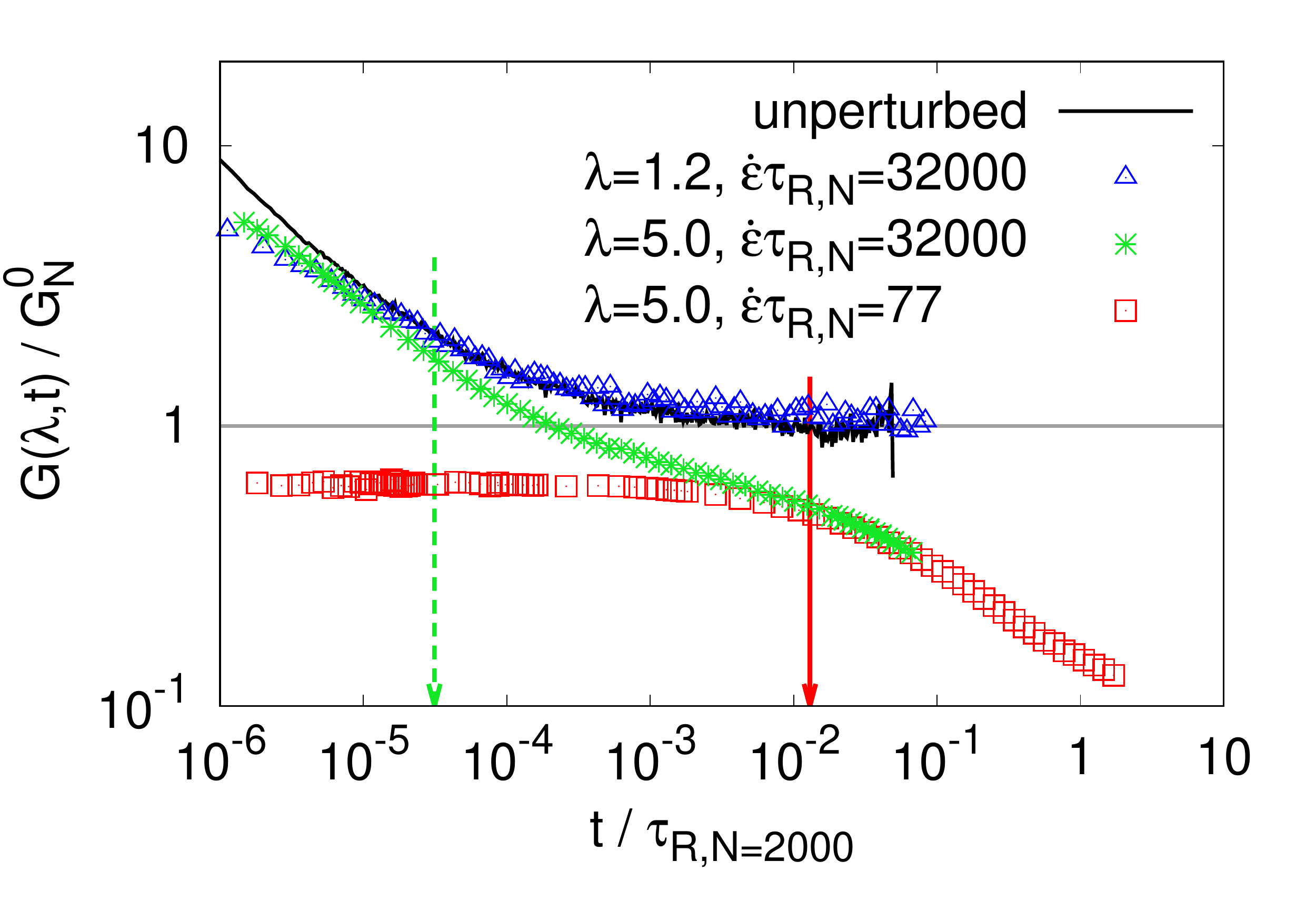}\hspace{0.2truecm}
(b)\includegraphics[width=0.32\textwidth,angle=270]{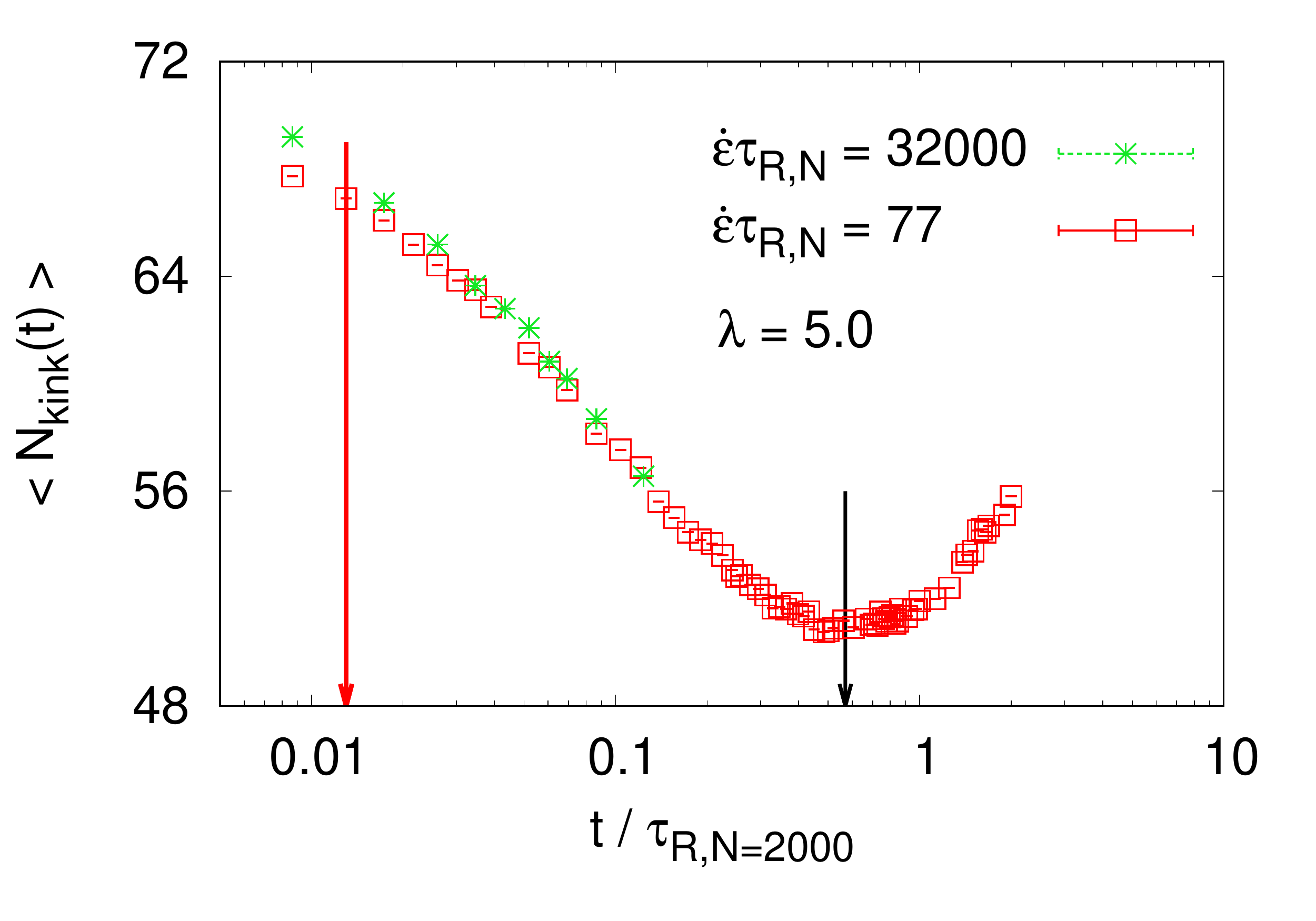}
\caption{(a) Rescaled stress relaxation modulus $G(\lambda,t)/G_N^0$,
and (b) average number of kinks per chain estimated from the curvature of the PPs of chains
, $\langle N_{\rm kink}(t) \rangle$, plotted as a
function of the rescaled relaxation time $t/\tau_{R,N}$ after deformation.
The corresponding strain $\lambda$ and the strain rate $\dot{\varepsilon}$ are shown
as indicated. Data are for $n_c=1000$ chains of chain size $N=2000$.
$t=\dot{\varepsilon}^{-1}=\tau_{R,N}/32000$ and $\tau_{R,N}/77$ in (a), and $t=\tau_{R,N}/77$ and 
$0.57\tau_{R,N}$ in (b) are indicated by arrows. 
Parts of data in (a) are taken from Refs.~\citenum{Hsu2016,Hsu2018a}.}
\label{fig-Gtr-Nkink}
\end{center}
\end{figure*}

Results of the stress relaxation modulus $G(\lambda,t)$ that characterize the viscoelasticity of entangled polymer melts in both linear (unperturbed and $\lambda \approx 1.2$) and non-linear regimes ($\lambda \approx 5.0$) are shown in Figure~\ref{fig-Gtr-Nkink} (parts of data have been shown in Refs.~\citenum{Hsu2016,Hsu2018a}).
For clarity only the case of $N=2000 (\approx 72 N_e)$ is discussed here.
For equilibrium polymer melts $G(\lambda=1.0,t)$ is calculated  from the stress autocorrelation function of off-diagonal elements of the stress tensor~\cite{Lee2009,Lee2010,Hsu2016}, using the Green-Kubo relationship.
$G(\lambda=1.0,t)$ reaches a plateau value $G_N^0$ with $N_e \approx 28$ for $\tau_{d,N} \gg t>\tau_e \approx 2\times 10^{-4}\tau_{R,N}$.
Alternatively $G(\lambda,t)$ can also be given by the normal stress difference after the simulation box is deformed, i.e. stretched along the $x$-direction,
\begin{equation}
       G(\lambda,t)=\sigma_{norm}(\lambda,t)/(\lambda^2-1/\lambda) \,
\end{equation}
where $\sigma_{norm}(\lambda,t)=\sigma_{xx}-\frac{1}{2}(\sigma_{yy}+\sigma_{zz})$.
For small and fast deformation ($\lambda \approx 1.2$ and $\dot{\varepsilon}\tau_{R,N}=32000$), $G(\lambda \approx 1.2,t)$ follows $t > \tau_e$ the data~\cite{Hsu2016} calculated from the Green-Kubo relation, and in good agreement with experiment~\cite{Liu2006} and the tube model~\cite{deGennes1979,Doi1980,Doi1986}. 
For the large but relatively slow deformation ($\lambda \approx 5.0$ and $\dot{\varepsilon}\tau_{R,N}=77$), $G(\lambda \approx 5.0,t)$ already from the very beginning displays a plateau modulus $G_N^0$ below that of the linear regime. 
 This is consistent with the experimental finding that one observes shear thinning in melt under such conditions~\cite{Luap2005,Auhl2008}. 
Only after a time corresponding to the inverse strain rate, i.e. $t=\dot{\varepsilon}^{-1}=\tau_{R,N}/77 \approx 10^{-2}\tau_{R,N}$ a further decay of $G(t)$ is observed. 
Keeping $\lambda \approx 5.0$, but choosing the same very fast strain rate  $\dot{\varepsilon}\tau_{R,N}=32000$ as for $\lambda \approx 1.2$, we see that $G(\lambda \approx 5.0,t)$ initially follows the data of $G(\lambda \approx 1.2,t)$ up to   $\dot{\varepsilon}^{-1}=\tau_{R,N}/32000 \approx 10^{-5}\tau_{R,N}$. 
Then it deviates and after the relaxation time reaches $t=\dot{\varepsilon}^{-1}=\tau_{R,N}/77$, both curves for $\lambda \approx 5.0$ follow the same softening pattern.
This indicates that relevant chain chain interpenetration did not change significantly during the slow stretching process
($\dot{\varepsilon}=77/\tau_{R,N}$ which is still quite fast comparing to $\tau_{d,N}^{-1}$), 
that is, topological constraints are not released up to some chain end effects. 
Furthermore this supports the concept that the long time behavior of the stress relaxation also in the non-linear regime seems to be just a function of the ultimate deformation and is independent of the original strain rate
$\tau^{-1}_{R,N}<\dot{\varepsilon}<\tau^{-1}_e$.
From the above we expect that estimates of $\langle N_{\rm kink}(t) \rangle$ for deformed polymer melts with two different strain rates at $\lambda \approx 5.0$ in Figure~\ref{fig-Gtr-Nkink}b should eventually coincide. 
Indeed, as expected for $t/\tau_{R,N}>1/77\approx 0.013$, the two sets of data perfectly coincide with each other, i.e. entanglement effects in both deformed polymer melts remain the same. 

\section{Conclusions}

In this paper, we have employed extensive molecular dynamics simulations to study  highly entangled polymer melts subject to an isochoric elongation in the non-linear viscoelastic regime.
Focusing on the analysis of topological constrains, as identified through a primitive path analysis~\cite{Everaers2004,Sukumaran2005,Hsu2018a}, our simulation results can serve as benchmarks and starting points for further experimental and theoretical studies of monodisperse polymer melts under a large step elongation.
An affine deformation of the overall conformations of polymer chains in a melt is to be expected by setting the strain rate $\tau_{R,N}^{-1}<\dot{\varepsilon}<\tau_e^{-1}$ (Figures~\ref{fig-Rgpp-el}).
The chain retraction mechanism predicted by the Doi-Edward tube model~\cite{Doi1986} and its refined GLaMM tube model~\cite{Graham2003} after
large step deformation are demonstrated to hold qualitatively (Figure~\ref{fig-Rgpp-rel}). 
Perpendicular to stretching, the signature of predicted chain retraction enhances with the increase of molecular weight, i.e. number of entanglements. 
However, the onset of delayed relaxation in the conformation of 
chains occurs earlier, and the duration of chain retraction process is 
shorter than the predicted Rouse time of chains in unperturbed melts~\cite{Doi1986,Graham2003}. 
Since the central assumption of an isotropic Gaussian chain conformation is not valid anymore here, entanglement points are identified as significant kinks along the PP by comparing the curvature and the tension force pattern along the PP (Figure~\ref{fig-fcosbb-N2000}).
The resulting average number of entanglement points (kinks) (Figure~\ref{fig-Nkink}) shows that the distribution of the entanglement points does not deform affinely upon an uniaxial elongation in the non-linear viscoelastic regime ($\lambda>1.5$). 
At subsequent relaxation times, the average number of kinks first decreases, and reaches its minimum value before it moves towards the value for unperturbed polymer melts again implying delayed relaxation. 
The distributions of ``effective entanglement length'' between two neighboring kinks along the same primitive path for the whole deformed polymer melt also show the similar behavior (Figure.~\ref{fig-plstr}).
Furthermore, we have observed a clustering and inhomogeneous distribution of entanglement points not only along the individual PPs of chains (Figure~\ref{fig-fcosbb-N2000}), but also for the whole melts (Figure~\ref{fig-Ents-N2000}). 
The distance between these jammed areas is found to be of the order of a few tube diameters of unperturbed melts.
Tracking the mean square displacement of inner monomers, and of center of mass right after polymer melts
are deformed, significant deviations from that for unperturbed polymer melts are 
discovered for $t>\tau_e$ (Figure~\ref{fig-g13}).
The relaxation retardation picture has also been confirmed by investigating the time dependent
intrinsic properties of PPs of stretched polymer melts characterized by 
the bond length, the orientational order parameter, and the entanglement length
estimated using the original recipe of the PPA~\cite{Everaers2004,Sukumaran2005} (Figures~\ref{fig-bpp}-\ref{fig-Ne}).
However, all these quantities do not follow affine deformation under strain.
Following the tube picture assumption, we also count the number of monomers confined in a tube-like regime of fixed tube diameter $d_T=d_T^{(0)}$ for the unperturbed polymer melt  to measure the transverse fluctuation of monomers during the elongation and relaxation processes (Figures~\ref{fig-Ntube},\ref{fig-pNtube}).
Our results again show a significantly delayed relaxation compared to the current theoretical 
considerations~\cite{Graham2003}. 
Finally, we see that the time evolutions of the stress relaxation modulus for deformed polymer melt 
only initially depends on the strain rates while they follow the same softening patterns ultimately in
the non-linear viscoelastic regime (Figure~\ref{fig-Gtr-Nkink}a). The same scenarios are also
seen from the time evolutions of the number of effective entanglement points (Figure~\ref{fig-Gtr-Nkink}b).
All our results indicate that in the non-linear viscoelastic regime, 
the topological constraints in highly strained large 
polymer melts are better described by the PP meshes than the corresponding
elongated OPs of chains. 
The resulting entanglement effects play an important role for the relaxation retardation.
However, it is still unclear how long the relaxation retardation would last,
and what could be the time for equilibrating the deformed polymer melts if 
it would happen.
These findings so far have not been considered in the current theoretical framework and similarities to knotted polymers are worth further considerations.
The clustering and inhomogeneous distribution of entanglement points offers interesting options for both experimental and simulation investigations, especially in the vicinity of the glass transition point.
Especially, it is important to understand whether the complex topological constraints in deformed polymer melts are the crucial polymer characteristic for understanding the dynamical and thermodynamic properties of polymer chains in a deformed melt.

\begin{acknowledgement}
We are grateful to M. Doi, G. S. Grest, D. Vlassopoulos, T.
Ohkuma for stimulating and helpful
discussions, and K. Ch. Daoulas for a critical
reading of the manuscript.
This work has been supported by European Research Council under the European
Union's Seventh Framework Programme (FP7/2007-2013)/ERC Grant Agreement
No.~340906-MOLPROCOMP.
We also gratefully acknowledge the computing time granted by the John von
Neumann Institute for Computing (NIC) and provided on the supercomputer JUROPA
at J\"ulich Supercomputing Centre (JSC),
and the Max Planck Computing and Data Facility (MPCDF).
\end{acknowledgement}

\bibliography{hp-kk-cluster1.bbl}

\providecommand{\latin}[1]{#1}
\makeatletter
\providecommand{\doi}
  {\begingroup\let\do\@makeother\dospecials
  \catcode`\{=1 \catcode`\}=2\doi@aux}
\providecommand{\doi@aux}[1]{\endgroup\texttt{#1}}
\makeatother
\providecommand*\mcitethebibliography{\thebibliography}
\csname @ifundefined\endcsname{endmcitethebibliography}
  {\let\endmcitethebibliography\endthebibliography}{}
\begin{mcitethebibliography}{74}
\providecommand*\natexlab[1]{#1}
\providecommand*\mciteSetBstSublistMode[1]{}
\providecommand*\mciteSetBstMaxWidthForm[2]{}
\providecommand*\mciteBstWouldAddEndPuncttrue
  {\def\EndOfBibitem{\unskip.}}
\providecommand*\mciteBstWouldAddEndPunctfalse
  {\let\EndOfBibitem\relax}
\providecommand*\mciteSetBstMidEndSepPunct[3]{}
\providecommand*\mciteSetBstSublistLabelBeginEnd[3]{}
\providecommand*\EndOfBibitem{}
\mciteSetBstSublistMode{f}
\mciteSetBstMaxWidthForm{subitem}{(\alph{mcitesubitemcount})}
\mciteSetBstSublistLabelBeginEnd
  {\mcitemaxwidthsubitemform\space}
  {\relax}
  {\relax}

\bibitem[de~Gennes(1979)]{deGennes1979}
de~Gennes,~P.~G. \emph{Scaling Concepts in polymer physics}; Cornell University
  Press: Itharca, New York, 1979\relax
\mciteBstWouldAddEndPuncttrue
\mciteSetBstMidEndSepPunct{\mcitedefaultmidpunct}
{\mcitedefaultendpunct}{\mcitedefaultseppunct}\relax
\EndOfBibitem
\bibitem[Doi and Edwards(1986)Doi, and Edwards]{Doi1986}
Doi,~M.; Edwards,~S. \emph{The theory of polymer dynamics}; Oxford University
  Press: New York, 1986\relax
\mciteBstWouldAddEndPuncttrue
\mciteSetBstMidEndSepPunct{\mcitedefaultmidpunct}
{\mcitedefaultendpunct}{\mcitedefaultseppunct}\relax
\EndOfBibitem
\bibitem[McLeish(2003)]{Mcleish2003}
McLeish,~T. C.~B. Present puzzles of entangled polymers. \emph{Rheol. Rev.}
  \textbf{2003}, 197--233\relax
\mciteBstWouldAddEndPuncttrue
\mciteSetBstMidEndSepPunct{\mcitedefaultmidpunct}
{\mcitedefaultendpunct}{\mcitedefaultseppunct}\relax
\EndOfBibitem
\bibitem[Everaers \latin{et~al.}(2004)Everaers, Sukumaran, Grest, Svaneborg,
  Sivasubramanian, and Kremer]{Everaers2004}
Everaers,~R.; Sukumaran,~S.~K.; Grest,~G.~S.; Svaneborg,~C.;
  Sivasubramanian,~A.; Kremer,~K. Rheology and microscopic topology of
  entangled polymeric liquids. \emph{Science} \textbf{2004}, \emph{303},
  823\relax
\mciteBstWouldAddEndPuncttrue
\mciteSetBstMidEndSepPunct{\mcitedefaultmidpunct}
{\mcitedefaultendpunct}{\mcitedefaultseppunct}\relax
\EndOfBibitem
\bibitem[Sukumaran \latin{et~al.}(2005)Sukumaran, Grest, Kremer, and
  Everaers]{Sukumaran2005}
Sukumaran,~S.~K.; Grest,~G.~S.; Kremer,~K.; Everaers,~R. Identifying the
  primitive path mesh in entangled polymer liquids. \emph{J. Polym. Sci. B}
  \textbf{2005}, \emph{43}, 917\relax
\mciteBstWouldAddEndPuncttrue
\mciteSetBstMidEndSepPunct{\mcitedefaultmidpunct}
{\mcitedefaultendpunct}{\mcitedefaultseppunct}\relax
\EndOfBibitem
\bibitem[Padding and Briels(2011)Padding, and Briels]{Padding2011}
Padding,~J.~T.; Briels,~W.~J. Systematic coarse-graining of the dynamics of
  entangled polymer melts: the road from chemistry to rheology. \emph{J.
  Phys.:Condens Matter} \textbf{2011}, \emph{23}, 233101 (17pp)\relax
\mciteBstWouldAddEndPuncttrue
\mciteSetBstMidEndSepPunct{\mcitedefaultmidpunct}
{\mcitedefaultendpunct}{\mcitedefaultseppunct}\relax
\EndOfBibitem
\bibitem[Qin and S.~T.~Milner(2012)Qin, and S.~T.~Milner]{Qin2012}
Qin,~J.; S.~T.~Milner,~V. G.~M.,~P. S.~Stephanou Effects of tube persistene
  length on dynamics of mildly entangled polymers. \emph{J. Rheol.}
  \textbf{2012}, \emph{56}, 707--723\relax
\mciteBstWouldAddEndPuncttrue
\mciteSetBstMidEndSepPunct{\mcitedefaultmidpunct}
{\mcitedefaultendpunct}{\mcitedefaultseppunct}\relax
\EndOfBibitem
\bibitem[Larson and Wang(2015)Larson, and Wang]{Terentjev2015}
Larson,~R.~G.; Wang,~Z. In \emph{The Oxford handbook of soft condensed matter};
  Terentjev,~E.~M., Weitz,~D.~A., Eds.; Oxford University Press: Oxford, 2015;
  Chapter 6, pp 233--269\relax
\mciteBstWouldAddEndPuncttrue
\mciteSetBstMidEndSepPunct{\mcitedefaultmidpunct}
{\mcitedefaultendpunct}{\mcitedefaultseppunct}\relax
\EndOfBibitem
\bibitem[Doi(1980)]{Doi1980}
Doi,~M. Molecular rheology of concentrated polymer systems. I. \emph{J. Polym.
  Sci. Polym. Phys. Ed.} \textbf{1980}, \emph{18}, 1005\relax
\mciteBstWouldAddEndPuncttrue
\mciteSetBstMidEndSepPunct{\mcitedefaultmidpunct}
{\mcitedefaultendpunct}{\mcitedefaultseppunct}\relax
\EndOfBibitem
\bibitem[Kremer \latin{et~al.}(1988)Kremer, Grest, and Carmesin]{Kremer1988}
Kremer,~K.; Grest,~G.~S.; Carmesin,~I. Crossover from Rouse to Reptation
  Dynamics: A Molecular-Dynamics Simulation. \emph{Phys. Rev. Lett.}
  \textbf{1988}, \emph{61}, 566--569\relax
\mciteBstWouldAddEndPuncttrue
\mciteSetBstMidEndSepPunct{\mcitedefaultmidpunct}
{\mcitedefaultendpunct}{\mcitedefaultseppunct}\relax
\EndOfBibitem
\bibitem[Kremer and Grest(1990)Kremer, and Grest]{Kremer1990}
Kremer,~K.; Grest,~G.~S. Dynamics of entangled linear polymer melts: a
  molecular-dynamics simulation. \emph{J. Chem. Phys.} \textbf{1990},
  \emph{92}, 5057\relax
\mciteBstWouldAddEndPuncttrue
\mciteSetBstMidEndSepPunct{\mcitedefaultmidpunct}
{\mcitedefaultendpunct}{\mcitedefaultseppunct}\relax
\EndOfBibitem
\bibitem[Paul \latin{et~al.}(1991)Paul, Binder, Heermann, and Kremer]{Paul1991}
Paul,~W.; Binder,~K.; Heermann,~D.~W.; Kremer,~K. Dynamiccs of polymer
  solutions and melts. Reptation predictions and scaling of relaxation times.
  \emph{J. Chem. Phys.} \textbf{1991}, \emph{95}, 7726\relax
\mciteBstWouldAddEndPuncttrue
\mciteSetBstMidEndSepPunct{\mcitedefaultmidpunct}
{\mcitedefaultendpunct}{\mcitedefaultseppunct}\relax
\EndOfBibitem
\bibitem[Wittmer \latin{et~al.}(1992)Wittmer, Paul, and Binder]{Wittmer1992}
Wittmer,~J.; Paul,~W.; Binder,~K. Rouse and Repatation Dynamics at Finite
  Temperatures: A Monte Carlo Simulation. \emph{Macromolecules} \textbf{1992},
  \emph{25}, 7211\relax
\mciteBstWouldAddEndPuncttrue
\mciteSetBstMidEndSepPunct{\mcitedefaultmidpunct}
{\mcitedefaultendpunct}{\mcitedefaultseppunct}\relax
\EndOfBibitem
\bibitem[Kremer and Grest(1992)Kremer, and Grest]{Kremer1992}
Kremer,~K.; Grest,~G.~S. Simulations for Structural and Dynamic Properties of
  Dense Polymer Systems. \emph{J. Chem. Soc. Faraday Trans} \textbf{1992},
  \emph{88}, 1707\relax
\mciteBstWouldAddEndPuncttrue
\mciteSetBstMidEndSepPunct{\mcitedefaultmidpunct}
{\mcitedefaultendpunct}{\mcitedefaultseppunct}\relax
\EndOfBibitem
\bibitem[Kopf \latin{et~al.}(1997)Kopf, D\"unweg, and Paul]{Kopf1997}
Kopf,~A.; D\"unweg,~B.; Paul,~W. Dynamics of polymer ``isotope" mixtures:
  Molecular dynamics simulation and Rouse model analysis. \emph{J. Chem. Phys.}
  \textbf{1997}, \emph{107}, 6945\relax
\mciteBstWouldAddEndPuncttrue
\mciteSetBstMidEndSepPunct{\mcitedefaultmidpunct}
{\mcitedefaultendpunct}{\mcitedefaultseppunct}\relax
\EndOfBibitem
\bibitem[P\"utz \latin{et~al.}(2000)P\"utz, Kremer, and Grest]{Puetz2000}
P\"utz,~M.; Kremer,~K.; Grest,~G.~S. What is the entanglement length in a
  polymer melt? \emph{Europhys. Lett.} \textbf{2000}, \emph{49}, 735--741\relax
\mciteBstWouldAddEndPuncttrue
\mciteSetBstMidEndSepPunct{\mcitedefaultmidpunct}
{\mcitedefaultendpunct}{\mcitedefaultseppunct}\relax
\EndOfBibitem
\bibitem[Harmandaris \latin{et~al.}(2003)Harmandaris, Mavrantzas, Theodorou,
  Kr\"oger, Ramirez, \"Ottinger, and Vlassopoulos]{Harmandaris2003}
Harmandaris,~V.~A.; Mavrantzas,~V.~G.; Theodorou,~D.~N.; Kr\"oger,~M.;
  Ramirez,~J.; \"Ottinger,~H.~C.; Vlassopoulos,~D. Crossover from the Rouse to
  the Entangled Polymer Melt Regime: Signals from Long, Detailed Atomistic
  Molecular Dynamics Simulations, Supported by Rheological Experiments.
  \emph{Macromolecules} \textbf{2003}, \emph{36}, 1376--1387\relax
\mciteBstWouldAddEndPuncttrue
\mciteSetBstMidEndSepPunct{\mcitedefaultmidpunct}
{\mcitedefaultendpunct}{\mcitedefaultseppunct}\relax
\EndOfBibitem
\bibitem[Tsolou \latin{et~al.}(2005)Tsolou, Mavrantzas, and
  Theodorou]{Tsolou2005}
Tsolou,~G.; Mavrantzas,~V.~G.; Theodorou,~D.~N. Detailed Atomistic Molecular
  Dynamics Simulation of cis-1,4-Poly(butadiene). \emph{Macromolecules}
  \textbf{2005}, \emph{38}, 1478--1492\relax
\mciteBstWouldAddEndPuncttrue
\mciteSetBstMidEndSepPunct{\mcitedefaultmidpunct}
{\mcitedefaultendpunct}{\mcitedefaultseppunct}\relax
\EndOfBibitem
\bibitem[Hsu and Kremer(2016)Hsu, and Kremer]{Hsu2016}
Hsu,~H.-P.; Kremer,~K. Static and dynamic properties of large polymer melts in
  equilibrium. \emph{J. Chem. Phys.} \textbf{2016}, \emph{144}, 154907\relax
\mciteBstWouldAddEndPuncttrue
\mciteSetBstMidEndSepPunct{\mcitedefaultmidpunct}
{\mcitedefaultendpunct}{\mcitedefaultseppunct}\relax
\EndOfBibitem
\bibitem[Hsu and Kremer(2017)Hsu, and Kremer]{Hsu2017}
Hsu,~H.-P.; Kremer,~K. Detailed analysis of Rouse mode and dynamic scattering
  function of highly entangled polymer melts in equilibrium. \emph{Eur. Phys.
  J. Special Topics} \textbf{2017}, \emph{226}, 693\relax
\mciteBstWouldAddEndPuncttrue
\mciteSetBstMidEndSepPunct{\mcitedefaultmidpunct}
{\mcitedefaultendpunct}{\mcitedefaultseppunct}\relax
\EndOfBibitem
\bibitem[Fetters \latin{et~al.}(1994)Fetters, Lohse, Richter, Witten, and
  Zirkel]{Fetters1994}
Fetters,~L.~J.; Lohse,~D.~J.; Richter,~D.; Witten,~T.~A.; Zirkel,~A. Connection
  between Polymer Molecular Weight, Density, Chain Dimensions, and Melt
  Viscoelastic Properties. \emph{Macromolecules} \textbf{1994}, \emph{27},
  4639--4647\relax
\mciteBstWouldAddEndPuncttrue
\mciteSetBstMidEndSepPunct{\mcitedefaultmidpunct}
{\mcitedefaultendpunct}{\mcitedefaultseppunct}\relax
\EndOfBibitem
\bibitem[Wischnewski \latin{et~al.}(2002)Wischnewski, Monkenbusch, Willner,
  Richter, Likhtman, McLeish, and Farago]{Wischnewski2002}
Wischnewski,~A.; Monkenbusch,~M.; Willner,~L.; Richter,~D.; Likhtman,~A.~E.;
  McLeish,~T. C.~B.; Farago,~B. Molecular Observation of Contour-Length
  Fluctuations Limiting Topological Confinement in Polymer Melts. \emph{Phys.
  Rev. Lett.} \textbf{2002}, \emph{88}, 058301\relax
\mciteBstWouldAddEndPuncttrue
\mciteSetBstMidEndSepPunct{\mcitedefaultmidpunct}
{\mcitedefaultendpunct}{\mcitedefaultseppunct}\relax
\EndOfBibitem
\bibitem[Wischnewski and Richter(2006)Wischnewski, and
  Richter]{Wischnewski2006}
Wischnewski,~A.; Richter,~D. In \emph{Soft Matter, Vol. 1:Polymer melts and
  mixtures}; Gompper,~G., Schick,~M., Eds.; Wiley-VCH: Weinheim, 2006; Chapter
  1, pp 17--85\relax
\mciteBstWouldAddEndPuncttrue
\mciteSetBstMidEndSepPunct{\mcitedefaultmidpunct}
{\mcitedefaultendpunct}{\mcitedefaultseppunct}\relax
\EndOfBibitem
\bibitem[Graessley(2008)]{Graessley2008}
Graessley,~W.~W. \emph{Polymeric liquids \& networks: structure and
  properties}; Garland Science, London and New York, 2008\relax
\mciteBstWouldAddEndPuncttrue
\mciteSetBstMidEndSepPunct{\mcitedefaultmidpunct}
{\mcitedefaultendpunct}{\mcitedefaultseppunct}\relax
\EndOfBibitem
\bibitem[Herrmann \latin{et~al.}(2012)Herrmann, Kresse, Wohlfahrt, Bauer,
  Privalov, Kruk, Fatkullin, Fujara, and R\"ossler]{Herrmann2012}
Herrmann,~A.; Kresse,~B.; Wohlfahrt,~M.; Bauer,~I.; Privalov,~A.~F.; Kruk,~D.;
  Fatkullin,~N.; Fujara,~F.; R\"ossler,~E.~A. Mean Square Displacement and
  Reorientational Correlation Function in Entangled Polymer Melts Revealed by
  Field Cycling $^1$H and $^2$H NMR Relaxometry. \emph{Macromolecules}
  \textbf{2012}, \emph{45}, 6516--6526\relax
\mciteBstWouldAddEndPuncttrue
\mciteSetBstMidEndSepPunct{\mcitedefaultmidpunct}
{\mcitedefaultendpunct}{\mcitedefaultseppunct}\relax
\EndOfBibitem
\bibitem[Iwata and Edwards(1989)Iwata, and Edwards]{Iwata1989}
Iwata,~K.; Edwards,~S.~F. New model of polymer entanglement: Localized Gauss
  integral model. Plateau modulus $G_N$, topological second virial coefficient
  $A^\theta_2$ and physical foundation of the tube model. \emph{J. Chem. Phys.}
  \textbf{1989}, \emph{90}, 4567--4581\relax
\mciteBstWouldAddEndPuncttrue
\mciteSetBstMidEndSepPunct{\mcitedefaultmidpunct}
{\mcitedefaultendpunct}{\mcitedefaultseppunct}\relax
\EndOfBibitem
\bibitem[Edwards(1967)]{Edwards1967}
Edwards,~S.~F. Statistical mechanics with topological constraints: I.
  \emph{Proc. Phys. Soc.} \textbf{1967}, \emph{91}, 513\relax
\mciteBstWouldAddEndPuncttrue
\mciteSetBstMidEndSepPunct{\mcitedefaultmidpunct}
{\mcitedefaultendpunct}{\mcitedefaultseppunct}\relax
\EndOfBibitem
\bibitem[Moreira \latin{et~al.}(2015)Moreira, Zhang, M{\"u}ller, Stuehn, and
  Kremer]{Moreira2015}
Moreira,~L.~A.; Zhang,~G.; M{\"u}ller,~F.; Stuehn,~T.; Kremer,~K. Direct
  equilibration and characterization of polymer melts for computer simulations.
  \emph{Macromol. Theor. Simul.} \textbf{2015}, \emph{24}, 419\relax
\mciteBstWouldAddEndPuncttrue
\mciteSetBstMidEndSepPunct{\mcitedefaultmidpunct}
{\mcitedefaultendpunct}{\mcitedefaultseppunct}\relax
\EndOfBibitem
\bibitem[Kr\"oger(2005)]{Kroeger2005}
Kr\"oger,~M. Shortest multiple disconnected path for the analysis of
  entanglements in two- and three-dimensional polymeric systems. \emph{Comput.
  Phys. Comm.} \textbf{2005}, \emph{168}, 209--232\relax
\mciteBstWouldAddEndPuncttrue
\mciteSetBstMidEndSepPunct{\mcitedefaultmidpunct}
{\mcitedefaultendpunct}{\mcitedefaultseppunct}\relax
\EndOfBibitem
\bibitem[Shanbhag and Kr\"oger(2007)Shanbhag, and Kr\"oger]{Shanbhag2007}
Shanbhag,~S.; Kr\"oger,~M. Primitive Path Networks Generated by Annealing and
  Geometrical Methods: Insights into Differences. \emph{Macromolecules}
  \textbf{2007}, \emph{40}, 2897--2903\relax
\mciteBstWouldAddEndPuncttrue
\mciteSetBstMidEndSepPunct{\mcitedefaultmidpunct}
{\mcitedefaultendpunct}{\mcitedefaultseppunct}\relax
\EndOfBibitem
\bibitem[Tzoumanekas and Theodorou(2006)Tzoumanekas, and
  Theodorou]{Tzoumanekas2006}
Tzoumanekas,~C.; Theodorou,~D.~N. Topological analysis of linear polymer
  melts:  a statistical approach. \emph{Macromolecules} \textbf{2006},
  \emph{39}, 4592--4604\relax
\mciteBstWouldAddEndPuncttrue
\mciteSetBstMidEndSepPunct{\mcitedefaultmidpunct}
{\mcitedefaultendpunct}{\mcitedefaultseppunct}\relax
\EndOfBibitem
\bibitem[Shanbhag and Larson(2005)Shanbhag, and Larson]{Shanbhag2005}
Shanbhag,~S.; Larson,~R.~G. Chain retraction potential in a fixed entanglement
  network. \emph{Phys. Rev. Lett.} \textbf{2005}, \emph{94}, 076001\relax
\mciteBstWouldAddEndPuncttrue
\mciteSetBstMidEndSepPunct{\mcitedefaultmidpunct}
{\mcitedefaultendpunct}{\mcitedefaultseppunct}\relax
\EndOfBibitem
\bibitem[Zhou and Larson(2005)Zhou, and Larson]{Zhou2005}
Zhou,~Q.; Larson,~R.~G. Primitive Path Identification and Statistics in
  Molecular Dynamics Simulations of Entangled Polymer Melts.
  \emph{Macromolecules} \textbf{2005}, \emph{38}, 5761--5765\relax
\mciteBstWouldAddEndPuncttrue
\mciteSetBstMidEndSepPunct{\mcitedefaultmidpunct}
{\mcitedefaultendpunct}{\mcitedefaultseppunct}\relax
\EndOfBibitem
\bibitem[Hoy and Grest(2007)Hoy, and Grest]{Hoy2007}
Hoy,~R.~S.; Grest,~G.~S. Entanglements of an End-Grafted Polymer Brush in a
  Polymeric Matrix. \emph{Macromolecules} \textbf{2007}, \emph{40},
  8389--8395\relax
\mciteBstWouldAddEndPuncttrue
\mciteSetBstMidEndSepPunct{\mcitedefaultmidpunct}
{\mcitedefaultendpunct}{\mcitedefaultseppunct}\relax
\EndOfBibitem
\bibitem[Everaers(2012)]{Everaers2012}
Everaers,~R. Topological versus rheological entanglement length in
  primitive-path analysis protocols, tube models, and slip-link models.
  \emph{Phys. Rev. E} \textbf{2012}, \emph{86}, 022801\relax
\mciteBstWouldAddEndPuncttrue
\mciteSetBstMidEndSepPunct{\mcitedefaultmidpunct}
{\mcitedefaultendpunct}{\mcitedefaultseppunct}\relax
\EndOfBibitem
\bibitem[Doi(1983)]{Doi1983}
Doi,~M. Explanation for the 3.4 power law for viscosity of polymeric liquids on
  the basis of the tube model. \emph{J. Polym. Sci. Polym. Phys. Ed.}
  \textbf{1983}, \emph{21}, 667\relax
\mciteBstWouldAddEndPuncttrue
\mciteSetBstMidEndSepPunct{\mcitedefaultmidpunct}
{\mcitedefaultendpunct}{\mcitedefaultseppunct}\relax
\EndOfBibitem
\bibitem[Likhtman and McLeish(2002)Likhtman, and McLeish]{Likhtman2002}
Likhtman,~A.~E.; McLeish,~T. C.~B. Quantitative theory for linear dynamics of
  linear entangled polymers. \emph{Macromolecules} \textbf{2002}, \emph{35},
  6332\relax
\mciteBstWouldAddEndPuncttrue
\mciteSetBstMidEndSepPunct{\mcitedefaultmidpunct}
{\mcitedefaultendpunct}{\mcitedefaultseppunct}\relax
\EndOfBibitem
\bibitem[Likhtman and Ponmurugan(2014)Likhtman, and Ponmurugan]{Likhtman2014}
Likhtman,~A.~E.; Ponmurugan,~M. Microscopic definition of polymer
  entanglements. \emph{Macromolecules} \textbf{2014}, \emph{47},
  1470--1481\relax
\mciteBstWouldAddEndPuncttrue
\mciteSetBstMidEndSepPunct{\mcitedefaultmidpunct}
{\mcitedefaultendpunct}{\mcitedefaultseppunct}\relax
\EndOfBibitem
\bibitem[Klein(1978)]{Klein1978}
Klein,~J. The onset of entangled behavior in semidilute and concentrated
  polymer solutions. \emph{Macromolecules} \textbf{1978}, \emph{11}, 852\relax
\mciteBstWouldAddEndPuncttrue
\mciteSetBstMidEndSepPunct{\mcitedefaultmidpunct}
{\mcitedefaultendpunct}{\mcitedefaultseppunct}\relax
\EndOfBibitem
\bibitem[Daoud and de~Gennes(1979)Daoud, and de~Gennes]{Daoud1979}
Daoud,~M.; de~Gennes,~P.~G. Some remarks on the dynamics of polymer melts.
  \emph{J. Polym. Sci. Polym. Phys. Ed.} \textbf{1979}, \emph{17}, 1971\relax
\mciteBstWouldAddEndPuncttrue
\mciteSetBstMidEndSepPunct{\mcitedefaultmidpunct}
{\mcitedefaultendpunct}{\mcitedefaultseppunct}\relax
\EndOfBibitem
\bibitem[Rubinstein and Colby(1988)Rubinstein, and Colby]{Rubinstein1988}
Rubinstein,~M.; Colby,~R.~H. Self-consistent theory of polydisperse entangled
  polymers: Linear viscoelasticity of binary blends. \emph{J. Chem. Phys.}
  \textbf{1988}, \emph{89}, 5291--5306\relax
\mciteBstWouldAddEndPuncttrue
\mciteSetBstMidEndSepPunct{\mcitedefaultmidpunct}
{\mcitedefaultendpunct}{\mcitedefaultseppunct}\relax
\EndOfBibitem
\bibitem[Hsu and Kremer(2018)Hsu, and Kremer]{Hsu2018a}
Hsu,~H.-P.; Kremer,~K. Primitive Path Analysis and Stress Distribution in
  Highly Strained Macromolecules. \emph{ACS Macro Lett.} \textbf{2018},
  \emph{7}, 107--111\relax
\mciteBstWouldAddEndPuncttrue
\mciteSetBstMidEndSepPunct{\mcitedefaultmidpunct}
{\mcitedefaultendpunct}{\mcitedefaultseppunct}\relax
\EndOfBibitem
\bibitem[Hsu and Kremer(2018)Hsu, and Kremer]{Hsu2018b}
Hsu,~H.-P.; Kremer,~K. Chain retraction in highly entangled stretched polymer
  melts. \emph{Phys. Rev. Lett.} \textbf{2018}, \emph{121}, 167801\relax
\mciteBstWouldAddEndPuncttrue
\mciteSetBstMidEndSepPunct{\mcitedefaultmidpunct}
{\mcitedefaultendpunct}{\mcitedefaultseppunct}\relax
\EndOfBibitem
\bibitem[Graham \latin{et~al.}(2003)Graham, Likhtman, McLeish, and
  Milner]{Graham2003}
Graham,~R.~S.; Likhtman,~A.~E.; McLeish,~T. C.~B.; Milner,~S.~T. Microscopic
  theory of linear, entangled polymer chains under rapid deformation including
  chain stretch and convective constraint release. \emph{J. Rheol.}
  \textbf{2003}, \emph{47}, 1171--1200\relax
\mciteBstWouldAddEndPuncttrue
\mciteSetBstMidEndSepPunct{\mcitedefaultmidpunct}
{\mcitedefaultendpunct}{\mcitedefaultseppunct}\relax
\EndOfBibitem
\bibitem[Blanchard \latin{et~al.}(2005)Blanchard, Graham, Heinrich,
  Pyckhout-Hintzen, Richter, Likhtman, McLeish, Read, Straube, and
  Kohlbrecher]{Blanchard2005}
Blanchard,~A.; Graham,~R.~S.; Heinrich,~M.; Pyckhout-Hintzen,~W.; Richter,~D.;
  Likhtman,~A.~E.; McLeish,~T. C.~B.; Read,~D.~J.; Straube,~E.; Kohlbrecher,~J.
  Small-angle neutron scattering observation of chain retraction after a large
  step deformation. \emph{Phys. Rev. Lett.} \textbf{2005}, \emph{95},
  166001\relax
\mciteBstWouldAddEndPuncttrue
\mciteSetBstMidEndSepPunct{\mcitedefaultmidpunct}
{\mcitedefaultendpunct}{\mcitedefaultseppunct}\relax
\EndOfBibitem
\bibitem[Graham \latin{et~al.}(2006)Graham, Bent, Hutchings, Richards, Groves,
  Embery, Nicholson, McLeish, Likhtman, Harlen, Read, Gough, Spares, Coates,
  and Grillo]{Graham2006}
Graham,~R.~S.; Bent,~J.; Hutchings,~L.~R.; Richards,~R.~W.; Groves,~D.~J.;
  Embery,~J.; Nicholson,~T.~M.; McLeish,~T. C.~B.; Likhtman,~A.~E.;
  Harlen,~O.~G.; Read,~D.~J.; Gough,~T.; Spares,~R.; Coates,~P.~D.; Grillo,~I.
  Measuring and predicting the dynamics of linear monodisperse entangled
  polymers in rapid flow through an abrupt contraction, a small angle neutron
  scattering study. \emph{Macromolecules} \textbf{2006}, \emph{39}, 2700\relax
\mciteBstWouldAddEndPuncttrue
\mciteSetBstMidEndSepPunct{\mcitedefaultmidpunct}
{\mcitedefaultendpunct}{\mcitedefaultseppunct}\relax
\EndOfBibitem
\bibitem[Wang \latin{et~al.}(2017)Wang, Lam, Chen, Wang, Liu, Liu, Porcar,
  Stanley, Zhao, Hong, and Wang]{Wang2017}
Wang,~Z.; Lam,~C.~N.; Chen,~W.-R.; Wang,~W.; Liu,~J.; Liu,~Y.; Porcar,~L.;
  Stanley,~C.~B.; Zhao,~Z.; Hong,~K.; Wang,~Y. Fingerprinting molecular
  relaxation in deformed polymers. \emph{Phys. Rev. X} \textbf{2017}, \emph{7},
  031003\relax
\mciteBstWouldAddEndPuncttrue
\mciteSetBstMidEndSepPunct{\mcitedefaultmidpunct}
{\mcitedefaultendpunct}{\mcitedefaultseppunct}\relax
\EndOfBibitem
\bibitem[Xu \latin{et~al.}(2018)Xu, Carrillo, Lam, Sumpter, and Wang]{Xu2018}
Xu,~W.-S.; Carrillo,~J.-M.~Y.; Lam,~C.~N.; Sumpter,~B.~G.; Wang,~Y. Molecular
  dynamics investigation of the relaxation mechanism of entangled polymers
  after a large step deformation. \emph{ACS Macro Lett.} \textbf{2018},
  \emph{7}, 190--195\relax
\mciteBstWouldAddEndPuncttrue
\mciteSetBstMidEndSepPunct{\mcitedefaultmidpunct}
{\mcitedefaultendpunct}{\mcitedefaultseppunct}\relax
\EndOfBibitem
\bibitem[Zhou and Schroeder(2018)Zhou, and Schroeder]{Zhou2018}
Zhou,~Y.; Schroeder,~C.~M. Dynamically Heterogeneous Relaxation of Entangled
  Polymer Chains. \emph{Phys. Rev. Lett.} \textbf{2018}, \emph{120},
  26781\relax
\mciteBstWouldAddEndPuncttrue
\mciteSetBstMidEndSepPunct{\mcitedefaultmidpunct}
{\mcitedefaultendpunct}{\mcitedefaultseppunct}\relax
\EndOfBibitem
\bibitem[Nielsen \latin{et~al.}(2009)Nielsen, Hassager, Rasmussen, and
  McKinley]{Nielsen2009}
Nielsen,~J.~K.; Hassager,~O.; Rasmussen,~H.~K.; McKinley,~G.~H. Observing the
  chain stretch transition in a highly entangled polyisoprene melt using
  transient extensional rheometry. \emph{J. Rheol.} \textbf{2009}, \emph{53},
  1327\relax
\mciteBstWouldAddEndPuncttrue
\mciteSetBstMidEndSepPunct{\mcitedefaultmidpunct}
{\mcitedefaultendpunct}{\mcitedefaultseppunct}\relax
\EndOfBibitem
\bibitem[Yaoita \latin{et~al.}(2012)Yaoita, Isaki, Masubuchi, Watanabe,
  Ianniruberto, and Marrucci]{Yaoita2012}
Yaoita,~T.; Isaki,~T.; Masubuchi,~Y.; Watanabe,~H.; Ianniruberto,~G.;
  Marrucci,~G. Chain network simulation of elongational flows of entangled
  linear chains: stretch/orientation-induced reduction of monomeric friction.
  \emph{Macromolecules} \textbf{2012}, \emph{45}, 2773--2782\relax
\mciteBstWouldAddEndPuncttrue
\mciteSetBstMidEndSepPunct{\mcitedefaultmidpunct}
{\mcitedefaultendpunct}{\mcitedefaultseppunct}\relax
\EndOfBibitem
\bibitem[Yaoita \latin{et~al.}(2011)Yaoita, Isaki, Masubuchi, H.~Watanabe, and
  Marrucci]{Yaoita2011}
Yaoita,~T.; Isaki,~T.; Masubuchi,~Y.; H.~Watanabe,~H. G.~I.; Marrucci,~G.
  Primitive chain network simulation of elongational flows of entangled linear
  chains: role of finite chain extensibility. \emph{Macromolecules}
  \textbf{2011}, \emph{44}, 9675--9682\relax
\mciteBstWouldAddEndPuncttrue
\mciteSetBstMidEndSepPunct{\mcitedefaultmidpunct}
{\mcitedefaultendpunct}{\mcitedefaultseppunct}\relax
\EndOfBibitem
\bibitem[Bhattacharjee \latin{et~al.}(2017)Bhattacharjee, Nguyen, Masubuchi,
  and Sridhar]{Bhattacharjee2017}
Bhattacharjee,~P.~K.; Nguyen,~D.~A.; Masubuchi,~Y.; Sridhar,~T. Extensional
  step strain rate experiments on an entangled polymer solution.
  \emph{Macromolecules} \textbf{2017}, \emph{50}, 386--395\relax
\mciteBstWouldAddEndPuncttrue
\mciteSetBstMidEndSepPunct{\mcitedefaultmidpunct}
{\mcitedefaultendpunct}{\mcitedefaultseppunct}\relax
\EndOfBibitem
\bibitem[Bach \latin{et~al.}(2003)Bach, Almdal, Rasmussen, and
  Hassager]{Bach2003}
Bach,~A.; Almdal,~K.; Rasmussen,~H.~K.; Hassager,~O. Elongational Viscosity of
  Narrow Molar Mass Distribution Polystyrene. \emph{Macromolecules}
  \textbf{2003}, \emph{36}, 5174--5179\relax
\mciteBstWouldAddEndPuncttrue
\mciteSetBstMidEndSepPunct{\mcitedefaultmidpunct}
{\mcitedefaultendpunct}{\mcitedefaultseppunct}\relax
\EndOfBibitem
\bibitem[Zhang \latin{et~al.}(2014)Zhang, Moreira, Stuehn, Daoulas, and
  Kremer]{Zhang2014}
Zhang,~G.; Moreira,~L.~A.; Stuehn,~T.; Daoulas,~K.~C.; Kremer,~K. Equilibration
  of high molecular weight polymer melts: a hierarchical strategy. \emph{ACS
  Macro Lett.} \textbf{2014}, \emph{3}, 198\relax
\mciteBstWouldAddEndPuncttrue
\mciteSetBstMidEndSepPunct{\mcitedefaultmidpunct}
{\mcitedefaultendpunct}{\mcitedefaultseppunct}\relax
\EndOfBibitem
\bibitem[Bird \latin{et~al.}(1977)Bird, Armstrong, and Hassager]{Bird1977}
Bird,~R.~B.; Armstrong,~R.~C.; Hassager,~O. \emph{Dynamics of Polymeric
  Liquids}; Wiley, New York, 1977; Vol. 1 and 2\relax
\mciteBstWouldAddEndPuncttrue
\mciteSetBstMidEndSepPunct{\mcitedefaultmidpunct}
{\mcitedefaultendpunct}{\mcitedefaultseppunct}\relax
\EndOfBibitem
\bibitem[Ceperley \latin{et~al.}(1978)Ceperley, Kalos, and
  Lebowitz]{Ceperly1978}
Ceperley,~D.; Kalos,~M.~H.; Lebowitz,~J.~L. Computer Simulation of the Dynamics
  of a Single Polymer Chain. \emph{Phys. Rev. Lett.} \textbf{1978}, \emph{41},
  313\relax
\mciteBstWouldAddEndPuncttrue
\mciteSetBstMidEndSepPunct{\mcitedefaultmidpunct}
{\mcitedefaultendpunct}{\mcitedefaultseppunct}\relax
\EndOfBibitem
\bibitem[Bishop \latin{et~al.}(1982)Bishop, Ceperley, Frisch, and
  Kalos]{Bishop1982}
Bishop,~M.; Ceperley,~D.; Frisch,~H.~L.; Kalos,~M.~H. Investigations of model
  polymers: Dynamics of melts and statics of a long chain in a dilute melt of
  shorter chains. \emph{J. Chem. Phys.} \textbf{1982}, \emph{76}, 1557\relax
\mciteBstWouldAddEndPuncttrue
\mciteSetBstMidEndSepPunct{\mcitedefaultmidpunct}
{\mcitedefaultendpunct}{\mcitedefaultseppunct}\relax
\EndOfBibitem
\bibitem[Grest and Kremer(1986)Grest, and Kremer]{Kremer1986}
Grest,~G.~S.; Kremer,~K. Molecular dynamics simulation for polymers in the
  presence of a heat bath. \emph{Phys. Rev. A} \textbf{1986}, \emph{33},
  3628\relax
\mciteBstWouldAddEndPuncttrue
\mciteSetBstMidEndSepPunct{\mcitedefaultmidpunct}
{\mcitedefaultendpunct}{\mcitedefaultseppunct}\relax
\EndOfBibitem
\bibitem[Halverson \latin{et~al.}(2013)Halverson, Brandes, Lenz, Arnold, Bevc,
  Starchenko, Kremer, Stuehn, and Reith]{Espressopp}
Halverson,~J.~D.; Brandes,~T.; Lenz,~O.; Arnold,~A.; Bevc,~S.; Starchenko,~V.;
  Kremer,~K.; Stuehn,~T.; Reith,~D. ESPResSo++: A modern multiscale simulation
  package for soft matter systems. \emph{Comput. Phys. Commun.} \textbf{2013},
  \emph{184}, 1129--1149\relax
\mciteBstWouldAddEndPuncttrue
\mciteSetBstMidEndSepPunct{\mcitedefaultmidpunct}
{\mcitedefaultendpunct}{\mcitedefaultseppunct}\relax
\EndOfBibitem
\bibitem[Archer \latin{et~al.}(1995)Archer, Chen, and Larson]{Archer1995}
Archer,~L.~A.; Chen,~Y.~L.; Larson,~R.~G. Delayed slip after step strains in
  highly entangled polystyrene mixtures. \emph{J. Rheol.} \textbf{1995},
  \emph{39}, 519--525\relax
\mciteBstWouldAddEndPuncttrue
\mciteSetBstMidEndSepPunct{\mcitedefaultmidpunct}
{\mcitedefaultendpunct}{\mcitedefaultseppunct}\relax
\EndOfBibitem
\bibitem[Archer \latin{et~al.}(2002)Archer, Sanchez-Reyes, and
  Juliani]{Archer2002}
Archer,~L.~A.; Sanchez-Reyes,~J.; Juliani, Relaxation Dynamics of Polymer
  Liquids in Nonlinear Step Shear. \emph{Macromolecules} \textbf{2002},
  \emph{35}, 10216--10224\relax
\mciteBstWouldAddEndPuncttrue
\mciteSetBstMidEndSepPunct{\mcitedefaultmidpunct}
{\mcitedefaultendpunct}{\mcitedefaultseppunct}\relax
\EndOfBibitem
\bibitem[Venerus and Nair(2006)Venerus, and Nair]{Venerus2006}
Venerus,~D.~C.; Nair,~R. Stress relaxation dynamics of an entangled polystyrene
  solution following step strain flow. \emph{J. Rheol.} \textbf{2006},
  \emph{50}, 59--75\relax
\mciteBstWouldAddEndPuncttrue
\mciteSetBstMidEndSepPunct{\mcitedefaultmidpunct}
{\mcitedefaultendpunct}{\mcitedefaultseppunct}\relax
\EndOfBibitem
\bibitem[Grosberg(2016)]{Grosberg2016}
Grosberg,~A.~Y. Do knots self-tighten for entropic reasons? \emph{Poly. Sci.
  Ser. A} \textbf{2016}, \emph{58}, 864--872\relax
\mciteBstWouldAddEndPuncttrue
\mciteSetBstMidEndSepPunct{\mcitedefaultmidpunct}
{\mcitedefaultendpunct}{\mcitedefaultseppunct}\relax
\EndOfBibitem
\bibitem[Narsimhan \latin{et~al.}(2016)Narsimhan, Renner, and
  Doyle]{Narsimhan2016}
Narsimhan,~V.; Renner,~C.~B.; Doyle,~P.~S. Jamming of knots along a tensioned
  chain. \emph{ACS Macro Lett.} \textbf{2016}, \emph{5}, 123--127\relax
\mciteBstWouldAddEndPuncttrue
\mciteSetBstMidEndSepPunct{\mcitedefaultmidpunct}
{\mcitedefaultendpunct}{\mcitedefaultseppunct}\relax
\EndOfBibitem
\bibitem[Ferry(1980)]{Ferry1980}
Ferry,~J.~D. \emph{Viscoelastic Properties of Polymers}; Wiley,New York, 1980;
  p 244\relax
\mciteBstWouldAddEndPuncttrue
\mciteSetBstMidEndSepPunct{\mcitedefaultmidpunct}
{\mcitedefaultendpunct}{\mcitedefaultseppunct}\relax
\EndOfBibitem
\bibitem[Milner and McLeish(1998)Milner, and McLeish]{Milner1998}
Milner,~S.~T.; McLeish,~T. C.~B. Reptation and contour-length fluctuations in
  melts of linear polymers. \emph{Phys. Rev. Lett.} \textbf{1998}, \emph{81},
  725\relax
\mciteBstWouldAddEndPuncttrue
\mciteSetBstMidEndSepPunct{\mcitedefaultmidpunct}
{\mcitedefaultendpunct}{\mcitedefaultseppunct}\relax
\EndOfBibitem
\bibitem[McLeish(2002)]{McLeish2002}
McLeish,~T. C.~B. Tube theory of entangled polymer dynamics. \emph{Adv. Phys.}
  \textbf{2002}, \emph{51}, 1379--1527\relax
\mciteBstWouldAddEndPuncttrue
\mciteSetBstMidEndSepPunct{\mcitedefaultmidpunct}
{\mcitedefaultendpunct}{\mcitedefaultseppunct}\relax
\EndOfBibitem
\bibitem[Lee and Kremer(2009)Lee, and Kremer]{Lee2009}
Lee,~W.~B.; Kremer,~K. Entangled polymer melts: relation between plateau
  modulus and stress autocorrelation function. \emph{Macromolecules}
  \textbf{2009}, \emph{42}, 6270\relax
\mciteBstWouldAddEndPuncttrue
\mciteSetBstMidEndSepPunct{\mcitedefaultmidpunct}
{\mcitedefaultendpunct}{\mcitedefaultseppunct}\relax
\EndOfBibitem
\bibitem[Lee \latin{et~al.}(2010)Lee, Halverson, and Kremer]{Lee2010}
Lee,~W.~B.; Halverson,~J.; Kremer,~K. Reply to Commnet on ``Entangled Polymer
  Melts: Relation between Plateau Modulus and Stress Autocorrelation Function".
  \emph{Macromolecules} \textbf{2010}, \emph{43}, 3984\relax
\mciteBstWouldAddEndPuncttrue
\mciteSetBstMidEndSepPunct{\mcitedefaultmidpunct}
{\mcitedefaultendpunct}{\mcitedefaultseppunct}\relax
\EndOfBibitem
\bibitem[Liu \latin{et~al.}(2006)Liu, He, van Ruymbeke, Keunings, and
  Bailly]{Liu2006}
Liu,~C.; He,~J.; van Ruymbeke,~E.; Keunings,~R.; Bailly,~C. Evaluation of
  different methods for the determination of the plateau modulus and the
  entanglement molecular weight. \emph{Polymer} \textbf{2006}, \emph{47},
  4461--4479\relax
\mciteBstWouldAddEndPuncttrue
\mciteSetBstMidEndSepPunct{\mcitedefaultmidpunct}
{\mcitedefaultendpunct}{\mcitedefaultseppunct}\relax
\EndOfBibitem
\bibitem[Luap \latin{et~al.}(2005)Luap, M\"uller, Schweizer, and
  Venerus]{Luap2005}
Luap,~C.; M\"uller,~C.; Schweizer,~T.; Venerus,~D.~C. Simultaneous stress and
  birefringence measurements during uniaxial elongation of polystyrene melts
  with narrow molecular weight distribution. \emph{Rheol Acta} \textbf{2005},
  \emph{45}, 83--91\relax
\mciteBstWouldAddEndPuncttrue
\mciteSetBstMidEndSepPunct{\mcitedefaultmidpunct}
{\mcitedefaultendpunct}{\mcitedefaultseppunct}\relax
\EndOfBibitem
\bibitem[Auhl \latin{et~al.}(2008)Auhl, Ramirez, Likhtman, Chambon, and
  Fernyhough]{Auhl2008}
Auhl,~D.; Ramirez,~J.; Likhtman,~A.~E.; Chambon,~P.; Fernyhough,~C. Linear and
  nonlinear shear flow behavior of monodisperse polyiosprene melts with a large
  range of molecular weights. \emph{J. Rheol.} \textbf{2008}, \emph{52},
  801--835\relax
\mciteBstWouldAddEndPuncttrue
\mciteSetBstMidEndSepPunct{\mcitedefaultmidpunct}
{\mcitedefaultendpunct}{\mcitedefaultseppunct}\relax
\EndOfBibitem
\end{mcitethebibliography}

\end{document}